\documentclass[fleqn,letterpaper,10pt]{article}

\usepackage{graphicx}
\usepackage{amsmath}
\usepackage{fancyhdr}
\usepackage{enumerate}
\usepackage{indentfirst}
\usepackage{xspace}

\fancypagestyle{plain}{
 \fancyhf{}

}

\newcommand{\pionpm}{\ensuremath{\pi^{\pm}}}
\newcommand{\pionm}{\ensuremath{\pi^{-}}}
\newcommand{\pionp}{\ensuremath{\pi^{+}}}

\newcommand{\pip}{\ensuremath{\pi^{+}}\xspace}

\newcommand{\pim}{\ensuremath{\pi^{-}}\xspace}

\newcommand{\ddpi}{\ensuremath{d^2\sigma^{\pi}/dpd\Omega}\xspace}

\newcommand{\rad}{\ensuremath{\mbox{rad}}\xspace}
\newcommand{\GeV}{\ensuremath{\mbox{GeV}}\xspace}

\newcommand{\GeVc}{\ensuremath{\mbox{GeV}/c}\xspace}
\newcommand{\MeVc}{\ensuremath{\mbox{MeV}/c}\xspace}

\newcommand{\mm}{\ensuremath{\mbox{mm}}\xspace}

\newcommand{\mrad}{\ensuremath{\mbox{mrad}}\xspace}

\newcommand{\ps}{\ensuremath{\mbox{ps}}\xspace}

\begin{document}

\title{\bf Measurement of the production cross-sections of 
$\pi^\pm$ \\
in p-C and $\pi^\pm$-C 
interactions
at 12~GeV/c}
\author{HARP Collaboration}
\maketitle

\begin{abstract}

The results of the measurements of the  double-differential production 
cross-sections of pions,\ \ddpi, in p-C and $\pi^\pm$-C interactions
using the forward spectrometer of the HARP experiment are presented.  
The incident particles are 12~\GeVc\ protons 
and
charged pions 
directed onto a carbon target with a 
thickness of  5\% of a nuclear interaction length.  
For p-C interactions the analysis is performed using 
100\,035 reconstructed secondary tracks, 
while the corresponding numbers of tracks for
$\pi^-$-C and $\pi^+$-C analyses are 106\,534 and 10\,122 respectively. 
Cross-section results are presented in the kinematic range
0.5~\GeVc  $\leq p_{\pi} <$ 8~\GeVc 
and 30~mrad $\leq \theta_{\pi} <$ 240~mrad in the laboratory frame.  
The measured cross-sections have a direct impact on the precise
calculation of atmospheric neutrino fluxes and on the improved 
reliability of extensive air shower simulations by reducing 
the uncertainties of hadronic interaction models in the low energy range.  

\end{abstract} 

\clearpage

\thispagestyle{plain}
\begin{center}
{\large HARP collaboration}\\
\newcommand{\afdoct}{{3}\xspace}
\vspace{0.1cm}
{\small
M.G.~Catanesi, 
E.~Radicioni
\\ 
{\bf Universit\`{a} degli Studi e Sezione INFN, Bari, Italy} 
\\
R.~Edgecock, 
M.~Ellis$^{1}$,          
S.~Robbins$^{2,3}$,      
F.J.P.~Soler$^{4}$
\\
{\bf Rutherford Appleton Laboratory, Chilton, Didcot, UK} 
\\
C.~G\"{o}\ss ling 
\\
{\bf Institut f\"{u}r Physik, Universit\"{a}t Dortmund, Germany} 
\\
S.~Bunyatov, 
A.~Krasnoperov, 
B.~Popov$^5$, 
V.~Tereshchenko
\\
{\bf Joint Institute for Nuclear Research, JINR Dubna, Russia} 
\\
E.~Di~Capua, 
G.~Vidal--Sitjes$^{6}$  
\\
{\bf Universit\`{a} degli Studi e Sezione INFN, Ferrara, Italy}  
\\
A.~Artamonov$^7$,   
S.~Giani, 
S.~Gilardoni, 
P.~Gorbunov$^{7}$,  
A.~Grant,  
A.~Grossheim$^{10}$, 
P.~Gruber$^{11}$,    
V.~Ivanchenko$^{12}$,  
A.~Kayis-Topaksu$^{13}$,
J.~Panman, 
I.~Papadopoulos,  
E.~Tcherniaev, 
I.~Tsukerman$^7$,   
R.~Veenhof, 
C.~Wiebusch$^{14}$,    
P.~Zucchelli$^{9,15}$ 
\\
{\bf CERN, Geneva, Switzerland} 
\\
A.~Blondel, 
S.~Borghi$^{16}$,  
M.~Campanelli,       
M.C.~Morone$^{17}$, 
G.~Prior$^{18}$,   
R.~Schroeter
\\
{\bf Section de Physique, Universit\'{e} de Gen\`{e}ve, Switzerland} 
\\
R.~Engel,
C.~Meurer
\\
{\bf Forschungszentrum Karlsruhe, Institut f\"{u}r Kernphysik, Karlsruhe, Germany}
\\
\newcommand{\afkyot}{{20}\xspace}
I.~Kato$^{10,\afkyot}$ 
\\
{\bf University of Kyoto, Japan} %
\\
U.~Gastaldi
\\
{\bf Laboratori Nazionali di Legnaro dell' INFN, Legnaro, Italy} 
\\
\newcommand{\aflanl}{{20}\xspace}
G.~B.~Mills$^{\aflanl}$  
\\
{\bf Los Alamos National Laboratory, Los Alamos, USA} %
\\
J.S.~Graulich$^{21}$, 
G.~Gr\'{e}goire 
\\
{\bf Institut de Physique Nucl\'{e}aire, UCL, Louvain-la-Neuve,
  Belgium} 
\\
M.~Bonesini, 
F.~Ferri
\\
{\bf Universit\`{a} degli Studi e Sezione INFN, Milano, Italy} 
\\
M.~Kirsanov
\\
{\bf Institute for Nuclear Research, Moscow, Russia} 
\\
A. Bagulya, 
V.~Grichine,  
N.~Polukhina
\\
{\bf P. N. Lebedev Institute of Physics (FIAN), Russian Academy of
Sciences, Moscow, Russia} 
\\
V.~Palladino
\\
{\bf Universit\`{a} ``Federico II'' e Sezione INFN, Napoli, Italy} 
\\
\newcommand{\afclmb}{{21}\xspace}
L.~Coney$^{\afclmb}$, 
D.~Schmitz$^{\afclmb}$
\\
{\bf Columbia University, New York, USA} %
\\
G.~Barr, 
A.~De~Santo$^{22}$, 
C.~Pattison, 
K.~Zuber$^{23}$  
\\
{\bf Nuclear and Astrophysics Laboratory, University of Oxford, UK} 
\\
F.~Bobisut, 
D.~Gibin,
A.~Guglielmi, 
M.~Mezzetto
\\
{\bf Universit\`{a} degli Studi e Sezione INFN, Padova, Italy} 
\\
J.~Dumarchez, 
F.~Vannucci 
\\
{\bf LPNHE, Universit\'{e}s de Paris VI et VII, Paris, France} 
\\
U.~Dore
\\
{\bf Universit\`{a} ``La Sapienza'' e Sezione INFN Roma I, Roma,
  Italy} 
\\
D.~Orestano, 
F.~Pastore, 
A.~Tonazzo, 
L.~Tortora
\\
{\bf Universit\`{a} degli Studi e Sezione INFN Roma III, Roma, Italy}
\\
C.~Booth, 
L.~Howlett
\\
{\bf Dept. of Physics, University of Sheffield, UK} 
\\
M.~Bogomilov, 
M.~Chizhov, 
D.~Kolev, 
R.~Tsenov
\\
{\bf Faculty of Physics, St. Kliment Ohridski University, Sofia,
  Bulgaria} 
\\
S.~Piperov, 
P.~Temnikov
\\
{\bf Institute for Nuclear Research and Nuclear Energy, 
Academy of Sciences, Sofia, Bulgaria} 
\\
M.~Apollonio, 
P.~Chimenti,  
G.~Giannini, 
G.~Santin$^{24}$  
\\
{\bf Universit\`{a} degli Studi e Sezione INFN, Trieste, Italy} 
\\
J.~Burguet--Castell, 
A.~Cervera--Villanueva, 
J.J.~G\'{o}mez--Cadenas, 
J. Mart\'{i}n--Albo,
P.~Novella,
M.~Sorel
\\
{\bf  Instituto de F\'{i}sica Corpuscular, IFIC, CSIC and Universidad de Valencia,
Spain} 
}
\end{center}
\thispagestyle{plain}
\vfill
\rule{0.3\textwidth}{0.4mm}
\newline
$^{~1}${Now at FNAL, Batavia, Illinois, USA.}
\newline
$^{~2}$Jointly appointed by Nuclear and Astrophysics Laboratory,
            University of Oxford, UK.
\newline
$^{~3}${Now at Codian Ltd., Langley, Slough, UK.}
\newline
$^{~4}${Now at University of Glasgow, UK.}
\newline
$^{~5}${Also supported by LPNHE, Universit\'{e}s de Paris VI et VII, Paris, France.} 
\newline
$^{~6}${Now at Imperial College, University of London, UK.}
\newline
$^{~7}${ITEP, Moscow, Russian Federation.}
\newline
$^{~8}${Permanently at Instituto de F\'{\i}sica de Cantabria,
            Univ. de Cantabria, Santander, Spain.} 
\newline
$^{~9}${Now at SpinX Technologies, Geneva, Switzerland.}
\newline
$^{10}${Now at TRIUMF, Vancouver, Canada.}
\newline
$^{11}${Now at University of St. Gallen, Switzerland.}
\newline
$^{12}${On leave of absence from Ecoanalitica, Moscow State University,
Moscow, Russia.}
\newline
$^{13}${Now at \c{C}ukurova University, Adana, Turkey.}
\newline
$^{14}${Now at III Phys. Inst. B, RWTH Aachen, Aachen, Germany.}
\newline
$^{15}$On leave of absence from INFN, Sezione di Ferrara, Italy.
\newline
$^{16}${Now at CERN, Geneva, Switzerland.}
\newline
$^{17}${Now at Univerity of Rome Tor Vergata, Italy.}
\newline
$^{18}${Now at Lawrence Berkeley National Laboratory, Berkeley, California, USA.}
\newline
$^{19}${K2K Collaboration.}
\newline
$^{20}${MiniBooNE Collaboration.}
\newline
$^{21}${Now at Section de Physique, Universit\'{e} de Gen\`{e}ve, Switzerland, Switzerland.}
\newline
$^{22}${Now at Royal Holloway, University of London, UK.}
\newline
$^{23}${Now at University of Sussex, Brighton, UK.}
\newline
$^{24}${Now at ESA/ESTEC, Noordwijk, The Netherlands.}

\clearpage

\section{Introduction}\label{sec:intro}

The HARP experiment~\cite{ref:harp-prop,ref:harpTech} at the CERN PS
was designed to make measurements of hadron yields from a large range
of nuclear targets and for incident particle momenta from 1.5~GeV/c to 15~GeV/c.
The main motivations are the measurement of pion yields for a quantitative
design of the proton driver of a future neutrino factory~\cite{ref:nufact}, 
a substantial improvement in the calculation of the atmospheric neutrino
fluxes~\cite{ref:atm_nu_flux}
and the measurement of particle yields as input for the flux
calculation of accelerator neutrino experiments,
such as K2K~\cite{ref:k2k,ref:k2kfinal},
MiniBooNE~\cite{ref:miniboone} and SciBooNE~\cite{ref:sciboone}.

The first HARP physics publication~\cite{ref:alPaper} reported measurements of the
$\pi^+$ production cross-section from an aluminum target 
at 12.9~GeV/c proton momentum. 
This corresponds to the energy of the KEK PS
and the target material used by the K2K experiment.  
The results obtained in Ref.~\cite{ref:alPaper} were
subsequently applied to the final neutrino oscillation analysis 
of K2K~\cite{ref:k2kfinal}, allowing a significant reduction 
of the dominant systematic error associated with the calculation of
the so-called far-to-near ratio 
(see~\cite{ref:alPaper} and~\cite{ref:k2kfinal} for a detailed discussion) 
and thus an increased K2K sensitivity to the oscillation signal. 

Our next goal was to contribute to the understanding of 
the MiniBooNE and SciBooNE neutrino fluxes. 
They are both produced by the Booster Neutrino Beam at 
Fermilab which originates from protons accelerated to 8.9~GeV/c by 
the booster before being collided against a beryllium target. 
As was the case for the K2K beam, a fundamental input for the calculation 
of the resulting 
neutrino
flux is the measurement of the $\pi^+$
production cross-sections 
from a 
beryllium target at 8.9~GeV/c proton momentum, which is presented
in~\cite{ref:bePaper}. 

We have also performed measurements with the HARP detector
of the double-differential cross-section
for $\pi^{\pm}$ production at large angles by
protons in the momentum range of 3--12.9~\GeVc impinging
on different thin 5\% nuclear interaction length ($\lambda_{\mathrm{I}}$) 
targets~\cite{ref:harp:tantalum,ref:harp:carboncoppertin,ref:harp:bealpb}. 
These measurements are of special interest for target materials used in conventional
accelerator neutrino beams and in neutrino factory designs.

In this paper we address one of the other main motivations of the HARP
experiment: the measurement of the yields of positive and negative
pions relevant for a precise calculation of the atmospheric neutrino
fluxes and improved modeling of extended air showers (EAS). 
We present measurements of the double-differential cross-section, 
$
{{\mathrm{d}^2 \sigma^{\pi}}}/{{\mathrm{d}p\mathrm{d}\Omega }}
$
for $\pi^{\pm}$ production (in the kinematic range 
0.5~\GeVc  $\leq p_{\pi} <$ 8~\GeVc and 
30~mrad $\leq \theta_{\pi} <$ 240~mrad) by
protons and 
charged pions
of 12~\GeVc momentum impinging
on a thin carbon target of 5\% 
$\lambda_{\mathrm{I}}$. 
These measurements are performed using the forward spectrometer of the
HARP detector.
HARP results on the measurement 
of the double-differential $\pi^{\pm}$ production
cross-section  in proton--carbon collisions  in the range of pion
momentum $100~\MeVc \leq p_{\pi} < 800~\MeVc$ 
and angle $0.35~\rad \le \theta_{\pi} <2.15~\rad$
obtained with the HARP large-angle spectrometer are presented in 
a separate article~\cite{ref:harp:carboncoppertin}. 

The existing world data for $\pi^{\pm}$ production on light targets 
in the low energy region of incoming beam ($\leq$25~{GeV}) are rather limited.
A number of fixed target measurements with a good phase space coverage
exist for beryllium targets and low energy proton 
beams~\cite{Baker61,Dekkers65,Allaby70,Cho71a,Eichten72,Antreasyan79}. 
However, in general these data are often restricted
to a few fixed angles and have limited statistics.
The work of Eichten et al.~\cite{Eichten72} 
has become a widely used standard reference dataset. 
This experiment used a proton beam with energy of 24~{GeV} 
and a beryllium target. The secondary
particles (pions, kaons, protons) were measured in a broad angular
range (17~{mrad} $<\theta<$ 127~{mrad}) and in momentum
region from 4~{\GeVc} up to 18~{\GeVc}. 
A measurement of inclusive pion production in proton-beryllium 
interactions at 6.4, 12.3, and 17.5~{\GeVc} proton beam momentum has
been published recently by the E910 experiment at BNL~\cite{E910}.
In this work the differential $\pip$ and $\pim$ production
cross-sections have been measured up to 400~{mrad} in $\theta_{\pi}$ 
and up to 6~{\GeVc} in $p_{\pi}$.
We should stress, however, that the data for pion projectiles are
still very scarce.

Carbon is isoscalar and so are nitrogen and
oxygen, so the extrapolation to air is the most straightforward.
Unfortunately, the existing data for a carbon target at low energies are very scarce.
The only measurement of p-C collisions, which was not
limited to a fixed angle, was the experiment done by 
Barton et al.~\cite{Barton83}. These data were collected using the Fermilab
Single Arm Spectrometer facility in the M6E beam line. A proton beam
with a momentum of 100~{GeV/c} and a thin 2\% $\lambda_{\mathrm{I}}$
(1.37~{g/cm$^{2}$}) 
carbon target was used. However, the phase space of the
secondary particles (pions, kaons, protons) covers only a very small
part of the phase space of interest to the calculation 
of the atmospheric neutrino fluxes and to EAS modeling.

Recently the p-C data at 158~{\GeVc} 
provided by the NA49 experiment at CERN SPS in a large acceptance
range have become available~\cite{Alt:2006fr}. 
The relevant data are expected also from 
the MIPP experiment at Fermilab~\cite{MIPP}.
We would like to mention that
the NA61 experiment~\cite{NA61} took first p-C data at 30~{\GeVc} 
in autumn of 2007.
The foreseen measurements of importance for astroparticle physics are
studies of p-C interactions at the incoming beam momenta 
30, 40, 50~{\GeVc} and
$\pi^\pm$-C interactions at 158 and 350~{\GeVc}.

\subsection{Experimental apparatus}\label{subsec:harp_det}

\begin{figure}[t]
\centering
\includegraphics[width=0.8\textwidth]{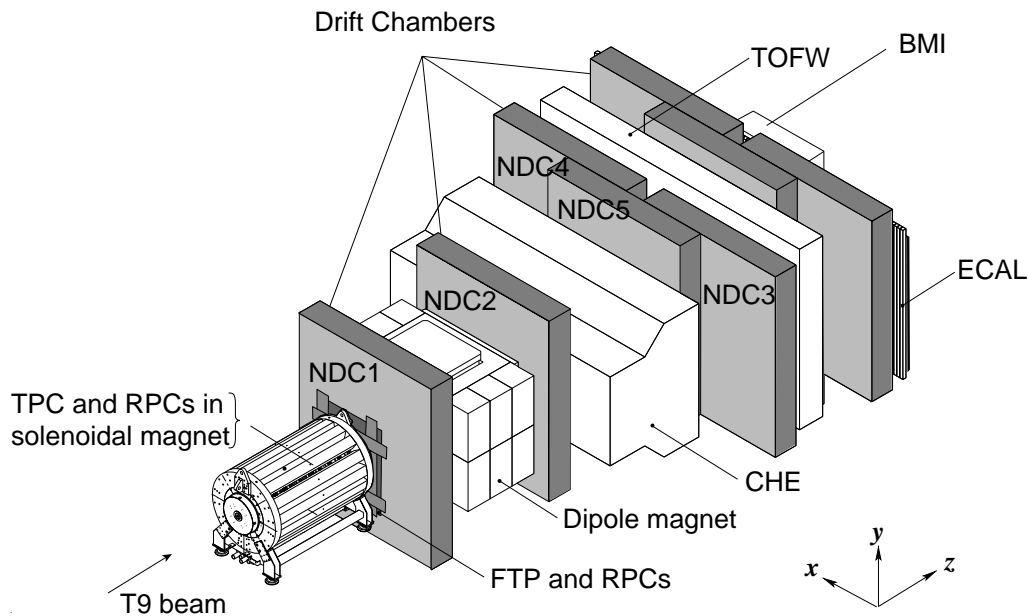} 
\caption{\label{fig:harp} 
Schematic layout of the HARP detector.
The convention for the coordinate system is shown in the lower-right
corner.
}
\end{figure}

 The HARP experiment~\cite{ref:harp-prop,ref:harpTech}
 makes use of a large-acceptance spectrometer consisting of a
 forward and large-angle detection system.
 The HARP detector is shown in Fig.~\ref{fig:harp}.
 A detailed
 description of the experimental apparatus can be found in Ref.~\cite{ref:harpTech}.
 The forward spectrometer is
 based on five modules of large area drift chambers
 (NDC1-5)~\cite{ref:NOMAD_NIM_DC} and a dipole magnet
 complemented by a set of detectors for particle identification (PID): 
 a time-of-flight wall (TOFW)~\cite{ref:tofPaper}, a large Cherenkov detector (CHE) 
 and an electromagnetic calorimeter (ECAL).
 It covers polar angles up to 250~mrad. 
 The muon contamination of the beam is measured with a muon identifier 
 consisting of thick iron absorbers and scintillation counters.
 The large-angle spectrometer -- based on a Time Projection Chamber (TPC) 
 and Resistive Plate Chambers (RPCs)
 located inside a solenoidal magnet --
 has a large acceptance in the momentum
 and angular range for the pions relevant to the production of the
 muons in a neutrino factory 
 (see the corresponding HARP 
 publications~\cite{ref:harp:tantalum,ref:harp:carboncoppertin,ref:harp:bealpb}).
 For the analysis described here we use the forward spectrometer and
 the beam instrumentation.

The HARP experiment, located in the T9 beam of the CERN PS, took data in 2001
and 2002.
The momentum definition of the T9 beam 
is known with a precision of the order of 1\%~\cite{ref:t9}. 

The target is placed inside the inner field cage (IFC) of the TPC.
The cylindrical carbon target used for the measurements reported here
has a purity of 99.99\%, a thickness of
18.94~{mm}, a diameter of 30.26~{mm} and a mass of
25.656~{g}. 
The corresponding
density of the target is
1.88~{g/cm$^{3}$} (for comparison the density of graphite is
2.27~{g/cm$^{3}$}).
The thickness of the carbon target is equivalent to 5\% of a nuclear
interaction length (3.56~{g/cm$^{2}$}).

\begin{figure}
\centering
\includegraphics[width=1.0\textwidth]{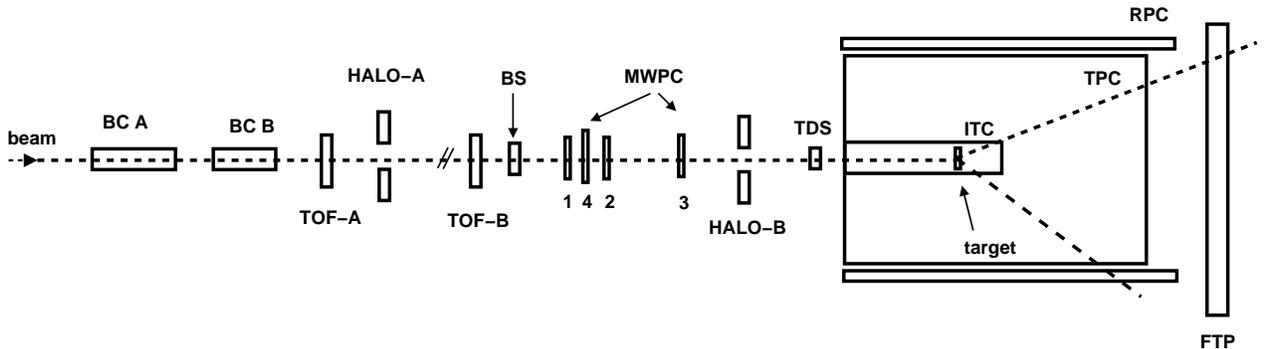} 
\caption{\label{fig:beamline} 
Schematic view of the trigger and beam
equipment. The description is given  in the text. The beam
enters from the left. The MWPCs are numbered: 1, 4, 2, 3 from left to
right. On the right, the position of the target inside the inner field
cage of the TPC is shown.
}
\end{figure}

A sketch of the equipment in the beam line is shown in
Fig.~\ref{fig:beamline}. 
A set of four multi-wire
proportional chambers (MWPCs) measures the position and direction of
the incoming beam particles with an accuracy of $\approx$1~\mm in
position and $\approx$0.2~\mrad in angle per projection.
A beam time-of-flight system (BTOF)
measures the time difference of particles over a $21.4$~m path-length. 
It is made of two
identical scintillation hodoscopes, TOFA and TOFB (originally built
for the NA52 experiment~\cite{ref:NA52}),
which, together with a small target-defining trigger counter (TDS,
also used for the trigger and described below), provide particle
identification at low energies. This provides separation of pions, kaons
and protons up to 5~\GeVc and determines the initial time at the
interaction vertex ($t_0$). 
The timing resolution of the combined BTOF system is about 70~\ps.
A system of two N$_2$-filled Cherenkov detectors (BCA and BCB) is
used to tag electrons at low energies and pions at higher energies. 
The electron and pion tagging efficiency is found to be close to
100\%.
At the beam energy used for this analysis the Cherenkov counters are used
to descriminate between protons and lighter particles, 
while the BTOF is used to reject ions. 

A set of trigger detectors completes the beam instrumentation: a
thin scintillator slab covering the full aperture of the last
quadrupole magnet in the beam line is used to start the trigger logic
decision (BS); a small scintillator disk, TDS mentioned above, positioned
upstream of the target to ensure that only 
particles hitting the target cause a trigger; and `halo' counters
(scintillators with a hole to let the beam particles pass) to veto
particles too far away from the beam axis. 
The TDS is designed to have a very high efficiency (measured to be
99.9\%~\cite{ref:grossheim}).
The trigger signal was formed by a logical OR of four photo-multipliers
which viewed the side of the disk from four sides through
light-guides. 
The distribution of multiplicity of the signals of the four
photo-multipliers could be used to infer the overall efficiency.
It is located as near as possible to the entrance of
the TPC and has a 20~mm diameter, smaller than the target diameter of
30~\mm. 
Its time resolution ($\sim 130 $~\ps) is sufficiently good to be used
as an additional detector for the BTOF system.

A downstream trigger in the forward trigger plane (FTP) 
was required to record the event. The FTP is a double plane of scintillation counters
covering the full aperture of the spectrometer magnet except 
a 60~mm central hole for allowing
non-interacting beam particles to pass.  
The efficiency is measured using tracks recognized by the pattern
recognition in the NDC's in a sample of events taken with a
beam-trigger only and with a trigger based on signals in the Cherenkov
detector. 
Accepting only tracks with a trajectory outside the central hole, the
efficiency of the FTP is measured to be $>$99.8\%.

The track reconstruction and particle identification algorithms as well as 
the calculation of reconstruction efficiencies are described in details 
in~\cite{ref:alPaper,ref:bePaper,ref:pidPaper}.

\subsection{Experimental techniques for the HARP forward spectrometer}

A detailed description of established experimental techniques 
for the data analysis in the HARP forward spectrometer 
can be found in Ref.~\cite{ref:alPaper,ref:pidPaper}.

With respect to our first published paper on pion production in
p--Al interactions~\cite{ref:alPaper}, a number of 
improvements to the analysis techniques 
and detector simulation have been made. 
The present results are based on the same event reconstruction as
described in Ref.~\cite{ref:bePaper}.
The most important improvements introduced in this analysis 
compared with the one presented in Ref.~\cite{ref:alPaper} are:
\begin{itemize}
\item An increase 
of the track reconstruction efficiency which is now constant over a
much larger kinematic range
and a better momentum resolution 
coming from improvements in the tracking algorithm. 
\item Better understanding of the momentum scale and resolution of the detector, 
based on data, which was then used to tune the simulation.
The empty-target data (which is used as a ``test
      beam'' exposure for the dipole spectrometer), elastic scattering
      data using a liquid hydrogen target and a method of comparison with the
      measurement of the particle velocity in the TOFW were used to
      study the momentum calibration. 
      This results in smaller systematic errors associated with the
      unsmearing corrections  
      determined from the Monte Carlo simulation. 
\item New 
particle identification hit selection algorithms both in the 
TOFW and in the 
CHE resulting in much reduced background and negligible efficiency
      losses.  
The PID algorithms developed for the HARP forward spectrometer are
described in details in Ref.~\cite{ref:alPaper,ref:pidPaper} and the recent
improvements are reported in Ref.~\cite{ref:bePaper}. In the kinematic
range of the current analysis, the pion identification efficiency is
about 98\%, while the background from mis-identified protons is well
below 1\%. 
\item Significant increases in Monte Carlo production  
have also reduced uncertainties from Monte Carlo statistics 
and allowed studies which have reduced certain systematics.
\end{itemize}

Further details of the improved analysis techniques can be found in~\cite{ref:bePaper}.
For the current analysis we have used identical reconstruction and PID
algorithms, while at the final stage of the analysis 
the unfolding technique introduced as UFO in~\cite{ref:alPaper} has been
applied. 
The application of this technique has already been described 
in Ref.~\cite{ref:harp:tantalum}.

The absolute normalization of the number of incident protons was
performed using `incident-proton' triggers. 
These are triggers where the same selection on the beam particle was
applied but no selection on the interaction was performed.
The rate of this trigger was down-scaled by a factor 64.

The muon contamination in the beam was measured by 
the beam muon identifier (BMI)
located downstream of the calorimeter (see Fig.~\ref{fig:harp}). 
The BMI is a 1.40~m wide structure placed in the horizontal
direction asymmetrically with respect to the beamline, in order to intercept
all the beam muons which are horizontally
deflected by the spectrometer magnet.
It consists of a passive 0.40~m layer
of iron followed by an iron-scintillator sandwich
with five planes of
six scintillators each, read out at both sides, giving a total of
$6.4 \ \lambda_{\mathrm{int}}$.
 
In section~\ref{Analysis} we present the analysis procedure.
Physics results are presented in section~\ref{sec:results}
together with a discussion on
the relevance of these results
to atmospheric neutrino flux calculations and to extensive air shower
simulations.
Finally, a summary is presented in section~\ref{sec:conclusions}.

\section{Analysis of charged pion production in p-C and {\pionpm}-C interactions}
\label{Analysis}

\subsection{Data selection}

The datasets used for the 
measurements
of the production cross-sections of positive and negative pions in p-C and
\pionpm-C interactions at 12~{\GeVc} were taken during two short run
periods (only two days long for each beam polarity) 
in June and September 2002. Over one
million events with positive beam and more than half a million events
with negative beam were collected. For detailed event statistics see
Table~\ref{eventstat}.

\subsubsection{Beam particle selection and interaction selection}

At the first stage of the analysis a favoured beam particle type is selected
using the beam time of flight system (TOF-A, TOF-B) and the Cherenkov
counters (BCA, BCB) as described in section~\ref{subsec:harp_det}.
A value of the pulse height consistent with the pedestal in both beam
Cherenkov detectors distinguishes protons from electrons and pions.
We also ask for time measurements in TOF-A, TOF-B and/or TDS 
which are needed for calculating 
the arrival time of the beam proton at the target. 
The beam TOF system is used to reject ions, such as deuterons, but at
12~\GeVc is not used to separate protons from pions.

The set of criteria for selecting beam protons for this analysis is as follows:
we require ADC counts to be less than 130 in BCA and less than 125 in BCB
(see~\cite{ref:harpTech,ref:alPaper} for more details).
The beam pions are selected by applying cuts on the ADC counts in BCA and BCB 
to be outside the range accepted for protons in
both Cherenkov counters.

In the 12~\GeVc beam setting the nitrogen pressure in the beam Cherenkov
counters was too low for kaons to be above the threshold.
Kaons are thus a background to the proton sample.
However, the fraction of kaons has been measured in the 12.9~\GeVc beam
configuration which is expected to be very similar to the beam used in
the present measurement.
In the 12.9~\GeVc beam the fraction of kaons compared to protons was
found to be 0.5\%.
Electrons radiate in the Cherenkov counters and would be counted as
pions. 
In the 3~\GeVc beam electrons are identified by both BCA and BCB, since
the pressure was such that pions remained below threshold.
In the 5~\GeVc beam electrons could be tagged by BCB only; in BCA pions
were already above threshold.
The $e/\pi$ fraction was measured to be 1\% in the 3~\GeVc beam and $<
10^{-3}$ in the  5~\GeVc beam.
By extrapolation from the lower-energy beam settings this electron
contamination can be estimated to be negligible ($< 10^{-3}$).

In addition to the momentum-selected beam of protons and pions originating
from the T9 production target one expects also the presence of muons
from pion decay both downstream and upstream of the beam momentum selection.  
Therefore, precise absolute knowledge of the pion rate incident on the
HARP targets is required when measurements 
of particle production with incident pions are performed. 
The particle identification detectors in the beam do not distinguish
muons from pions. 
A separate measurement of the muon component has been performed
using datasets without target (``empty-target datasets'')
both for Monte Carlo and real data.  
Since the empty-target data were taken with the same beam parameter
settings as the data taken with targets, the beam composition can be
measured in the empty-target runs and then used as an overall correction
for the counting of pions in the runs with targets. 

Muons are recognized by their longer range in the BMI.
The punch-through background in the BMI is measured counting the
protons (identified with the beam detectors) thus mis-identified as muons
by the BMI. 
A comparison of the punch-through rate between simulated incoming pions
and protons was used to determine a correction for the difference
between pions and protons and to determine the systematic error.
This difference is the dominant systematic error in the beam composition
measurement. 
The aim was to determine the composition of the beam as it strikes the
target, thus muons produced in pion decays after the HARP target
should be considered as a background to the measurement of muons in the
beam. 
The rate of these background muons, which depends mainly on the total 
inelastic cross-section and pion decay,  was calculated by a Monte Carlo 
simulation using GEANT4~\cite{ref:geant4}.
The muon fraction in the beam (at the target) is obtained taking
into account the efficiency of the BMI selection criteria as well as the
punch-through and decay backgrounds.
The result of this analysis for the 12~\GeVc beam is
$R=\mu/(\mu+\pi) = (2.8 \pm 1.0)\%$, 
where the quoted error includes both statistical and systematic errors. 

Summarizing, the purity of the proton beam is better than 99\%, with the
main background formed by kaons estimated to be 0.5\%.
This impurity is neglected in the analysis.
The pion beam has a negligible electron contamination and a muon
contamination of almost 3\%.
The muon contamination is taken into account in the normalization of the
pion beam.

\begin{figure}
\centering
\includegraphics[width=0.495\textwidth]{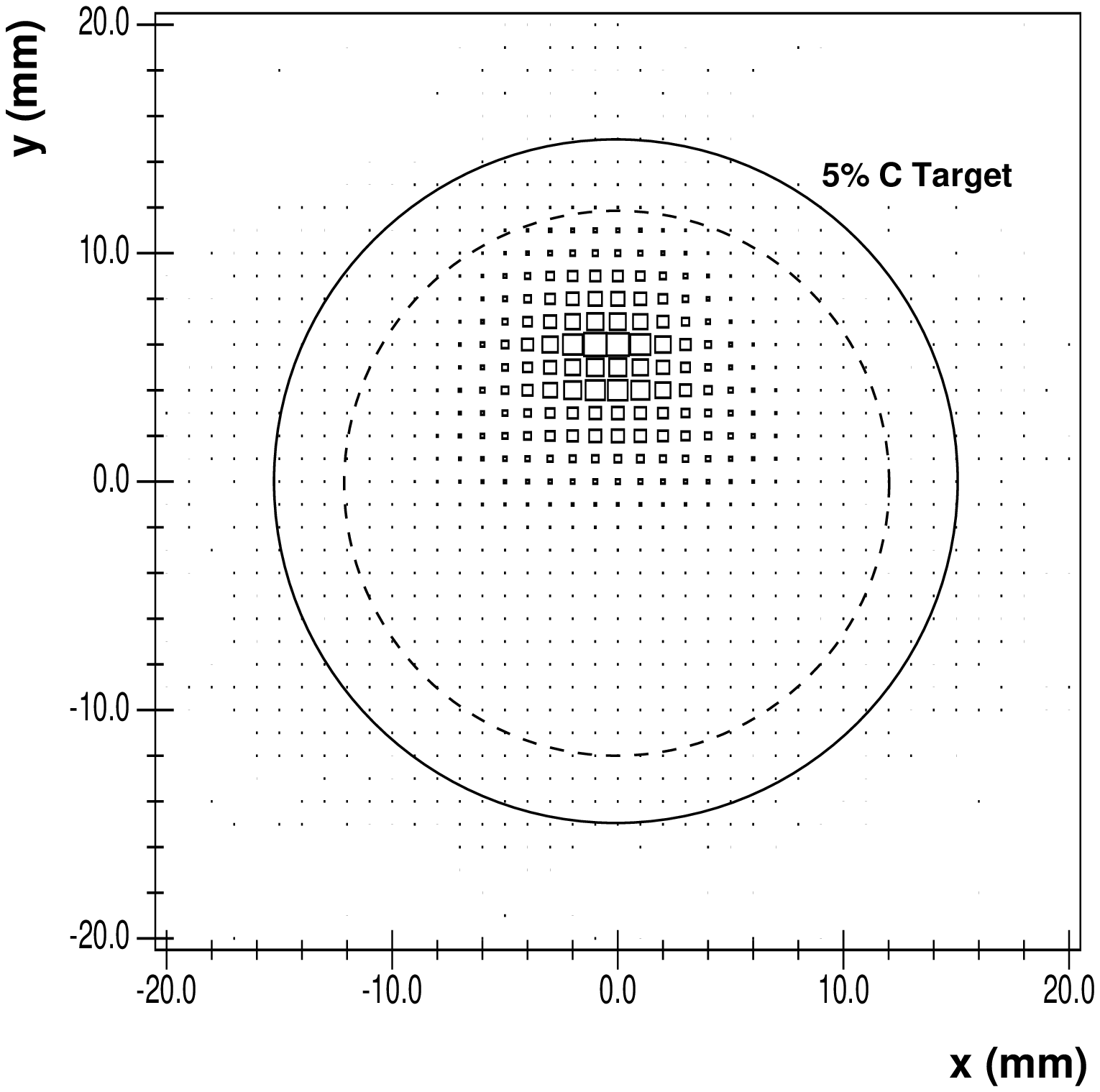} 
\includegraphics[width=0.495\textwidth]{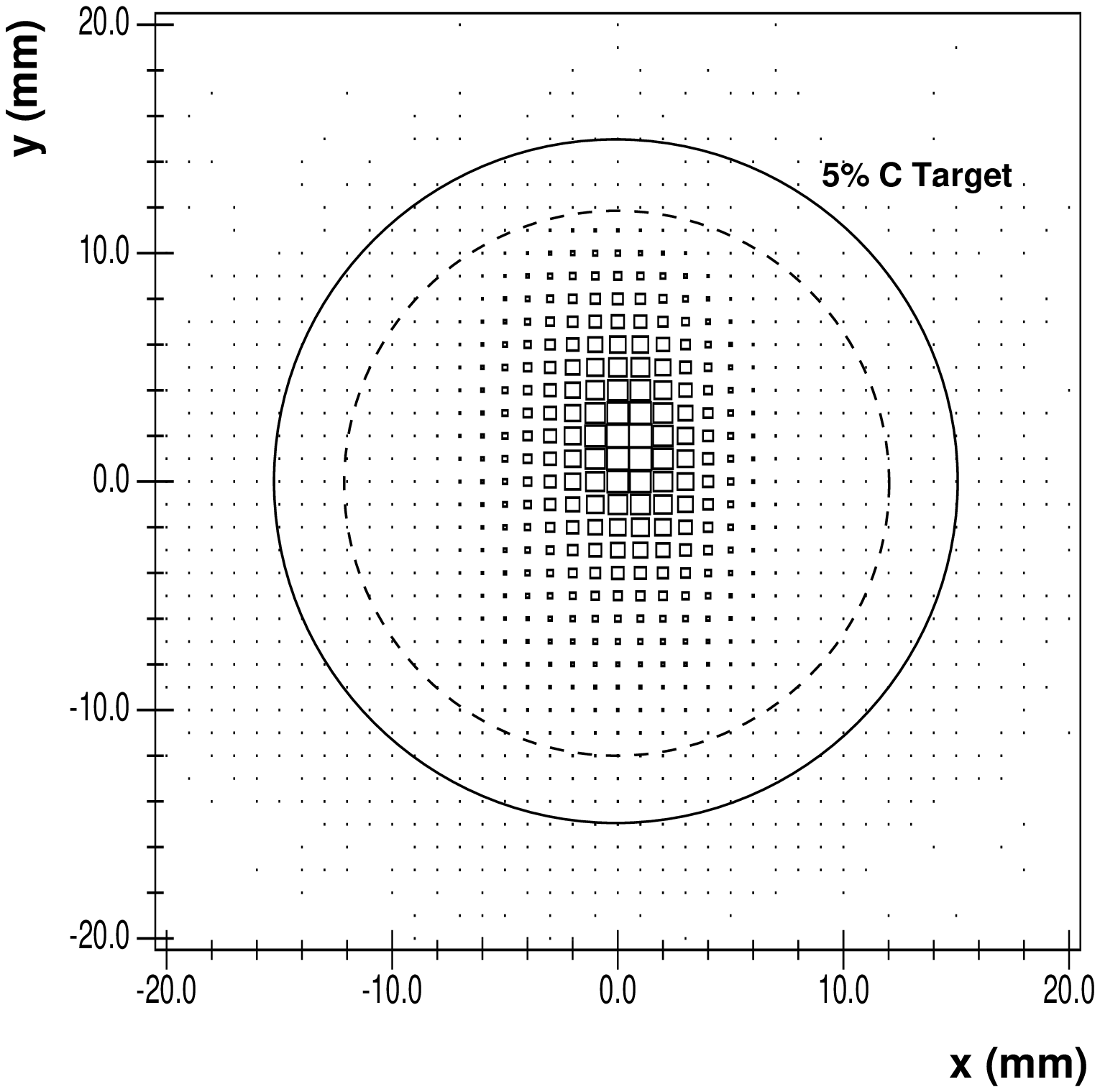}\begin{picture}(0,0)
\put(-390,220){{\large \bf Positive beam}}
\put(-150,220){{\large \bf Negative beam}}
\end{picture} 
\caption{\label{evtxy} Reconstructed position of positively (left
  panel) and negatively (right panel) charged beam particles at the
  target plane. The solid circle gives the position and size of the
  carbon target (diameter: 30.26~mm), the dashed circle indicates the
  region which corresponds to the accepted beam particles (diameter: 24~mm).}
\end{figure}

The distribution of the position of beam particles 
reconstructed in MWPCs and
extrapolated to the
target is shown in Fig.~\ref{evtxy}. The position of the positive-charge
selected beam is shifted by about 5~{mm} in the $y$-direction with 
respect to the nominal position ($x=0; y=0$) and covers a circular area
of about 8~{mm} in diameter. In the case of negatively charged beam
particles the beam hits the target more centrally but it has a broader
distribution of about 14~{mm} width in the $y$-direction. 
The distributions shown in Fig.~\ref{evtxy} are obtained using
``unbiased'' beam triggers where the requirement of the TDS hit and the
veto in the halo counters are not applied.
Also no requirement on an interaction seen in the spectrometers was
made. 
Under these conditions the full width of the beam is recorded including
particles which would not hit the target.
The latter are removed by the standard selection criteria.
To keep the
selection efficiency high and 
to exclude interactions
at the target edge only the beam particles within a radius of 12~{mm} 
with respect to the nominal beam axis
are accepted for the analysis. 
In addition, the MWPC track is required to have a
measured direction within 5~{mrad} of the nominal beam direction to
further reduce halo particles. After these criteria the remaining
number of events for datasets with positive and negative beam are
summarized in Table~\ref{eventstat}.
At 12~{\GeVc} the negative beam consists only of $e^-$ and $\pi^-$
(with a dominant fraction of $\pi^-$), 
while the positive beam is dominated by protons 
(with a small admixture of $\pi^+$). This explains a significantly 
different statistics of the $\pionm$-C and $\pionp$-C datasets.
Note that in the analysis the measured beam profiles are used 
in the MC simulations.

\begin{table}[t]
\centering
\caption{\label{eventstat} 
Total number of events in 12~{\GeVc} carbon target and empty target
datasets 
and in corresponding Monte Carlo simulations (see section~\ref{ufo}).
The total number also includes triggers taken for normalization,
 calibration and for cross-section measurements in the large-angle
 spectrometer.  
} 
\begin{tabular}{ l r r r }
\hline  
\multicolumn{1}{c}{\bf Selection} & \multicolumn{1}{c}{\bf Carbon
 data} &  \multicolumn{1}{c}{\bf Empty target data} &
 \multicolumn{1}{c}{\bf Monte Carlo} \\   
\hline  
Positive beam           &  1\,062 k  &   886 k       &   
\\

p-C        &   467 k  &  287 k       &   20.3 M   \\
$\pionp$-C  &    40 k  &   25 k       &   20.8 M   \\
\hline  
Negative beam           &    646 k  &  531 k       &   
\\

$\pionm$-C  &   350 k  &   214 k       &   20.8 M   \\

\hline
\end{tabular}
\end{table}

\subsubsection{Secondary track selection}

Secondary track selection criteria are optimized to ensure the quality
of momentum reconstruction and a clean time-of-flight measurement
while maintaining a high reconstruction efficiency. There are two
kinds of acceptance criteria concerning the track reconstruction
quality and the 
characteristics of the tracks relative to
the geometry of the forward spectrometer. These criteria are described in what
follows and a summary of track statistics for the three different datasets 
(p-C, $\pi^+$-C, $\pi^-$-C) is given in
Table~\ref{trackstat}. About 5\% to 6\% of all reconstructed tracks 
in accepted events are used for the final analysis. 
The sample of reconstructed tracks contains also large-angle and/or
low momentum tracks which are only seen in
the drift chamber module upstream of the dipole magnet.

The following reconstruction quality criteria have been applied:
\begin{itemize}
\item Successful momentum reconstruction of secondary particle
  (momentum estimator $p_2$, see Ref.~\cite{ref:alPaper} for details). 
      The above momentum measurement is obtained by extrapolating the
      segment of the track  downstream of the dipole magnet to the point
      defined by the position where the beam particle track traverses
      the longitudinal mid-plane of the target.
      Thus the position of the hits measured in the upstream drift
      chamber (NDC1) is not used for the momentum reconstruction.

\item More than three hits on the track in NDC2 and at least five hits
      in a road around the particle 
      trajectory~\footnote{The algorithm looks 
	for drift chamber hits in a tube
      around the trajectory and places a cut on the matching $\chi^2$.} in one of 
      the drift chamber modules NDC3, 4, or 5 or at least three hits on
      the track in one of 
      the modules NDC3, 4, 5 and more than five hits in a road around
      the particle trajectory in NDC2. 

\item A loose criterion requiring more than three hits in a road around
      the trajectory in NDC1 and average $\chi^2 \leq 
      30$ for these hits with respect to the track in NDC1 in order to
      reduce non-target interaction backgrounds. 

\item The track has a matched TOFW hit. 
      Hits are matched based on the $\chi^2$ of the extrapolation of the
      trajectory to the TOFW.
      When more particles share the same TOFW hit, the hit is assigned
      to the track with the best matching $\chi^2$.
      When more TOFW hits are consistent with the trajectory, the one
      with the earliest time measurement is chosen.
      Hits have to pass a minimum pulse height requirement in the
      photo-multipliers on both ends of the scintillator to be
      accepted.

\end{itemize}

\begin{table}[t]
\centering
\caption{\label{trackstat} 
Number of tracks in accepted events before and after the selection
criteria for secondary tracks are applied. About 5\% to 6\% of all
tracks are used for the final analysis.} 
\begin{tabular}{ c r r }
\hline  
\multicolumn{1}{c}{\bf Selection } & \multicolumn{1}{c}{\bf 
 Number of reconstructed tracks} &  \multicolumn{1}{c}{\bf Number of selected tracks}  \\ 
\hline
p-C        &   2\,057\,420 \hspace{0.5cm} &  100\,035 \hspace{0.5cm}  \\
$\pi^+$-C  &     192\,976 \hspace{0.5cm} &   10\,122 \hspace{0.5cm}  \\
$\pi^-$-C  &   1\,701\,041 \hspace{0.5cm} &  106\,534 \hspace{0.5cm}  \\
\hline
\end{tabular}
\end{table}

\begin{figure}
\centering
\includegraphics[width=0.495\textwidth]{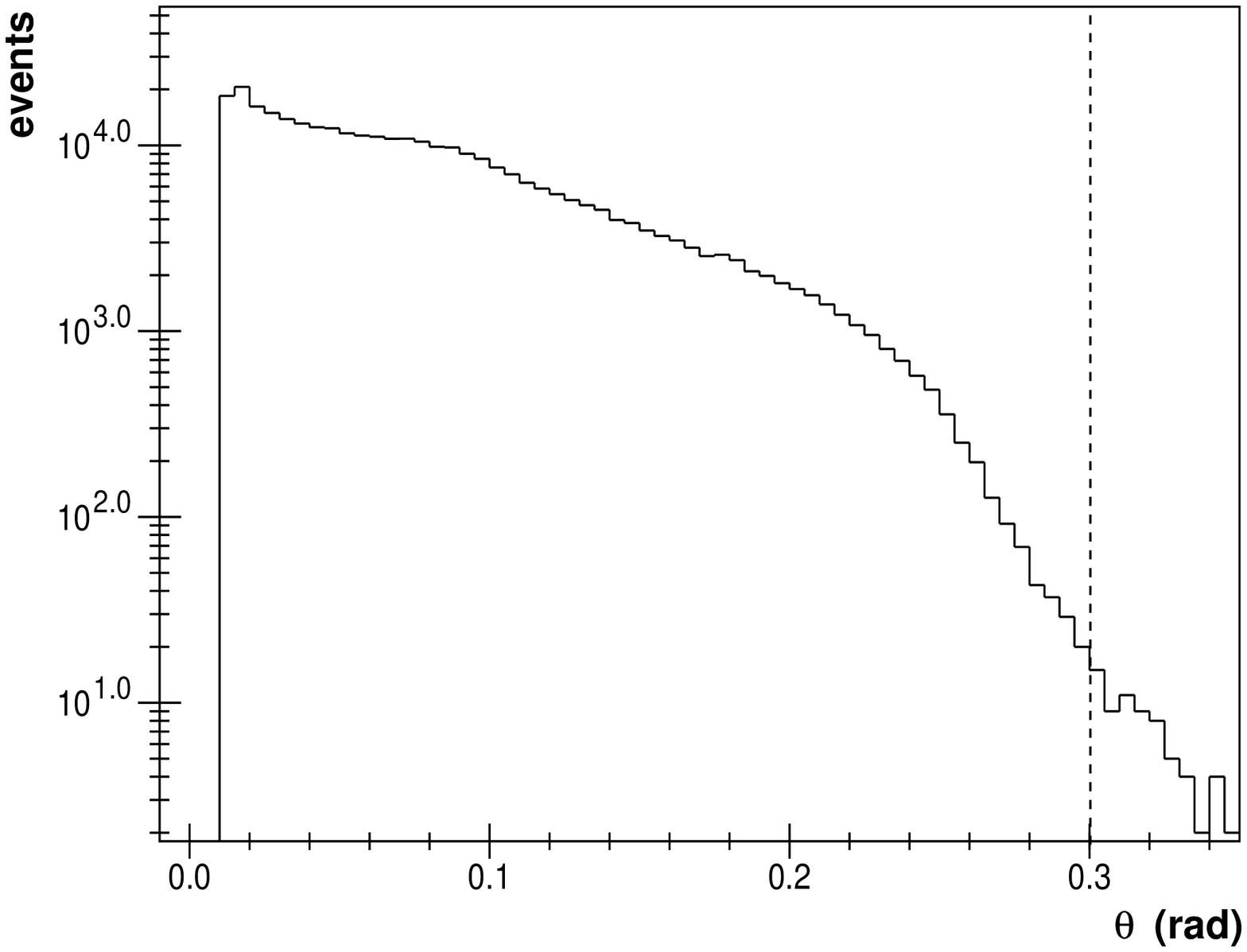} 
\includegraphics[width=0.495\textwidth]{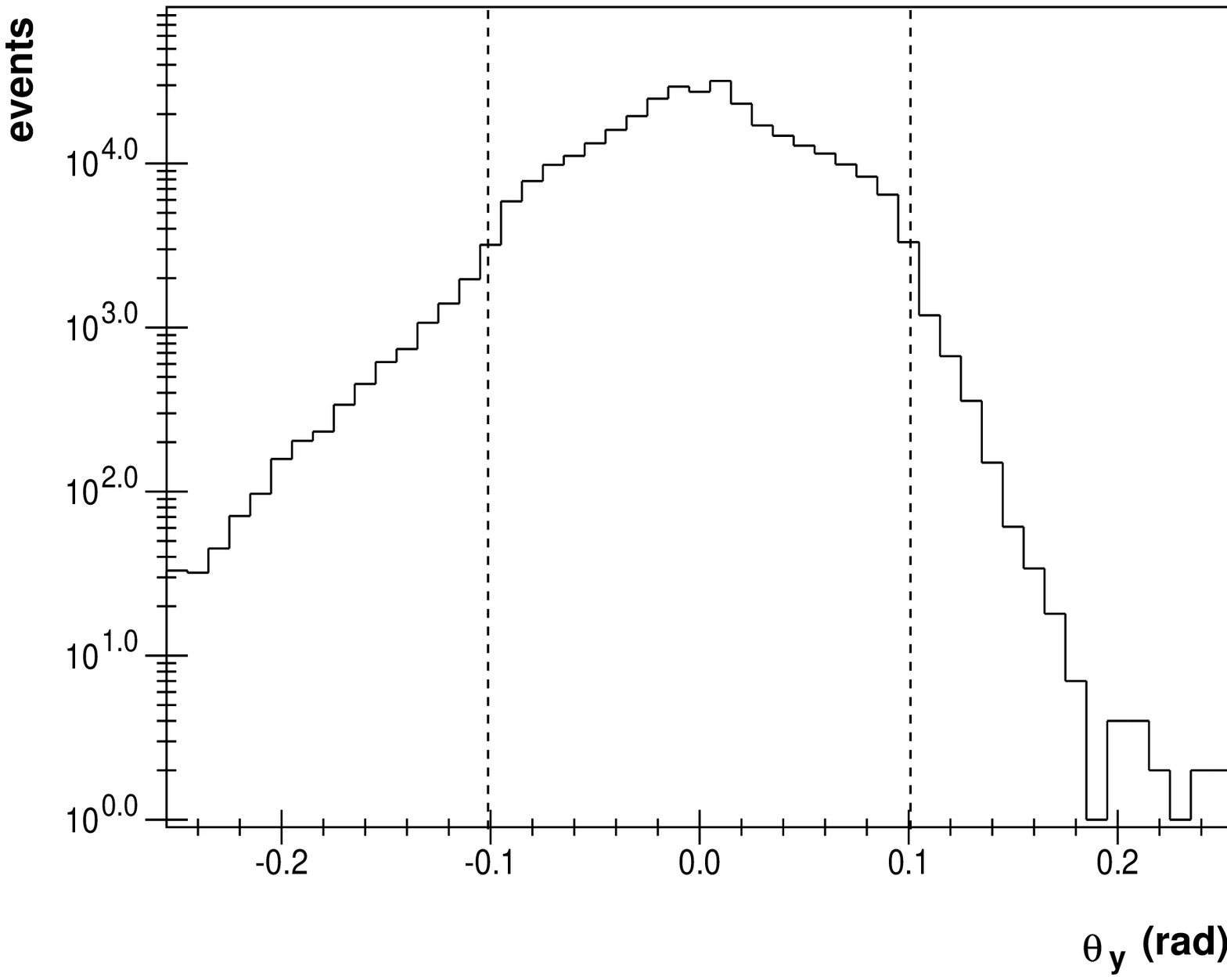} 
\caption{\label{theta} Distribution of $\theta$ (left panel) and
  $\theta_y$ (right panel) for reconstructed tracks. The acceptance criteria for these
  observables are indicated by dashed lines.}
\end{figure}

\begin{figure}
\centering
\includegraphics[width =0.9\textwidth]{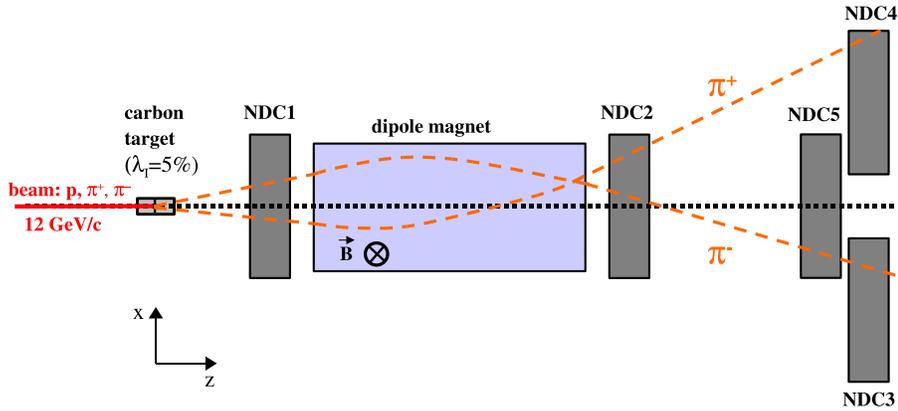}
\caption{\label{tracking4} 
Top view of the HARP forward spectrometer.
Only tracks which bend towards the beam axis are accepted. This
means the product of charge and $\theta_x$ must be negative.
}
\end{figure}

The criteria on track geometry are:
\begin{itemize}

\item The angle $\theta$ of a secondary particle with respect to the
  beam axis is required to be less than 300~{mrad}. The distribution
  of $\theta$ is shown in Fig.~\ref{theta} (left panel).
      Only tracks with $\theta < 240~\mrad$ are retained in the final
      analysis.  

\item The $y$-component $\theta_y$ of the angle $\theta$ is required
  to be between $-100$~{mrad} and 100~{mrad}, see Fig.~\ref{theta} (right
  panel).
      This cut is imposed by the vertical dipole magnet
      aperture\footnote{In previous publications, the more conservative requirement
      $-80~\mrad \leq \theta_y \leq 80~\mrad$ was applied.  No
      degradation of efficiency, momentum resolution and PID performance
      was observed in the larger vertical angle acceptance region.}.

\item The extrapolation of a secondary track should point to the
  nominal beam axis
  on the target plane within a radius of
  200~{mm}. 

\item Only tracks which bend towards the beam axis are accepted
  as shown in Fig.~\ref{tracking4}. This is the case if the product of
  charge and $\theta_x$ is negative. This criterion is applied to avoid
  the positive $\theta_x$ region for positively charged secondary
  particles and the negative  $\theta_x$ region for negatively charged
  particles where the efficiency is momentum dependent due to the
  defocusing effect of the dipole magnet (see~\cite{ref:alPaper}
  for more details). 

\end{itemize}

\subsection{Empty target subtraction}
\label{emptytarget}

There is a background induced by
interactions of beam particles in the materials outside the target.
This background is measured experimentally by taking data without a
target in the target holder. 
These measurements are called ``empty target data''. 
The ``empty target data'' are also subject to the event and track
selection criteria like the standard carbon datasets. The event
statistics of these data 
samples
are given in Table~\ref{eventstat}. 

To take into account this background the number of particles of the
observed 
type (\pionp, \pionm) in the ``empty target data''
are subtracted bin-by-bin (momentum and angular bins) from the number
of particles of the same type in the carbon data. 
The average empty-target subtraction amounts to $\approx$20\%.
The uncertainty
induced by
this method is discussed in section~\ref{errorest} and
labeled ``empty target subtraction''. 

\subsection{Calculation of cross-section}

The goal of this analysis is to measure the double-differential
inclusive production cross-section of negative and positive pions in p-C,
\pionp-C and \pionm-C interactions at 12~{\GeVc} in a broad range of
secondary pion momentum and angle. The cross-section is calculated as follows
\begin{eqnarray}
\frac{d^2 \sigma^{\alpha}}{dp d\Omega}(p_i,\theta_j) & = & 
\frac{A}{N_A \rho t} \cdot \frac{1}{N_{\rm pot}} \cdot \frac{1}{\Delta p_i \Delta \Omega_j} \cdot 
\sum_{p'_i,\theta'_j,\alpha'} \mathcal{M}^{\rm cor}_{p_i\theta_j\alpha p'_i\theta'_j\alpha'} \cdot 
N^{\alpha'}(p'_i,\theta'_j)\hspace{0.1cm},
\end{eqnarray} 
where 
\begin{itemize}
\item $\frac{d^2 \sigma^{\alpha}}{dp d\Omega}(p_i,\theta_j)$ is the
  cross-section in mb/(\GeVc sr) for the particle type $\alpha$ (p,
  \pionp or \pionm) for each momentum and angle bin ($p_i,\theta_j$)
  covered in this analysis;
\item $N^{\alpha'}(p'_i,\theta'_j)$  is the number of particles of
  type $\alpha$ in bins of reconstructed momentum $p'_i$ and angle
  $\theta_j'$ in the raw data after empty target subtraction;
\item $\mathcal{M}^{\rm cor}_{p\theta\alpha p'\theta'\alpha'}$ is the
  correction matrix which accounts for efficiency and resolution of
  the detector;
\item $\frac{A}{N_A \rho t}$, $\frac{1}{N_{\rm pot}}$ and
  $\frac{1}{\Delta p_i \Delta \Omega_j}$ are normalization factors,
  namely:
\subitem $\frac{N_A \rho t}{A}$ is the number of target nuclei per unit area 
\footnote{$A$ - atomic  mass, $N_A$ - Avogadro number, $\rho$ - target
  density and $t$ - target thickness};
\subitem $N_{\rm pot}$ is the number of incident beam particles on
  target (particles on target);
\subitem $\Delta p_i $ and $\Delta \Omega_j $ are the bin sizes in
  momentum and solid angle, respectively 
\footnote{$\Delta p_i = p^{\rm max}_i-p^{\rm min}_i$,\hspace{0.2cm}
  $\Delta \Omega_j = 2 \pi (\cos(\theta^{\rm min}_j)- 
  \cos(\theta^{\rm max}_j))$}.
\end{itemize}
We do not make a correction for the attenuation
of the proton beam in the target, so that strictly speaking the
cross-sections are valid for a $\lambda_{\mathrm{I}}=5\%$ target.

\subsection{Calculation of the correction matrix}
\label{ufo}

A calculation of the
correction matrix $M^{\rm cor}_{p_i\theta_j\alpha
  p'_i\theta'_j\alpha'}$ is 
a rather difficult task.
Various techniques are
described in the literature to obtain this matrix. As 
discussed 
in Ref.~\cite{ref:alPaper} for the p-Al analysis of HARP data at 12.9~{\GeVc}, two
complementary analyses have been performed to cross-check internal
consistency and possible biases in the respective procedures.
A comparison of both analyses shows that the results are consistent
within the overall systematic error~\cite{ref:alPaper}.

In the first method -- called ``Atlantic'' in~\cite{ref:alPaper} -- 
the correction matrix $M^{\rm
  cor}_{p_i\theta_j\alpha p'_i\theta'_j\alpha'}$ is decomposed into
distinct independent contributions, which are computed mostly using
the data themselves.
The second method -- called UFO in~\cite{ref:alPaper} -- 
is the unfolding method introduced 
by D'Agostini~\cite{ref:DAgostini}. 
It is based on the Bayesian unfolding technique.
In this case a simultaneous (three dimensional) unfolding of
momentum $p$, angle $\theta$ and particle type $\alpha$ is
performed. The correction matrix is computed using a Monte Carlo
simulation. This method has been used in recent HARP 
publications~\cite{ref:harp:tantalum,ref:harp:carboncoppertin,ref:harp:bealpb} 
and it is also applied in the analysis described here 
(see~\cite{ref:christine_phd} for additional information).

\subsubsection{Unfolding technique}

Caused by various error sources (biases and resolutions) 
and limited acceptance and efficiency
of an experiment, no measured observable represents the ``true'' physical
value. The unfolding method tries to solve this problem and to find the
corresponding true distribution from a distribution in the
measured observable. The main assumption 
is that the probability distribution function in the ``true'' physical
parameters can be approximated by a histogram with discrete bins. Then
the relation between the vector $\vec{x}$ of the true physical
parameter and the vector $\vec{y}$ of the measured observable can be
described by a matrix $\mathcal{M}_{\rm mig}$ which represents the
mapping from the true value to the measured one. This matrix is called
the migration (or smearing) matrix
\begin{equation}
\label{ufoproblem}
\vec{y}  =  \mathcal{M}_{\rm mig} \cdot \vec{x}\hspace{0.3cm}.
\end{equation}  
In our case these $\vec{x}$ and $\vec{y}$ vectors contain particle
momentum, polar angle and particle type.

The goal of the unfolding procedure is to determine a transformation
for the measurement to obtain the expected values for $\vec{x}$ using
the relation~(\ref{ufoproblem}), see e.g.~\cite{ref:Blobel}. The most simple and
obvious solution is the matrix inversion. But this method often provides
unstable
results. Large correlations between bins lead to large
off-diagonal elements in the migration matrix $\mathcal{M}_{\rm mig}$
and, thus, the result is dominated by very large variances and strong
negative correlation between neighbouring bins.
 
In the method of D'Agostini~\cite{ref:DAgostini}, 
the unfolding is performed by the
calculation of the unfolding matrix 
$\mathcal{M}^{\rm UFO}=\mathcal{M}^{\rm cor}$ 
in an iterative way which is used instead of 
$\mathcal{M}^{-1}_{\rm mig}$. Here $\mathcal{M}^{\rm UFO}$ is a
two-dimensional matrix connecting the measurement space (effects) with
the space of the true values (causes). Expected causes and measured
effects are represented by one-dimensional vectors with entries
$x_{\rm exp}(C_i)$ and $y(E_j)$ for each cause and effect bin $C_i$
and $E_j$, respectively:
\begin{equation}
\label{ufotechnic}
x_{\rm exp}(C_i)  =  \sum_j \mathcal{M}_{ij}^{\rm UFO} \hspace{0.2cm} y(E_j)\hspace{0.3cm}.
\end{equation}  
The Bayes' theorem provides the conditional probability $P(C_i|E_j)$
for effect $E_j$ to be caused by cause $C_i$ 
\begin{equation}
\label{probability}
P(C_i|E_j) 
=
P(E_j|C_i) \cdot P(C_i)  \hspace{0.3cm},
\end{equation} 
where $  P(E_j|C_i)$ is the probability for cause $C_i$ to produce
effect $E_j$ which corresponds to the migration matrix and could be
calculated from Monte Carlo, $P(C_i)$ is the probability for cause
$C_i$ to happen. The Eq.~(\ref{probability}) is solved in an iterative
process. The initial probability $P_0(C_i)$ could be assumed to be a
uniform distribution. The $P(C_i|E_j)$ found is used as the unfolding
matrix in the first interaction step and leads to a first estimation of
the expected values for causes 
\begin{equation}
\label{ufosolve}
x_{\rm exp}(C_i) = \sum_j P(C_i|E_j) \hspace{0.2cm} y(E_j) \hspace{0.3cm}.
\end{equation}
From $x_{\rm exp}(C_i)$ a new probability $P_1(C_i)$ for cause $C_i$
is calculated and inserted in Eq.~(\ref{probability}) for the next
iteration step. Before this, the distribution of $P_1(C_i)$ can
optionally be smoothed to reduce oscillations due to statistical
fluctuations. Between two consecutive iteration steps a $\chi^2$-test
is applied. The iteration process is terminated when the difference of $\chi^2$ 
between consecutive iteration steps is small. 
This procedure was tested on distributions obtained with simulated data
and verified to yield results consistent with the ``true input''
distributions. 
The final result of this method is
the unfolded distribution of $x_{\rm exp}(C_i)$ and its covariance
matrix. 
We have also checked that starting with flat priors at the first iteration 
does not introduce any biases in the final result.

The original unfolding program provided by D'Agostini is used 
in this analysis: $P_0(C_i)$ is assumed to be a uniform
distribution, while $P(E_j|C_i)$ is calculated from the Monte Carlo
simulation. 
In~\cite{ref:grossheim} it is shown  that smoothing the distribution of
$P_n(C_i)$ before inserting in the next iteration step does not lead
to better (smoother) results than without smoothing. Therefore the smoothing
process is not applied in this analysis.
The process converges and the iterations are stopped 
when the changes are smaller than the errors
(which typically happens after about four iterations). 
The entries of the one-dimensional vectors $\vec{x}$ and $\vec{y}$ as
well as the entries of the two-dimensional matrix 
$\mathcal{M}^{\rm UFO}$ carry the information on angle, momentum and particle type.

The Monte Carlo simulation of the HARP setup is based on 
GEANT4~\cite{ref:geant4}. 
The detector
materials are accurately 
described
in this simulation as well as the
relevant features of the detector response and the digitization
process. All relevant physics processes are considered, including
multiple scattering, energy loss, absorption and
re-interactions. 
The simulation is independent of the beam particle type
because it only generates for each event
exactly one secondary particle of a specific particle type inside the
target material and propagates it through the 
complete detector. Owing to this fact the
same simulation can be used for the three analyses of p-C, \pionp-C
and \pionm-C at 12~{\GeVc}.
A small difference (at the few percent level) is observed between the
efficiency calculated for 
events simulated with the single-particle Monte Carlo and with a
simulation using a multi-particle hadron-production model.
A similar difference is seen between the single-particle Monte Carlo and
the efficiencies measured directly from the data.
A momentum-dependent correction factor determined using the efficiency
measured with the  data is applied to take this into account. 
The track reconstruction used in this analysis and the simulation are
identical to the ones used for the \pionp\ production in p-Be
collisions~\cite{ref:bePaper}. 
A detailed description of the corrections and their magnitude can be
found there. 

\begin{figure}
\centering
\includegraphics[width=0.95\textwidth]{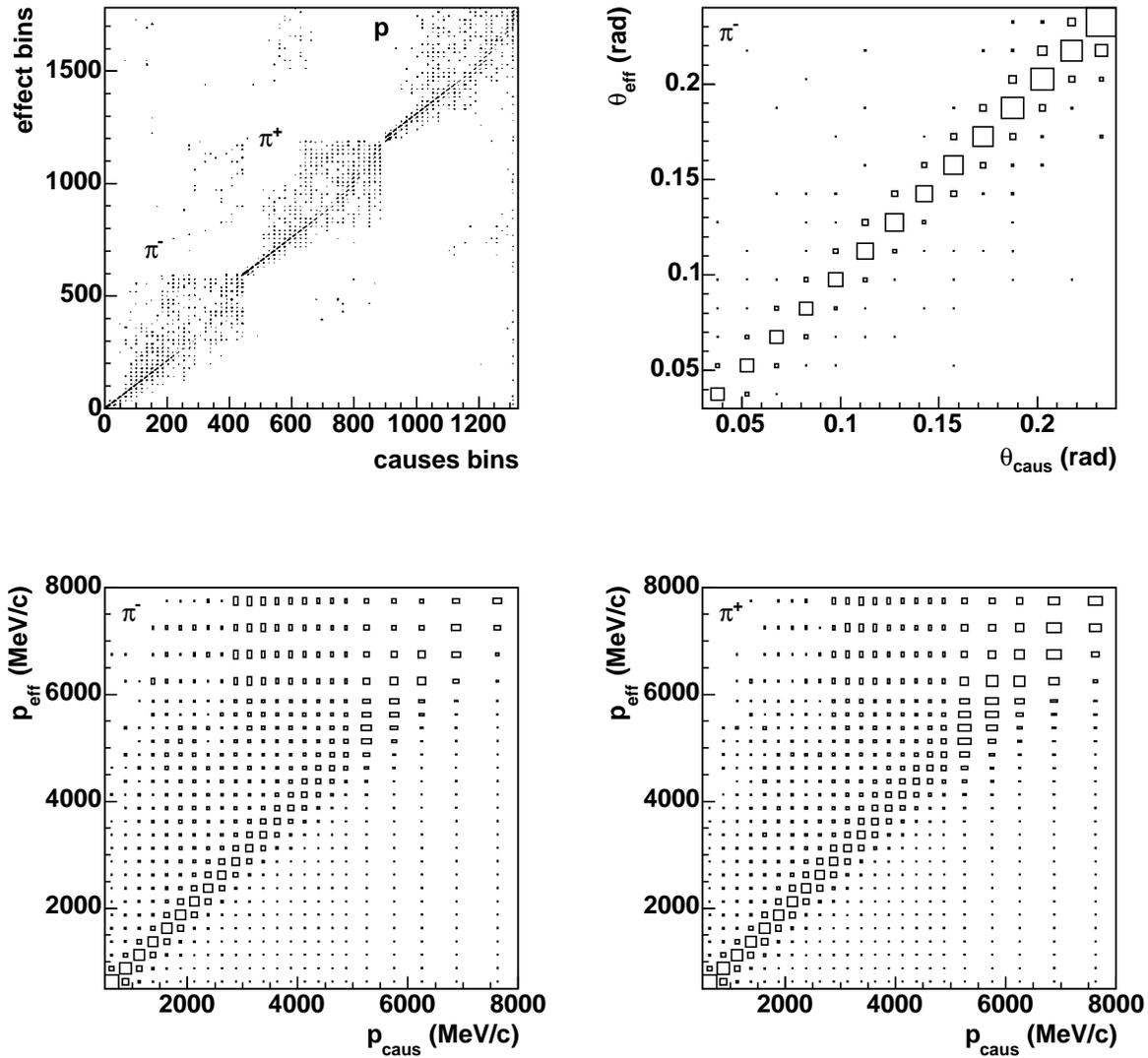}
\caption{\label{ufo_matrix_pr} 
Graphical representation of 
migration matrices calculated for p-C analysis. The left upper panel
shows the original migration matrix where the three dimensions (angle,
momentum and particle type) are merged into one dimension as 
$n_{\theta,p,\alpha} = n_\theta + n_p \cdot n_\theta^{\rm max} 
+ n_\alpha \cdot n_\theta^{\rm max} \cdot n_p^{\rm max} \hspace{0.3cm},$
where $n_{\theta,p,\alpha}$ is the bin number in the final vectors and
in the unfolding matrix; $n_\theta $, $n_p$ and $n_\alpha$ are the bin
numbers in the three dimensions $\theta$, $p$ and $\alpha$,
respectively; $n_\theta^{\rm max}$ and $n_p^{\rm max}$ are the total
number of bins in the observables $p$ and $\alpha$. 
The upper right panel shows an example of the angular migration matrix
for \pionm\ in one momentum causes-effects cell. 
The momentum migration matrices 
integrated
over $\theta$ for \pionm\ (left)
and \pionp\ (right) are shown in the two lower panels.
}
\end{figure}

The reconstruction efficiency (inside the geometrical acceptance) is
larger than 95\% above 1.5~\GeVc and drops to 80\% at 0.5~\GeVc. 
The requirement of a match with a TOFW hit has an efficiency between
90\% and 95\% independent of momentum.
The electron veto rejects about 1\% of the pions and protons below
3~\GeVc with a remaining background of less than 0.5\%.
Below Cherenkov threshold the TOFW separates pions and protons with
negligible background and an efficiency of $\approx$98\% for pions.
Above Cherenkov threshold the efficiency for pions is greater than 99\%
with only 1.5\% of the protons mis-identified as a pion.
The kaon background in the pion spectra is smaller than 1\% and were 
subtracted assuming a similar angular and momentum distribution of 
kaons and pions.

The absorption and decay of particles is simulated by the Monte Carlo.
The generated single particle can re-interact and produce background
particles by hadronic or electromagnetic processes, thus giving rise to
tracks in the dipole spectrometer.
In such cases also the additional measurements are entered into the
migration matrix thereby taking into account the combined effect of the
generated particle and any secondaries it creates.
The absorption correction is on average 20\%, approximately independent
of momentum.
Uncertainties in the absorption of secondaries in the dipole
spectrometer material are taken into account by
a variation of 10\% of this effect in the simulation. 
The effect of pion decay is treated in the same way as the absorption
and is 20\% at 500~\MeVc and negligible at 3~\GeVc. 

The uncertainty in the production of background due to tertiary
particles is larger. 
The average correction is $\approx$10\% and up to 20\% at
1~\GeVc. 
The correction includes reinteractions in the detector material as well
as a small component coming from reinteractions in the target.
The validity of the generators used in the simulation was checked by an
analysis of HARP data with incoming protons, and charged pions on
aluminium and carbon targets at lower momenta (3~\GeV/c and 5~\GeVc).
A 30\% uncertainty of the secondary production was considered.

The unfolding matrix for the p-C analysis calculated this way is shown
in Fig.~\ref{ufo_matrix_pr} in the left upper panel. The very good
separation in the three particle types (\pionm, \pionp \ and proton) can
be clearly seen. The angular (right upper panel) and momentum (lower
panels) unfolding matrices have a nearly diagonal structure as
expected. The binning chosen for these matrices is the same as the one
used for the particle spectra (see section~\ref{sec:results}). The
unfolding matrices for the two other analyses (\pionp-C and \pionm-C)
are by construction very similar as the same Monte Carlo tracks are
used, only the binning is different.

Owing to the large redundancy of the tracking system downstream of the
target the detection efficiency is very robust under the usual
variations of the detector performance during the long data taking
periods. 
Since the momentum is reconstructed without making use of the upstream
drift chamber (which is more sensitive in its performance to the beam
intensity) the reconstruction efficiency is uniquely determined by the
downstream system.
No variation of the overall efficiency has been observed.
The performance of the TOFW and CHE system have been monitored to be
constant for the data taking periods used in this analysis.
The calibration of the detectors was performed on a day-by-day basis.

\subsection{Error estimation}
\label{errorest}

\label{errstat}

The total statistical error of the corrected data is composed of the
statistical error of the raw data, but also of the statistical error
of the unfolding procedure, because the unfolding matrix is obtained
from the data themselves and hence contributes also to the statistical
error. The statistical error provided by the unfolding program is
equivalent to the propagated statistical error of the raw data. In
order to calculate the statistical error of the unfolding procedure a
separate analysis following~\cite{ref:grossheim} is applied. 
It is briefly described below. 
The p-C dataset is divided into two independent data samples
$a$ and $b$, one sample contains all events with odd and the other all
events with even event numbers. These data samples are unfolded in
three different ways:
 1) both samples are unfolded separately using the individually
  calculated unfolding matrix for each sample (set1);
 2) each of the two samples are unfolded with the unfolding matrix
 calculated by using the whole dataset (set2); 
 3) the whole dataset is unfolded twice, using the unfolding
  matrices generated for each part of the split dataset (set3). 
For all three sets the same Monte Carlo input is applied. 
Since
the statistics of the Monte Carlo sample is 
much larger compared
to the statistics of the raw data, 
the
statistical error related to the Monte Carlo is negligible. 
Set1 leads to the total statistical error of the unfolding
result, set2 - to the statistical error of the raw data and set3 - to
the statistical error of the unfolding matrix. For all sets the
difference between the unfolded result of data sample $a$ and $b$ is
calculated and divided by the propagated statistical error of the raw
data $a$ and $b$ for each bin $i$ in the effects space,
 \begin{equation}
\label{cross-check}
\Delta_{{ab}_i} = \frac{a_i-b_i}{\sqrt{\sigma_{a_i}^2 + \sigma_{b_i}^2}} \hspace{0.3cm}.
\end{equation} 
The distribution of $\Delta_{{ab}_i}$ shows for all three sets a Gaussian
shape with a mean close to zero. The width of the distribution of
$\Delta_{ab}$ for set1 is $k(\sigma_{\rm stat})=2.0$, for set2
$k(\sigma_{\rm stat}^{\rm data})=0.98$ and for set3 
$k(\sigma_{\rm stat}^{\rm UFO})=1.77$.
A consistency check gives
\begin{equation}
\label{consis-check}
k(\sigma_{\rm stat}) =  \sqrt{k^2(\sigma_{\rm stat}^{\rm data}) + k^2(\sigma_{\rm stat}^{\rm UFO})}
\nonumber \hspace{0.5cm} \longrightarrow \hspace{0.5cm}
2.0 \simeq \sqrt{0.98^2 + 1.77^2} \hspace{0.3cm}.
\end{equation}

In conclusion, the statistical error provided by the unfolding
procedure has to be multiplied globally by a factor of 2, which is done
for the three analyses (p-C, \pionp-C and \pionm-C) described here.
This factor is somewhat dependent on the shape of the distributions.
For example a value 1.7 was found for the analysis reported in
Ref.~\cite{ref:harp:tantalum}.

\begin{figure}
\centering
\includegraphics[width=0.495\textwidth,height=0.25\textheight]{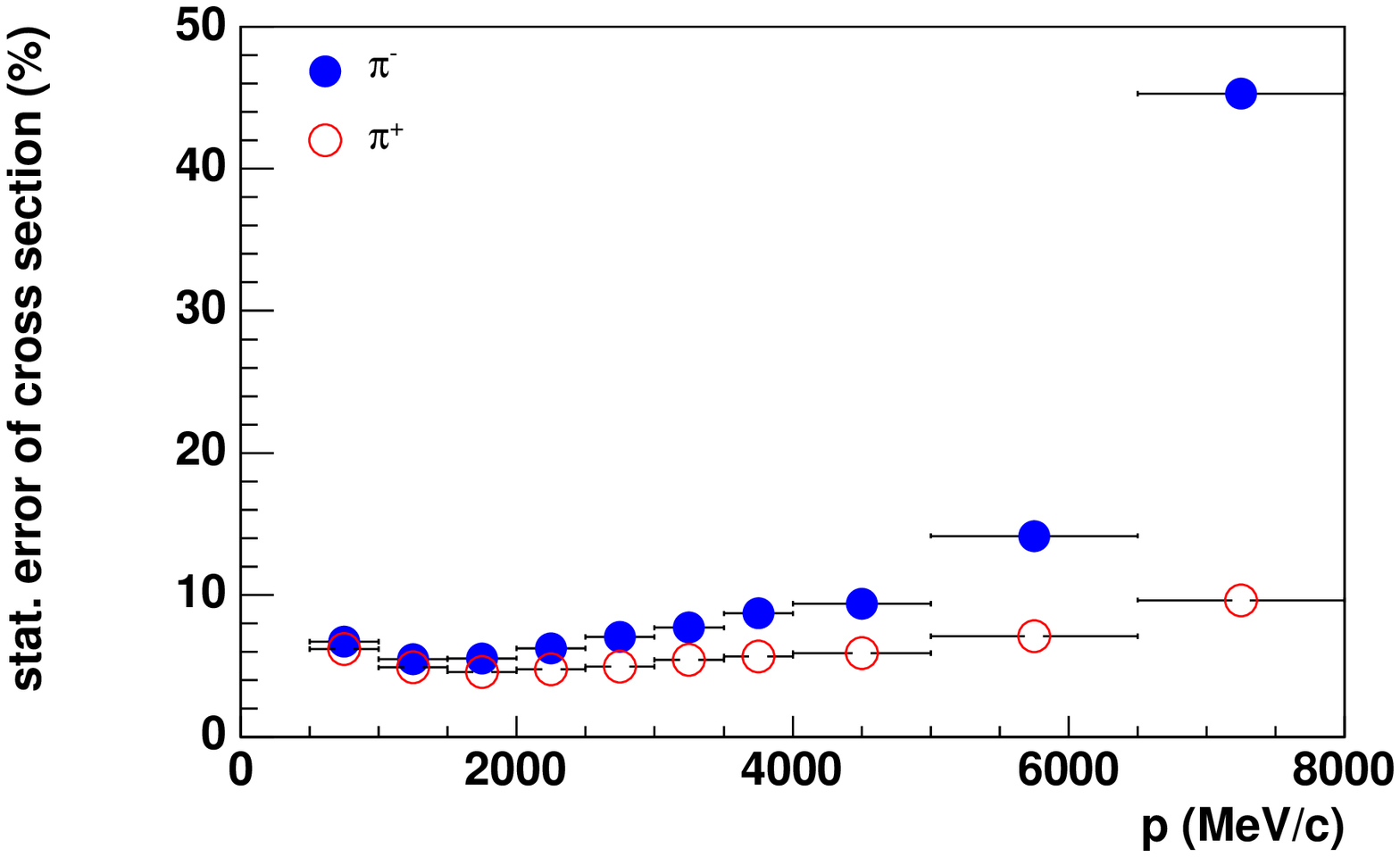}
\includegraphics[width=0.495\textwidth,height=0.25\textheight]{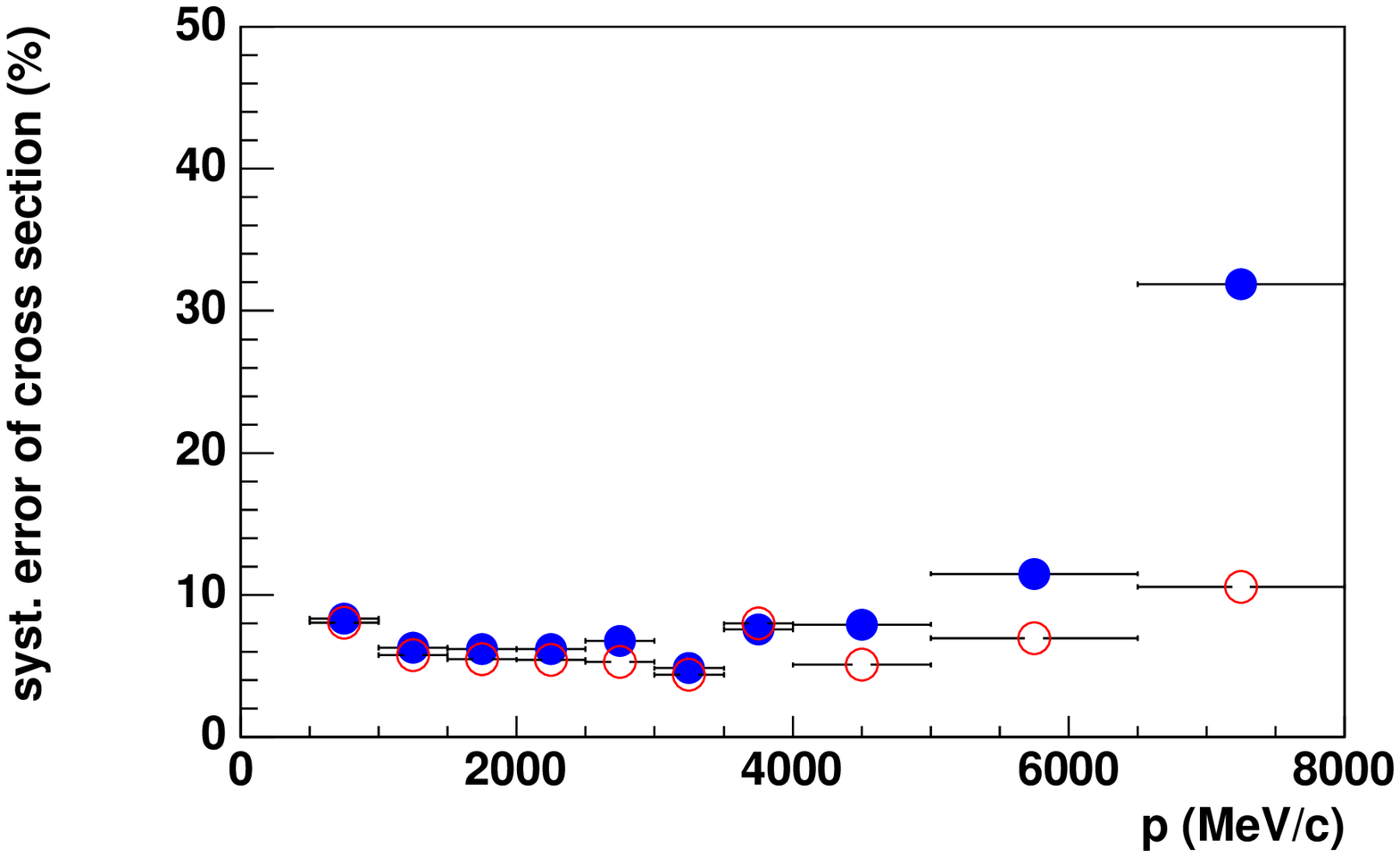} \\
\includegraphics[width=0.495\textwidth,height=0.25\textheight]{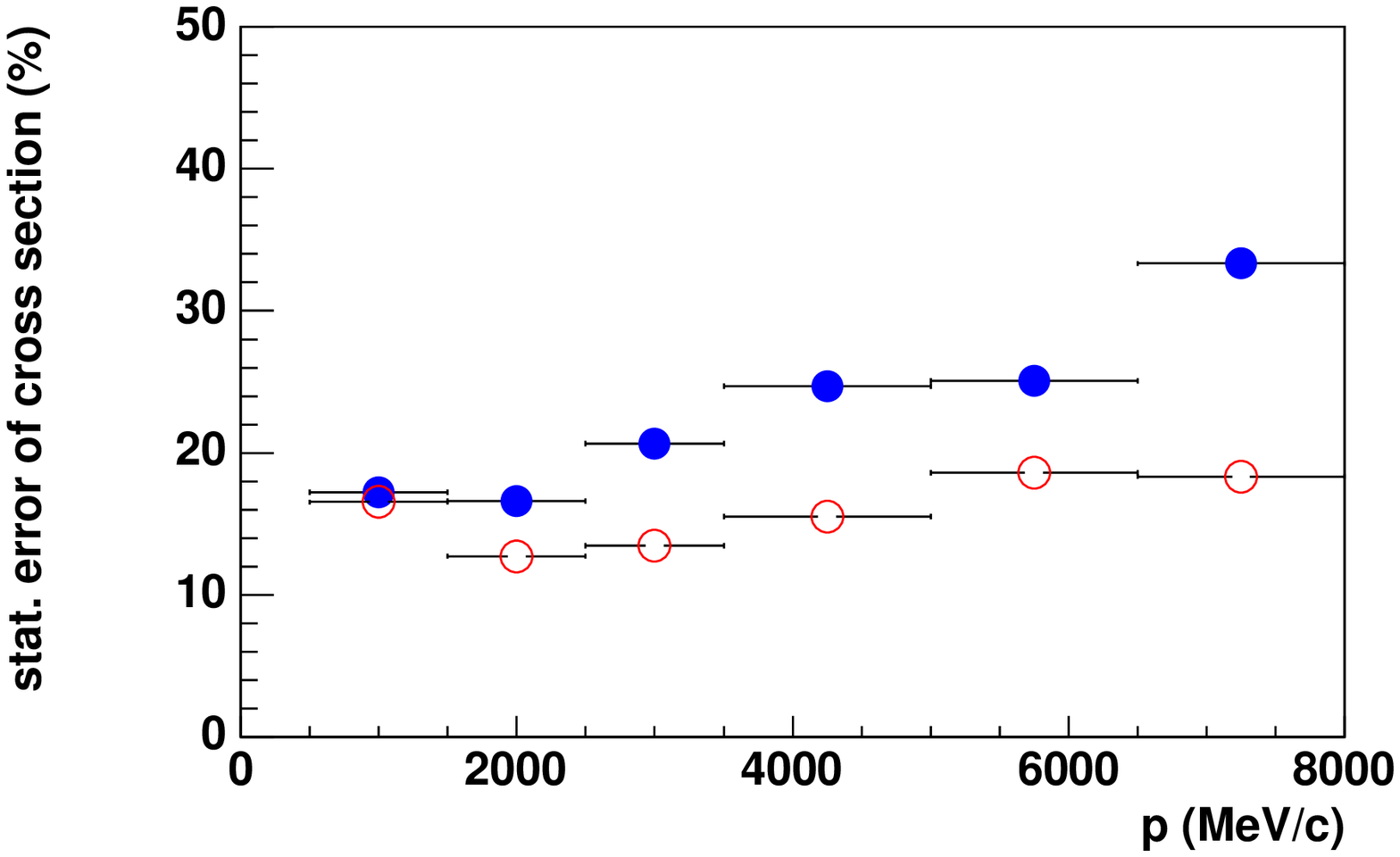}
\includegraphics[width=0.495\textwidth,height=0.25\textheight]{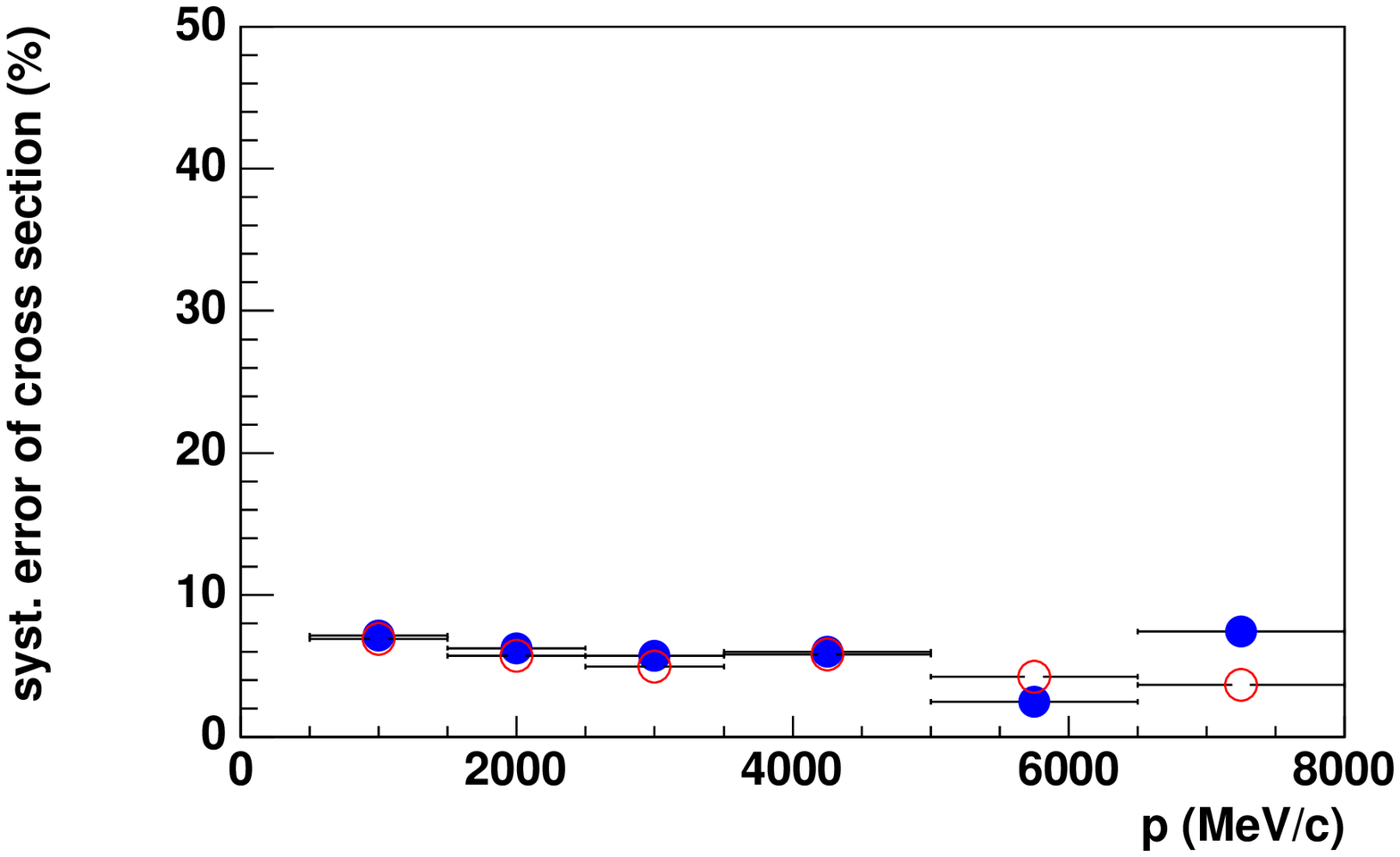} \\
\includegraphics[width=0.495\textwidth,height=0.25\textheight]{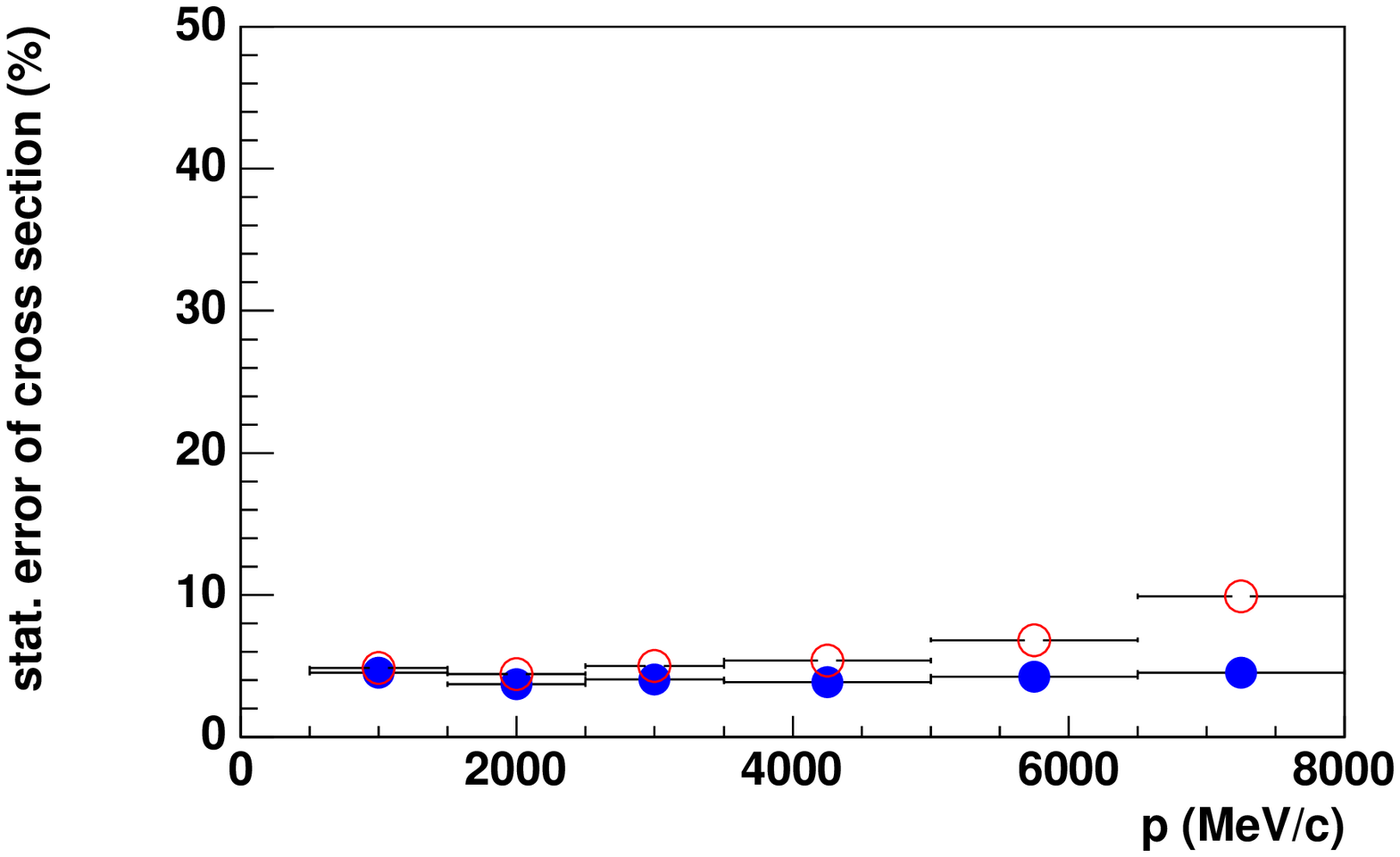} 
\includegraphics[width=0.495\textwidth,height=0.25\textheight]{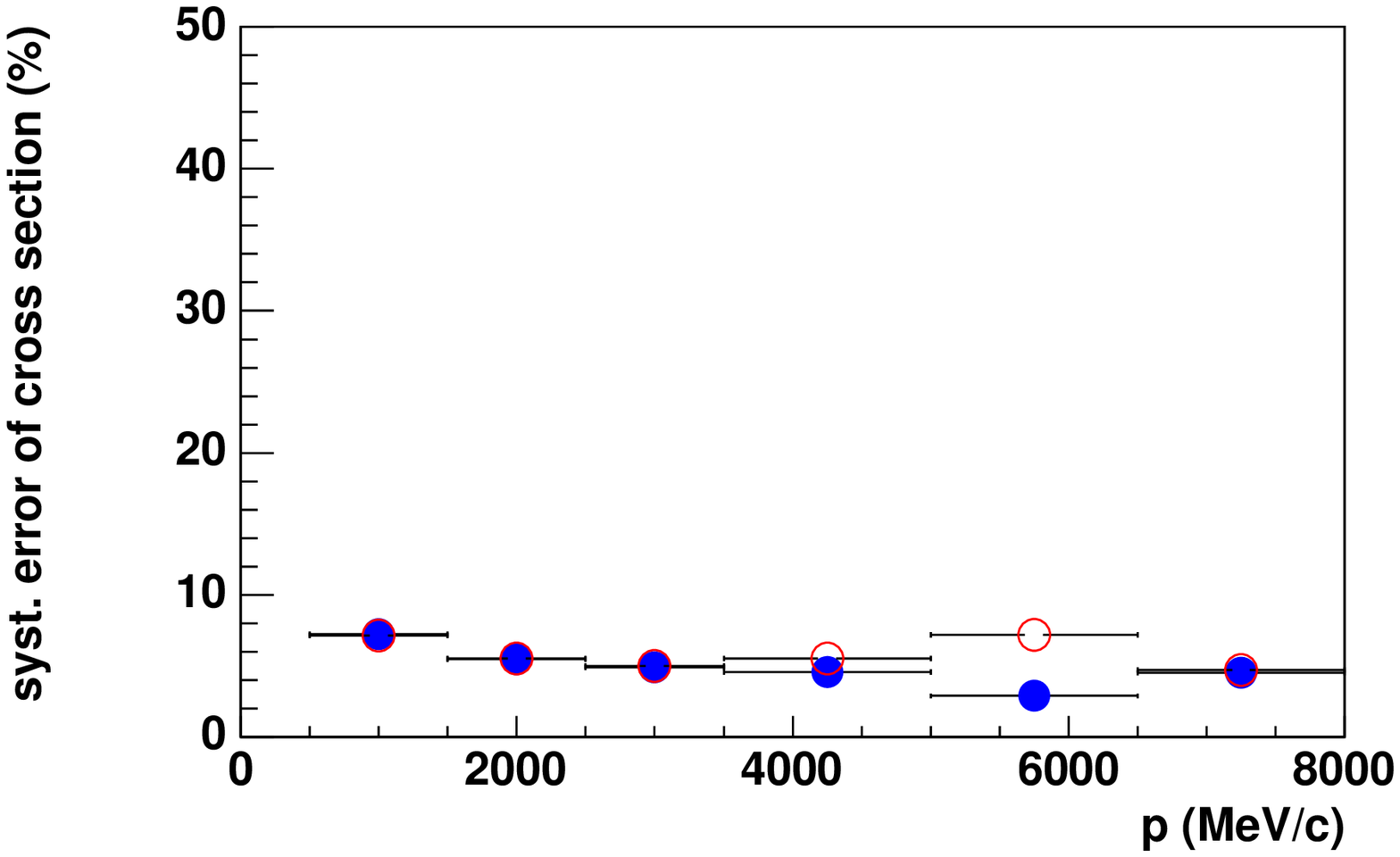} 
\begin{picture}(0,0)
\put(-160,520){{\large \bf Statistical errors}}
\put( 80,520){{\large \bf Systematic errors}}
\put( -120,490){{\large \bf p-C}}
\put(  120,490){{\large \bf p-C}}
\put( -120,320){{\large \bf $\pip$-C}}
\put(  120,320){{\large \bf $\pip$-C}}
\put( -120,150){{\large \bf $\pim$-C}}
\put(  120,150){{\large \bf $\pim$-C}}
\end{picture} 
\caption{\label{err_th_stat_syst} 
Statistical (left) and total systematic (right) 
errors of \pionm (filled circles) and \pionp (open
circles) as a function of momentum integrated over $\theta$ from
0.03~{rad} to 0.24~{rad}. Top: p-C, middle: \pionp-C, bottom:
\pionm-C. 
}
\end{figure}

The calculated statistical errors 
for each momentum--angle bin for all three datasets and separately for
secondary \pionm\ and \pionp\ 
are given in~\cite{ref:christine_phd}. 
Due to the high statistics of the dataset, 
the momentum binning for the p-C dataset is chosen finer than for the
other datasets. The limited statistics of the \pionp-C data is
reflected in a relatively large statistical error. Generally, the
statistical error increases slightly with larger angle and
significantly with increasing momentum. 
The binning for the \pionm-C dataset is chosen to be the same as for
the \pionp-C data to make a direct comparison possible.
The behaviour of statistical
error as a function of momentum is shown in Fig.~\ref{err_th_stat_syst}(left).

\label{errsyst}

Different sources 
of systematic errors are considered in the analysis. 
Namely they are track yield
corrections, particle identification, momentum and angular
reconstruction. Following mainly~\cite{ref:harp:tantalum}, the strategy to
calculate these systematic errors is to find different solutions of
the unfolding problem, i.e. different 'causes' result vectors.
The difference vector is used to create a covariance matrix for a
specific systematic error. Three different methods are applied to
calculate these different causes vectors: 
1) variation of the normalization of the causes vector; 
2) variation of the unfolding matrix; 
3) variation of the raw data. 
The first method is used for the estimation of the systematic error of
the track reconstruction efficiency. 
The uncertainties in the efficiency are estimated from the small
differences observed between the data and the simulation.

The second method is applied for most of the systematic error
estimations. The loss of secondary particles has to be considered due
to particle decay and absorption in the detector materials as well as
additional background particles generated in secondary
reactions. These effects are simulated by Monte Carlo: two
single-particle Monte Carlo simulations are generated, in the first
simulation these effects are taken into account 
while not in the second one.
Both Monte Carlo simulations are used for unfolding data, 
then
the results are compared. The uncertainties in the absorption are
estimated by a variation of 10\% and the uncertainty in the production
of background particle due to tertiary particles by a 30\%
variation~\cite{ref:bePaper}. The performance of particle
identification, momentum and angular measurements are correlated due
to the simultaneous unfolding process of these observables as
described in section~\ref{ufo}. The calculation of systematic errors
of particle identification, angular and momentum resolution as well as
of momentum scale is done by varying the acceptance criteria for these
observables in the raw data and in the Monte Carlo. For the momentum
resolution possible discrepancies up to 10\% of the resolution are
taken into account~\cite{ref:bePaper}.  The systematic uncertainty in the
momentum determination is estimated to be of the order of 2\% using
the elastic scattering analysis~\cite{ref:bePaper}. 
The angular scale was varied by 1\%.

The third method is introduced for the estimation of the systematic
error of the empty target subtraction. In addition to the standard
empty target subtraction only 95\% of the calculated empty target
value is subtracted from the raw data\footnote{the maximum effect of 
the 5\%~\ $\lambda_{\mathrm{I}}$ target is to ``absorb'' 5\% of the beam particles}.
The systematic error is taken from the difference of these two results. 
The statistical error of the empty target subtraction is taken into account as a 
diagonal statistical error in $N^{\alpha'}(p'_i,\theta'_j)$ by simple error propagation. 

Due to the fact that kaons are not 
considered by the particle
identification method in the current analysis~\cite{ref:pidPaper} 
misidentified secondary kaons form an additional
error source. To reduce this effect a specific 
Monte Carlo simulation only with secondary kaons is generated. 
Simulated kaons are classified as pions or protons according to the same
PID criteria as applied to the data.
The remaining mis-identified kaons are then subtracted 
assuming a 50\% uncertainty on the K/$\pi$ ratio.
The central value of the  K/$\pi$ ratio was taken from Ref.~\cite{kliemant:2004}.
This procedure also takes into account that decay muons from kaons
produced in the target can be identified as pions in the spectrometer;
these are subtracted by this procedure.
We do not make an explicit correction for pions coming from decays of
other particles created in the target.  
Pions created in strong decays are considered to be part of the
inclusive production cross-sections.
A small background coming from weak decays other than from charged kaons
is neglected (such as K$^0$'s and $\Lambda^0$'s).
These pions have a very small efficiency given the cuts applied in this
analysis. 

Following Ref.~\cite{ref:harp:tantalum} the overall normalization 
of the results is calculated  
relative to the number of incident beam particles accepted by the selection. 
The uncertainty is 2\% and 3\% for incident protons and pions, respectively.

As a result of these systematic error studies each error source can be
represented by a covariance matrix. The sum of these matrices
describes the total systematic error. 
Detailed information about
the diagonal elements of the
covariance matrix of the total systematic error 
for each momentum--angular bin 
can be found in~\cite{ref:christine_phd}. 
In Fig.~\ref{err_th_stat_syst}(right) the total systematic error 
integrated over angle is shown as
a function of momentum. For the \pionp-C
and \pionm-C datasets the systematic error has a nearly flat
distribution and is approximately 6\%. For the p-C dataset the
systematic error increases for higher momenta but also stays nearly
constant around 8\% below 6~{\GeVc}.

The
dimensionless quantity $\delta_{\rm diff}$, expressing the
typical error on the double-differential cross-section, is defined as
follows
\begin{equation}
\label{errdiff}
\delta_{\rm diff} = \frac{\sum_i(\delta[d^2 \sigma^\pi/(d p d \Omega)])_i}{\sum_i (d^2 \sigma^\pi / (d p d \Omega))_i} \hspace{0.3cm} ,
\end{equation}
where $i$ labels a given  momentum--angular bin $(p,\theta)$, $(d^2
\sigma^\pi / (d p d \Omega))_i$ is the central value for the
double-differential cross-section measurement in that bin, and 
$(\delta[d^2 \sigma^\pi/(d p d \Omega)])_i$ is the error associated
with this measurement.

The
dimensionless quantity $\delta_{\rm int}$ is defined,
expressing the fractional error on the integrated pion cross-section
$\sigma^\pi$ in the momentum range 0.5~{\GeVc}$<p<$8.0~{\GeVc} and the
angular range 0.03~{rad}$<\theta<$0.24~{rad} for the p-C data and in
the angular range 0.03~{rad}$<\theta<$0.21~{rad} for the \pionpm-C
data\footnote{The binning of the \pionpm\ data was chosen to accommodate
the lower statistics of the \pionp\ data and is only determined for
$\theta<$0.21~{rad}.}, as 
follows
\begin{equation}
\label{errint}
\delta_{\rm int} = \frac{\sqrt{\sum_{i,j}(\Delta p \Delta \Omega)_i 
C_{ij} (\Delta p \Delta \Omega)_j}}{\sum_i (d^2 \sigma^\pi/d p d \Omega)_i 
(\Delta p \Delta \Omega)_i } \hspace{0.3cm} ,
\end{equation}
where $(d^2 \sigma^\pi/d p d \Omega)_i$ is the double-differential
cross-section in bin $i$, $(\Delta p \Delta \Omega)_i$  is the
corresponding phase space element, and $C_{ij}$ is the covariance
matrix of the double-differential cross-section. Then $\sqrt{C_{ii}}$
corresponds to the error $(\delta[d^2 \sigma^\pi/(d p d \Omega)])_i$
in Eq.~(\ref{errdiff}).

\begin{table}
\centering
\caption{\label{errtabpC} 
Summary of the uncertainties affecting the double-differential and
integrated cross-section measurements of p-C data.} %
\vspace{0.15 cm}
\begin{tabular}{|l|l|r|r|r|r|}
\hline {\bf Error category}  & {\bf Error source} & $\delta^{\pi^{-}}_{\text{diff}}$(\%) & $\delta^{\pi^{-}}_{\text{int}}$(\%) & $\delta^{\pi^{+}}_{\text{diff}}$(\%) & $\delta^{\pi^{+}}_{\text{int}}$(\%)  \\
\hline Statistical                & Data statistics           & 12.8 & 3.2 & 10.8 & 2.5 \\
\hline Track yield corrections    & Reconstruction efficiency &  1.6 & 1.3 &  1.1 & 0.5 \\
                                  & Pion, proton absorption   &  4.2 & 3.7 &  3.7 & 3.2 \\
                                  & Tertiary subtraction      &  9.8 & 4.2 &  8.6 & 3.7 \\
                                  & Empty target subtraction  &  1.2 & 1.2 &  1.2 & 1.2 \\
                                  & Subtotal                  & 10.8 & 5.9 &  9.5 & 5.1 \\
\hline Particle identification    & Electron veto             & $<0.1$& $<0.1$ & $<0.1$ & $<0.1$ \\
                                  & Pion, proton ID correction& $<0.1$ & 0.1 &  0.1 & 0.1 \\
                                  & Kaon subtraction          & $<0.1$ & $<0.1$ & $<0.1$ & $<0.1$ \\
                                  & Subtotal                  &  0.1 & 0.1 &  0.1 & 0.1 \\
\hline Momentum reconstruction    & Momentum scale            &  2.6 & 0.4 &  2.8 & 0.3 \\
                                  & Momentum resolution       &  0.7 & 0.2 &  0.8 & 0.3 \\
                                  & Subtotal                  &  2.7 & 0.5 &  2.9 & 0.4 \\
\hline Angle reconstruction       & Angular scale             &  0.5 & 0.1 &  1.3 & 0.5 \\
\hline Systematic error           & Subtotal                  & 11.2 & 5.9 & 10.0 & 5.1 \\
\hline Overall normalization      & Subtotal                  &  2.0 & 2.0 &  2.0 & 2.0  \\
\hline \hline All                        & Total                     & 17.1 & 7.0 & 14.9 & 6.1 \\
\hline
\end{tabular}
\end{table}

\begin{table}
\centering
\caption{\label{errtabpipC} 
Summary of the uncertainties affecting the double-differential and
integrated cross-section measurements of \pionp-C data.} %
\vspace{0.15 cm}
\begin{tabular}{|l|l|r|r|r|r|}
\hline {\bf Error category}  & {\bf Error source} & $\delta^{\pi^{-}}_{\text{diff}}$(\%) & $\delta^{\pi^{-}}_{\text{int}}$(\%) & $\delta^{\pi^{+}}_{\text{diff}}$(\%) & $\delta^{\pi^{+}}_{\text{int}}$(\%)  \\
\hline Statistical                & Data statistics           & 41.8 & 6.4 & 34.5 & 7.2 \\
\hline Track yield corrections    & Reconstruction efficiency &  1.4 & 0.7 &  0.9 & 0.5 \\
                                  & Pion, proton absorption   &  4.0 & 2.1 &  3.3 & 2.7 \\
                                  & Tertiary subtraction      &  9.3 & 4.7 &  7.6 & 6.3 \\
                                  & Empty target subtraction  &  1.0 & 0.7 &  1.0 & 1.0 \\
                                  & Subtotal                  & 10.3 & 5.2 &  8.4 & 6.9 \\
\hline Particle identification    & Electron veto             & $<0.1$ & $<0.1$ & $<0.1$ & $<0.1$ \\
                                  & Pion, proton ID correction&  0.1 & $<0.1$ &  0.2 & 0.2 \\
                                  & Kaon subtraction          & $<0.1$ & $<0.1$ & $<0.1$ & $<0.1$ \\
                                  & Subtotal                  &  0.1 & 0.1 &  0.2 & 0.2 \\
\hline Momentum reconstruction    & Momentum scale            &  3.2 & 0.2 &  3.6 & 0.5 \\
                                  & Momentum resolution       &  0.9 & 0.2 &  1.1 & 0.3 \\
                                  & Subtotal                  &  3.3 & 0.3 &  3.8 & 0.6 \\
\hline Angle reconstruction       & Angular scale             &  1.7 & 0.1 &  1.3 & 0.5 \\
\hline Systematic error           & Subtotal                  & 10.9 & 5.3 &  9.2 & 7.0 \\
\hline Overall normalization      & Subtotal                  &  3.0 & 3.0 &  3.0 & 3.0  \\
\hline \hline All                        & Total                     & 43.7 & 8.5 & 35.8 & 10.2 \\
\hline
\end{tabular}

\end{table}

\begin{table}
\centering
\caption{\label{errtabpimC} 
Summary of the uncertainties affecting the double-differential and
integrated cross-section measurements of \pionm-C data.} %
\vspace{0.15 cm}
\begin{tabular}{|l|l|r|r|r|r|}
\hline {\bf Error category}  & {\bf Error source} & $\delta^{\pi^{-}}_{\text{diff}}$(\%) & $\delta^{\pi^{-}}_{\text{int}}$(\%) & $\delta^{\pi^{+}}_{\text{diff}}$(\%) & $\delta^{\pi^{+}}_{\text{int}}$(\%)  \\
\hline Statistical                & Data statistics           &  8.5 & 2.2 & 10.0 & 1.9 \\
\hline Track yield corrections    & Reconstruction efficiency &  1.3 & 1.1 &  0.7 & 0.4 \\
                                  & Pion, proton absorption   &  3.5 & 3.1 &  3.8 & 2.3 \\
                                  & Tertiary subtraction      &  7.9 & 6.8 &  9.0 & 5.3 \\
                                  & Empty target subtraction  &  0.9 & 0.8 &  0.9 & 0.6 \\
                                  & Subtotal                  &  8.8 & 7.6 &  9.8 & 5.8 \\
\hline Particle identification    & Electron veto             & $<0.1$ &$<0.1$ & $<0.1$ &$<0.1$ \\
                                  & Pion, proton ID correction&  0.1 & 0.1 &  0.1 & 0.1 \\
                                  & Kaon subtraction          & $<0.1$ &$<0.1$ & $<0.1$ &$<0.1$ \\
                                  & Subtotal                  &  0.1 & 0.1 &  0.1 & 0.1 \\
\hline Momentum reconstruction    & Momentum scale            &  2.3 & 0.7 &  2.7 & 0.3 \\
                                  & Momentum resolution       &  0.6 & 0.2 &  0.5 & 0.2 \\
                                  & Subtotal                  &  2.4 & 0.7 &  2.7 & 0.4 \\
\hline Angle reconstruction       & Angular scale             &  0.6 & 0.3 &  0.7 & $<0.1$ \\
\hline Systematic error           & Subtotal                  &  9.1 & 7.6 & 10.2 & 5.8 \\
\hline Overall normalization      & Subtotal                  &  3.0 & 3.0 &  3.0 & 3.0  \\
\hline \hline All                        & Total                     & 12.6 & 8.2 & 14.4 & 6.5 \\
\hline
\end{tabular}
\end{table}

The values of $\delta_{\rm diff}$ 
and
$\delta_{\rm int}$ are summarized for
all specific systematic error sources in Table~\ref{errtabpC} for p-C
data, in Table~\ref{errtabpipC} for \pionp-C data and in
Table~\ref{errtabpimC} for \pionm-C data. The systematic errors are of
the same order for all three datasets, $\delta_{\rm diff}=$~9\%-11\%
and  $\delta_{\rm int}=$~5\%-8\%. The 
dominant error sources are given by 
particle absorption and the subtraction of tertiary particles. 
The decay correction is technically made as part of the absorption
correction and reported under ``absorption''.
The
errors of momentum and angular reconstruction are less important and
the errors caused by the particle misidentification are
negligible. For the datasets with positively charged beam the
systematic error is smaller for \pionp\ and for \pionm-C dataset it is
smaller for \pionm.

Systematic and statistical errors are of the same order for the p-C
and the \pionm-C data. For the \pionp-C dataset the statistical error
is dominating the total error. The \pionm-C data have the smallest
total error due to the data statistics and chosen bin width. 

There is a certain amount of correlation between the systematic errors
in the different spectra.
In the comparison of production spectra of the same secondary particle
type by different incoming particles, the absorption and decay errors
cancel. 
One also expects the tertiary subtraction uncertainty to cancel
partially, although this depends on the details of the production
models. (For example, the uncertainty in the background in the \pip
spectra measured in the \pip beam is expected to be correlated to the
background for \pim in the \pim beam, but less so for opposite charges.) 
Of the other relatively important errors the systematic component of the
empty target subtraction and the momentum scale error cancel between the
datasets.
The overall normalization errors are largely independent.

\section{Results}
\label{sec:results}

\begin{figure}
\centering
\includegraphics[height=0.8\textheight]{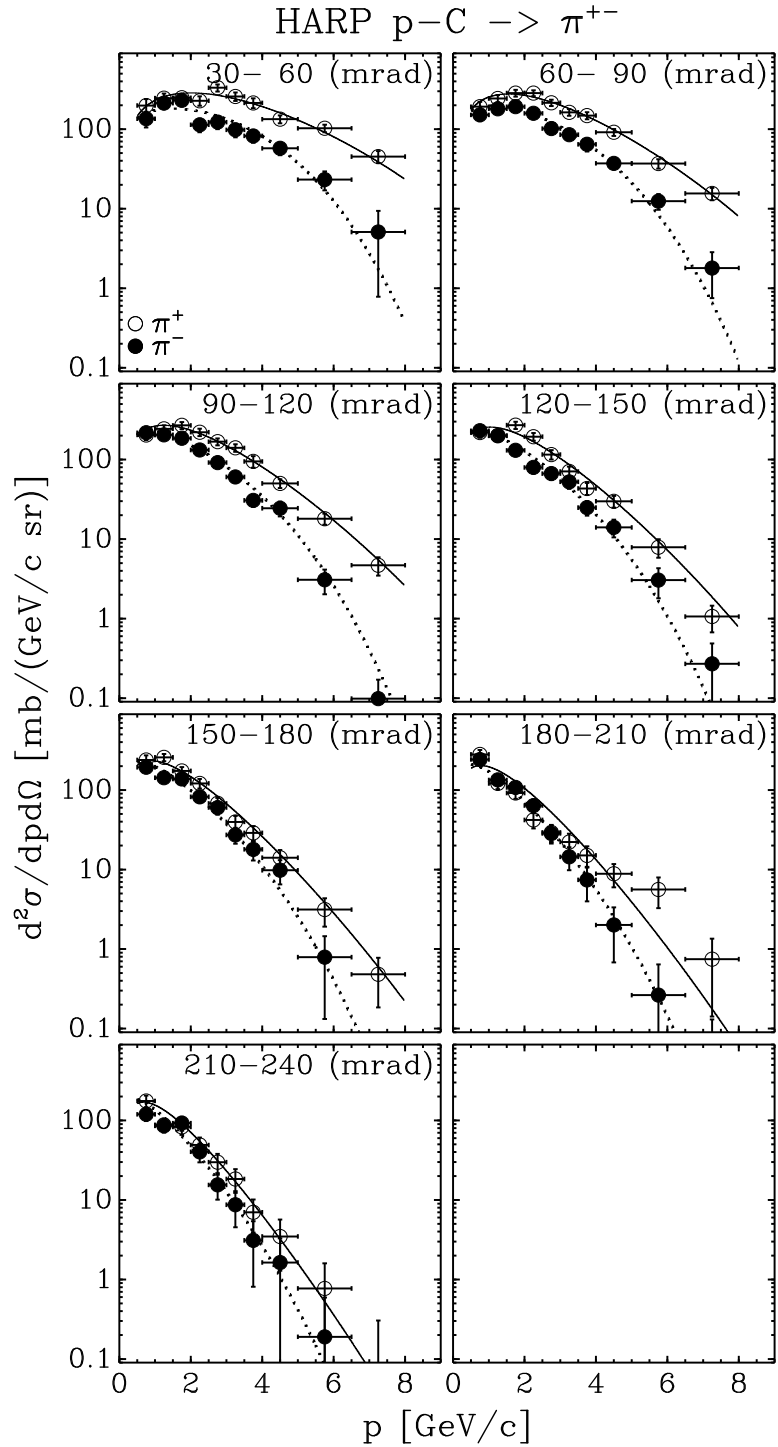}
\caption{\label{pC12GeV_pi_20_7_log} 
  Measurement of the double-differential production cross-section of
  positive (open circles) and negative (filled circles) pions
  from 12~\GeVc protons on carbon as a function of pion momentum, p, in
  bins of pion angle, $\theta$, in the laboratory frame. 
  Seven panels show different angular bins from 30~{mrad}
  to 240~{mrad} (the corresponding angular interval is printed on each panel).
  The error bars shown include statistical errors and all (diagonal) systematic errors.
 The curves show the Sanford-Wang parametrization 
 of Eq.~\ref{SWfunc} with parameter values given in Table~\ref{SWpar_prot}.
}
\end{figure}

\begin{figure}
\centering
\includegraphics[height=0.8\textheight]{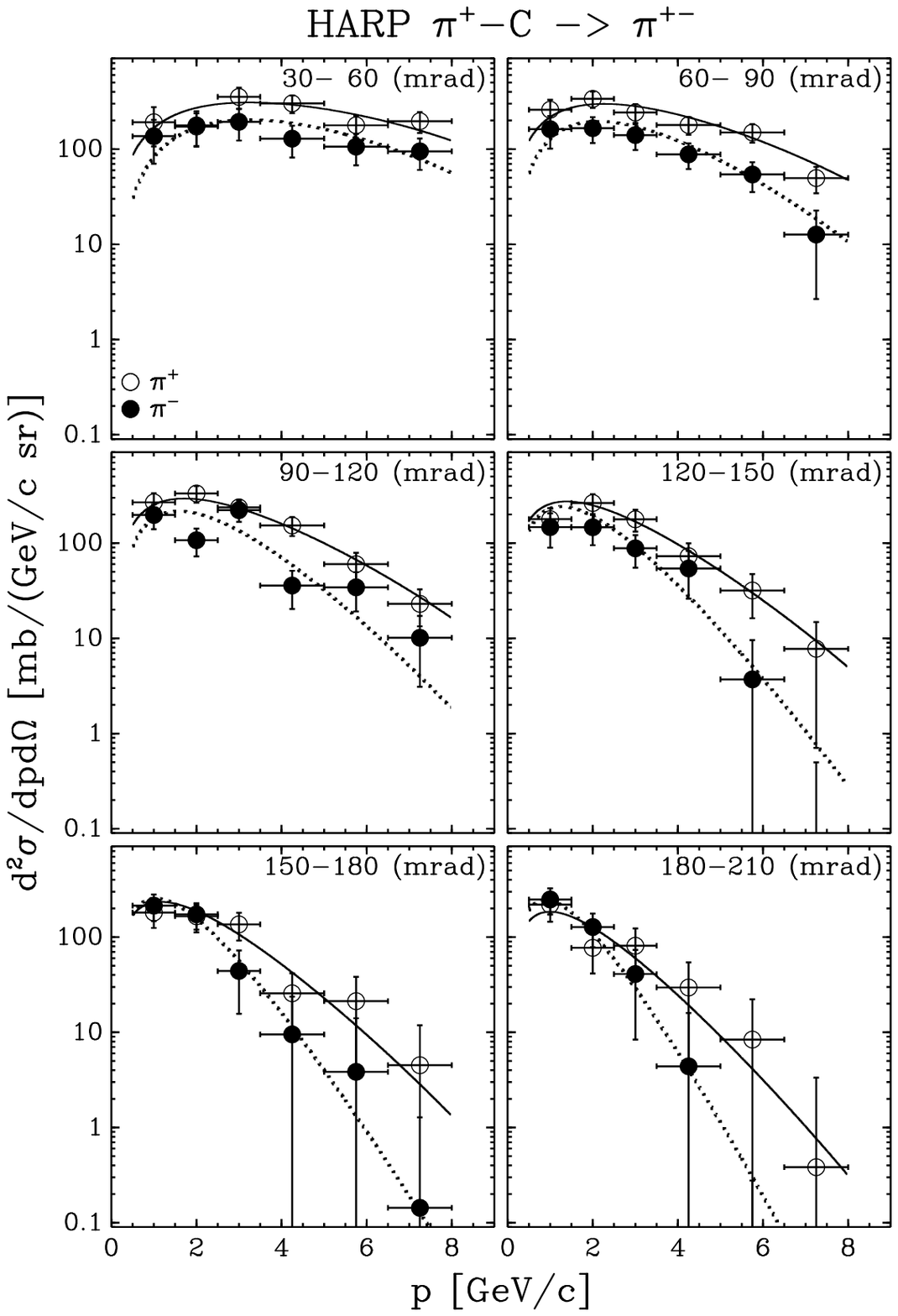}
\caption{\label{pipC12GeV_pi_log_nrb} 
  Measurement of the double-differential production cross-section of
  positive (open circles) and negative (filled circles) pions
  from 12~\GeVc $\pi^{+}$ on carbon as a function of pion momentum, p, in
  bins of pion angle, $\theta$, in the laboratory frame. 
  Six panels show different angular bins from 30~{mrad}
  to 210~{mrad} (the corresponding angular interval is printed on each panel).
  The error bars shown include statistical errors and all (diagonal) systematic errors.
 The curves show the Sanford-Wang parametrization 
 of Eq.~\ref{SWfunc} with parameter values given in Table~\ref{SWpar_pis}.
}
\end{figure}

\begin{figure}
\centering
\includegraphics[height=0.8\textheight]{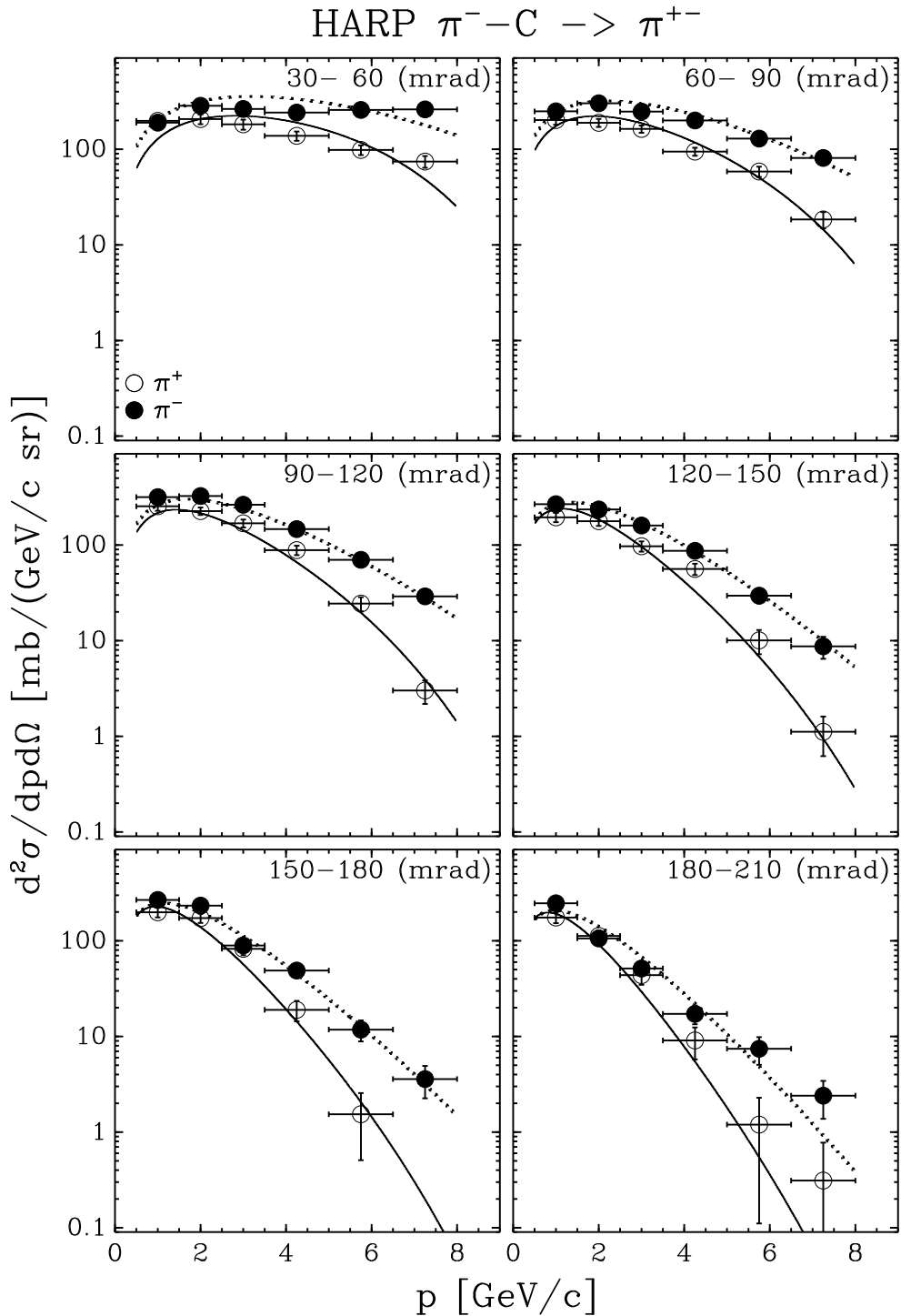}
\caption{\label{pimC12GeV_004_log_nrb} 
  Measurement of the double-differential production cross-section of
  positive (open circles) and negative (filled circles) pions
  from 12~\GeVc $\pi^{-}$ on carbon as a function of pion momentum, p, in
  bins of pion angle, $\theta$, in the laboratory frame. 
  Six panels show different angular bins from 30~{mrad}
  to 210~{mrad} (the corresponding angular interval is printed on each panel).
  The error bars shown include statistical errors and all (diagonal) systematic errors.
 The curves show the Sanford-Wang parametrization 
 of Eq.~\ref{SWfunc} with parameter values given in Table~\ref{SWpar_pis}.
}
\end{figure}

The results of the measurements of the double-differential
cross-sections for positive
and negative pions 
in p-C, \pionp-C and \pionm-C 
interactions
at 12~{\GeVc} in the laboratory system are presented as a function of
momentum for various angular bins in 
Figs.~\ref{pC12GeV_pi_20_7_log},~\ref{pipC12GeV_pi_log_nrb} 
and~\ref{pimC12GeV_004_log_nrb}, respectively. 
The central values
and square-root of the diagonal elements of the covariance matrix are listed in
Tables~\ref{tab:xsec_results_pC12_pip_pim_2}-\ref{tab:xsec_results_pimC12_pip_pim_1} 
in Appendix~\ref{app:data}.
The kinematic range of the measurements covers the
momentum region from 0.5~{\GeVc} to 8.0~{\GeVc} and the angular
range from 0.03~{rad} to 0.24~{rad} for p-C and from 0.03~{rad} to
0.21~{rad} for \pionp-C and \pionm-C data. The error bars 
correspond to
the combined statistical and systematic
errors as described in section~\ref{errorest}.
The overall normalization error of 2\% and 3\% for the normalization of
incident protons and pions, respectively, is not shown.

The shapes of the 
production cross-sections
are similar for secondary \pionp\
and \pionm\ as well as for different datasets. For larger angles
the spectra are softer and show a leading particle effect for produced
\pionp\ in p-C and \pionp-C reactions and for \pionm\ in \pionm-C
reactions. The distribution of secondary \pionp\ in \pionp-C reactions
show a very similar behaviour as the distribution of secondary \pionm\
in \pionm-C reactions as expected because of the isospin symmetry of
$\pionp+{\rm C}\rightarrow\pionp+X$ and $\pionm+{\rm
  C}\rightarrow\pionm+X$ reactions. 
The corresponding behaviour can be seen for
\pionm\ in \pionp-C interactions and for \pionp\ in \pionm-C interactions.
The \pionp/\pionm\ ratio is larger than unity in the positive particle beams 
and smaller than unity in the \pionm\ beam.

In section~\ref{sec:sw} the
measured cross-sections are fitted to a Sanford-Wang parametrization
while in section~\ref{sec:compare} a comparison of HARP p-C data with predictions of
different hadronic interaction models 
is shown.

\subsection{Sanford-Wang parametrization}
\label{sec:sw}

Sanford and Wang~\cite{SanfordWang1967} have developed an empirical
parametrization for describing the production cross-sections of mesons
in proton-nucleus interactions. This parametrization has the
functional form:
\begin{eqnarray}
\label{SWfunc}
\frac{d^2\sigma^\pi}{dpd\Omega}\left(p,\theta\right) & = & c_1p^{c_2}\left(1-\frac{p}{p_{\rm beam}}\right)\exp\left[-c_3\frac{p^{c_4}}{p_{\rm beam}^{c_5}}-c_6\theta\left(p-c_7p_{\rm beam}\cos^{c_8}\theta\right)\right] \hspace{0.3cm},
\end{eqnarray}
where 
\begin{itemize}
\item $\frac{d^2\sigma^\pi}{dpd\Omega}\left(p,\theta\right)$ is the
cross-section in mb/(\GeVc sr) for secondary pions as a function of
momentum $p$ (in \GeVc) and angle $\theta$ (in radians) of the
secondary particles;
\item $p_{\rm beam}$ is the beam momentum in \GeVc;
\item $c_1$, ..., $c_8$ are free parameters obtained from fits to meson production data.
\end{itemize}

The parameter $c_1$ is an overall normalization factor, the four
parameters $c_2,c_3,c_4,c_5$ can be interpreted as describing the
momentum distribution of the secondary pions in the forward direction, 
and the three parameters
$c_6,c_7,c_8$ as describing the angular distribution for fixed
secondary and beam momenta, $p$ and $p_{\rm beam}$.

\begin{table}[tb]
\centering
\caption{
\label{SWpar_prot} 
Sanford-Wang parameters and errors obtained by fitting
the p--C dataset.
} \vspace{0.5 cm}
\begin{minipage}{\linewidth}
\renewcommand{\footnoterule}{}
\centering
\begin{tabular}{|c|c|c|}
\hline & \multicolumn{2}{|c|}{p--C} \\
\hline \multicolumn{1}{|c}{\bf Param} &
\multicolumn{1}{|c}{\pionm} & \multicolumn{1}{c|}{\pionp} \\
\hline  $c_1$ &  144.46 $\pm$ 65.593    &  214.92 $\pm$ 93.307  \\
\hline  $c_2$ &  0.60749  $\pm$ 0.34902 &  0.95748  $\pm$ 0.44512 \\
\hline  $c_3$ &  16.947  $\pm$ 10.876   &  3.0906  $\pm$ 1.2601   \\
\hline  $c_4$=$c_5$ &  3.2512  $\pm$ 1.3657 &  1.6876  $\pm$ 1.5230 \\
\hline  $c_6$ &  5.9304  $\pm$ 1.2561     &  5.5728  $\pm$ 0.71771  \\
\hline  $c_7$ &  0.17152  $\pm$ 0.074772  &  0.15597  $\pm$ 0.06683 \\
\hline  $c_8$ &  27.241 $\pm$ 12.232    &  30.873 $\pm$ 13.388  \\
\hline  $\chi^2$/NDF & 95.6/63 & 147.7/63  \\
\hline
\end{tabular}
\end{minipage}
\end{table}

\begin{table}[tb]
\begin{center}
{
\caption{\label{tab:swpar_correlations_prot}
Correlation coefficients among the Sanford-Wang parameters, obtained
 by fitting the p--C dataset.
}
{
\begin{tabular}{ |c r r r r r r r|}
\hline
\multicolumn{8}{|c|}{\pionm}\\
\hline
{\bf Parameter} & $c_1$  & $c_2$  & $c_3$  & $c_4=c_5$  & $c_6$  & $c_7$  & $c_8$ \\
\hline
$c_1$     &  1.000 &        &        &        &        &        &        \\
$c_2$     & -0.433 &  1.000 &        &        &        &        &        \\
$c_3$     & -0.041 & -0.548 &  1.000 &        &        &        &        \\
$c_4=c_5$ & -0.113 & -0.535 &  0.950 &  1.000 &        &        &        \\
$c_6$     & -0.535 &  0.622 & -0.035 &  0.127 &  1.000 &        &        \\
$c_7$     & -0.837 &  0.121 &  0.024 &  0.050 &  0.214 &  1.000 &        \\
$c_8$     & -0.206 & -0.316 &  0.028 & -0.025 & -0.360 &  0.611 & 1.000  \\
\hline
\multicolumn{8}{|c|}{\pionp}\\
\hline
{\bf Parameter} & $c_1$  & $c_2$  & $c_3$  & $c_4=c_5$  & $c_6$  & $c_7$  & $c_8$ \\
\hline
$c_1$     &  1.000 &        &        &        &        &        &        \\
$c_2$     &  0.151 &  1.000 &        &        &        &        &        \\
$c_3$     &  0.061 & -0.151 &  1.000 &        &        &        &        \\
$c_4=c_5$ & -0.461 & -0.860 &  0.351 &  1.000 &        &        &        \\
$c_6$     & -0.544 &  0.248 & -0.373 &  0.065 &  1.000 &        &        \\
$c_7$     & -0.790 & -0.004 & -0.168 &  0.115 &  0.333 &  1.000 &        \\
$c_8$     & -0.083 & -0.275 &  0.092 &  0.080 & -0.416 &  0.488 & 1.000  \\
\hline
\end{tabular}
}
}
\end{center}
\end{table}

This empirical formula has been fitted to the measured
\pip\ and \pim\ production spectra in p--C, \pip--C and \pim--C reactions at
12~{\GeVc} reported here. As initial values for these
fits the parameters of the Sanford-Wang fit of the p--Al HARP analysis
at 12.9~{\GeVc} are taken from~\cite{ref:alPaper}.
The original Sanford-Wang parametrization has been proposed to describe
incoming proton data. 
We apply the same parametrization also to the \pip--C and \pim--C datasets. 

\begin{table}[tb]
\centering
\caption{
\label{SWpar_pis} 
Sanford-Wang parameters and errors obtained by fitting
the \pip--C and \pim--C datasets.
} \vspace{0.5 cm}
\begin{minipage}{\linewidth}
\renewcommand{\footnoterule}{}
\centering
\begin{tabular}{|c|c|c|c|c|c|c|}
\hline & \multicolumn{2}{|c|}{\pip--C} & \multicolumn{2}{|c|}{\pim--C}\\
\hline \multicolumn{1}{|c}{\bf Param} &
\multicolumn{1}{|c}{\pionm} & \multicolumn{1}{c|}{\pionp} & 
\multicolumn{1}{|c}{\pionm} & \multicolumn{1}{c|}{\pionp}\\

\hline  $c_1$ &  41.448
$\pm$  45.572  &   109.24$\pm$  114.73   &  156.49 $\pm$  56.132      &
78.963 $\pm$ 34.332  \\
\hline  $c_2$ & 1.8316 
$\pm$ 0.61113     &  1.2130 $\pm$ 0.57892   &   1.1673  $\pm$  0.17019      &
1.3561 $\pm$ 0.21690   \\
\hline  $c_3$ &  0. (fixed) 
     &  0. (fixed)  &  0. (fixed)      &
7.1493 $\pm$ 28.024   \\
\hline  $c_4$=$c_5$ &  ---  & --- & --- &
5.1098 $\pm$ 7.2508  \\
\hline  $c_6$ & 10.074
$\pm$ 1.8426     &  5.7823 $\pm$ 1.9875  &   5.6525 $\pm$ 0.54217      &
8.0965 $\pm$ 0.73121 \\
\hline  $c_7$ & 0.22877 
$\pm$ 0.098638     &  0.25667 $\pm$ 0.17396  &   0.19908  $\pm$ 0.06052      &
0.21960 $\pm$ 0.055566 \\
\hline  $c_8$ &  18.056 
$\pm$ 15.934    &  36.139 $\pm$ 25.437  &   30.368 $\pm$ 9.9403    &
25.561 $\pm$ 9.1022 \\
\hline  $\chi^2$/NDF &  37.4/31 & 18.5/31 &  133.6/31 & 136.7/29 \\

\hline
\end{tabular}
\end{minipage}
\end{table}

\begin{table}[htb]
\begin{center}
{
\caption{\label{tab:swpar_correlations_pis}
Correlation coefficients among the Sanford-Wang parameters, obtained
 by fitting the \pip--C and \pim--C datasets.
}
{
\begin{tabular}{|c r r r r r|}
\hline

\multicolumn{6}{|c|}{\pip--C $\to$ \pionm}\\
\hline
{\bf Parameter} & $c_1$  & $c_2$  & $c_6$  & $c_7$  & $c_8$ \\
\hline
$c_1$     &  1.000 &        &        &        &        \\
$c_2$     & -0.680 &  1.000 &        &        &        \\
$c_6$     & -0.592 &  0.891 &  1.000 &        &        \\
$c_7$     & -0.821 &  0.199 &  0.200 &  1.000 &        \\
$c_8$     & -0.445 & -0.134 & -0.093 &  0.819 & 1.000  \\
\hline

\multicolumn{6}{|c|}{\pip--C $\to$ \pionp}\\
\hline
{\bf Parameter} & $c_1$  & $c_2$ & $c_6$  & $c_7$  & $c_8$ \\
\hline
$c_1$     &  1.000 &        &        &        &        \\
$c_2$     & -0.753 &  1.000 &        &        &        \\
$c_6$     & -0.638 &  0.909 &  1.000 &        &        \\
$c_7$     & -0.804 &  0.263 &  0.205 &  1.000 &        \\
$c_8$     & -0.129 & -0.372 & -0.372 &  0.626 & 1.000  \\
\hline

\multicolumn{6}{|c|}{\pim--C $\to$ \pionm}\\
\hline
{\bf Parameter} & $c_1$  & $c_2$  & $c_6$  & $c_7$  & $c_8$ \\
\hline
$c_1$     &  1.000 &        &        &        &        \\
$c_2$     & -0.765 &  1.000 &        &        &        \\
$c_6$     & -0.489 &  0.796 &  1.000 &        &        \\
$c_7$     & -0.834 &  0.374 &  0.259 &  1.000 &        \\
$c_8$     & -0.240 & -0.218 & -0.240 &  0.611 & 1.000  \\
\hline

\end{tabular}

\begin{tabular}{|c r r r r r r r|}
\hline
\multicolumn{8}{|c|}{\pim--C $\to$ \pionp}\\
\hline
{\bf Parameter} & $c_1$  & $c_2$  & $c_3$  & $c_4=c_5$  & $c_6$  & $c_7$  & $c_8$ \\
\hline
$c_1$     &  1.000 &        &        &        &        &        &        \\
$c_2$     & -0.584 &  1.000 &        &        &        &        &        \\
$c_3$     & -0.024 & -0.250 &  1.000 &        &        &        &        \\
$c_4=c_5$ & -0.088 & -0.254 &  0.973 &  1.000 &        &        &        \\
$c_6$     & -0.545 &  0.668 & -0.018 &  0.097 &  1.000 &        &        \\
$c_7$     & -0.849 &  0.195 &  0.013 &  0.077 &  0.314 &  1.000 &        \\
$c_8$     & -0.429 & -0.168 & -0.024 &  0.000 & -0.116 &  0.753 & 1.000  \\
\hline
\end{tabular}
}
}
\end{center}
\end{table}

In the $\chi^2$ minimization procedure the full error matrix is
used. 
For these fits the Sanford-Wang parametrization 
has been integrated over momentum and angular
bin widths of the data. However, the results are nearly identical to
the fit results without integration over individual bins. 
Concerning the parameters estimation, the best-fit values of the
Sanford-Wang parameter set discussed above are reported in
Tables~\ref{SWpar_prot} and~\ref{SWpar_pis}, together with their errors. 
Since for some fits the $c_3$ parameter tends to zero, we decided to fix 
this parameter and to set it to zero. 
For these fits the $c_4$ and $c_5$ parameters are irrelevant 
(see Eq.~\ref{SWfunc}).
The correlation coefficients among the Sanford-Wang parameters
are shown in Tables~\ref{tab:swpar_correlations_prot} 
and~\ref{tab:swpar_correlations_pis}.
The fit parameter errors are estimated
by requiring $\Delta\chi^2\equiv \chi^2-\chi^2_{\hbox{\footnotesize {{min}}}} =
8.18$ (5.89), 
corresponding to the 68.27\% confidence level region for seven (five)
variable parameters.
Some parameters are strongly correlated resulting in large errors
of the extracted parameters.

The measurements for \pim\ and \pip\ in p--C,
\pip--C and \pim--C reactions are compared to the Sanford-Wang
parametrizations in 
Figs.~\ref{pC12GeV_pi_20_7_log},~\ref{pipC12GeV_pi_log_nrb}
and~\ref{pimC12GeV_004_log_nrb}, respectively.
One notes that 
the Sanford-Wang parametrization is not able to describe some of the
data spectral features especially at low and high momenta.
The goodness-of-fit of the Sanford-Wang parametrization hypothesis
can be assessed by considering $\chi^2$ per number of
degrees of freedom (NDF) given in Tables~\ref{SWpar_prot} and~\ref{SWpar_pis}. 
Especially for the \pim data one finds a high value of $\chi^2$.  
This may not be surprising since the parametrization was developed for
pion production by incoming protons rather than by incoming pions.
The \pim data with their high statistics are more likely to reveal
discrepancies than the \pip data which have much lower statistical
significance.

For tuning and modifying
models, often a parametrization of data like the Sanford-Wang
formula is used. This can be a suitable method to
interpolate between measured energy and phase space regions. However,
this method has some shortcomings. 
By construction, the reliability of parametrizations
for extrapolating to energy and phase space regions where no data are
available is limited (see~\cite{ref:christine_phd} for a more detailed discussion).

Detailed inspection of 
Figs.~\ref{pC12GeV_pi_20_7_log},~\ref{pipC12GeV_pi_log_nrb}
and~\ref{pimC12GeV_004_log_nrb} allows us to conclude that at high
momenta and in particular at large angles the parametrization does
not describe the data well enough. Especially for \pip\ momentum
spectra at angles larger than 0.18~{rad}, the Sanford-Wang fit
deviates considerably from the data and it should not be used in the
angular range above 0.18~{rad}.

\subsection{Comparison of p--C HARP data at 12~{GeV/c} with model predictions}
\label{sec:compare}

\begin{figure}[tbph]
\centering
\includegraphics[width =0.6\textwidth]{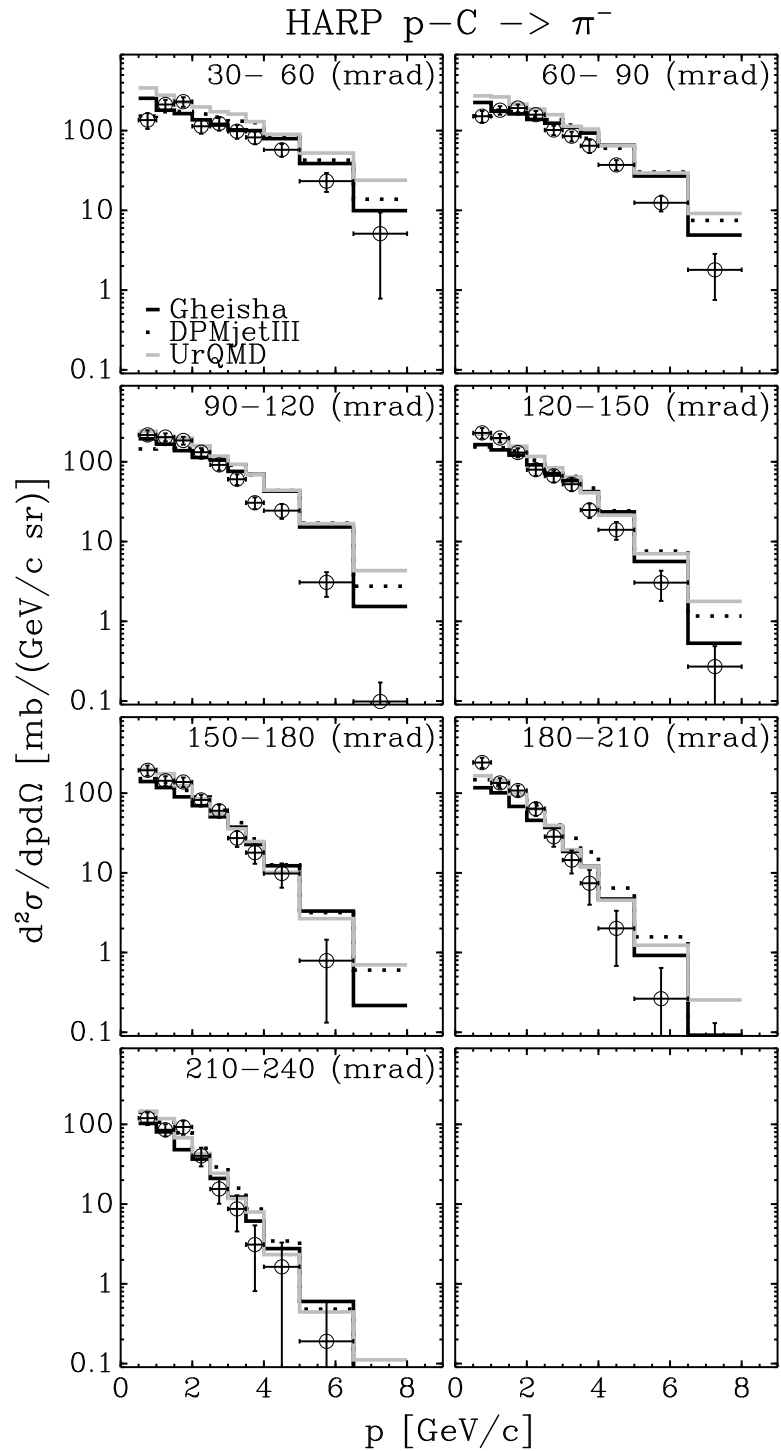}
\caption{
\label{pCpimModelslog}  
Comparison 
of the measured double-differential production cross-section of $\pi^{-}$
in p--C reactions at 12~{\GeVc} (points with error bars) 
with GHEISHA, UrQMD and DPMJET-III model predictions.
Seven panels show different angular bins from 30~{mrad}
to 240~{mrad} (the corresponding angular interval is printed on each panel).
}
\end{figure}

\begin{figure}[tbph]
\centering
\includegraphics[width =0.6\textwidth]{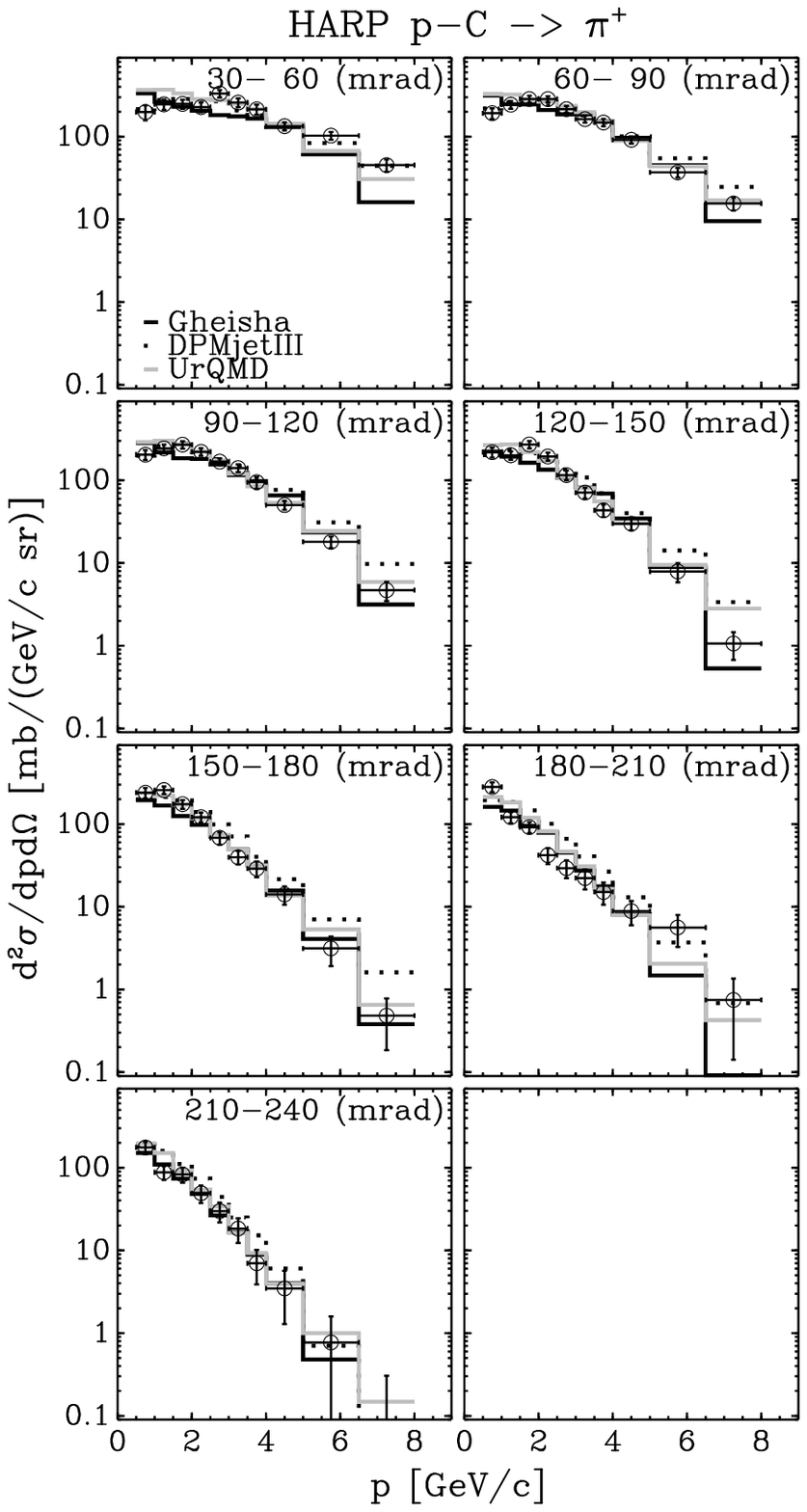}
\caption{
\label{pCpipModelslog}  
Comparison 
of the measured double-differential production cross-section of $\pi^{+}$
in p--C reactions at 12~{\GeVc} (points with error bars)
with GHEISHA, UrQMD and DPMJET-III model predictions.
Seven panels show different angular bins from 30~{mrad}
to 240~{mrad} (the corresponding angular interval is printed on each panel).
}
\end{figure}

A comparison of \pim\ and \pip\ production in p--C
reactions at 12~{\GeVc} with different model predictions is shown in
Figs.~\ref{pCpimModelslog} and~\ref{pCpipModelslog}. 
The three hadronic interaction models used for this comparison 
are GHEISHA~\cite{Fesefeldt85a},
UrQMD~\cite{Bleicher99a} and DPMJET-III~\cite{Roesler00a}. These are
the models typically used in air shower simulations. The GHEISHA and UrQMD are
implemented in CORSIKA~\cite{ref:CORSIKA} as low energy models (below 80~{\GeV}), whereas
the DPMJET-III is mostly used at higher energies but it is also able to make
predictions at lower energies. Comparing the predictions of these
models to the measured data, distinct discrepancies at low and high
momenta become visible. Especially the decrease of the cross-section
at very low momenta is not well described by the models. For \pip,
the prediction of the DPMJET-III seems relatively good, however,
this model
underestimates the \pim\ production at low momenta. At
large momenta the predictions of the three models are similar to each other, but
none of 
them
provides an acceptable description of the data.

\begin{figure}[tbph]
\centering
\includegraphics[width=0.6\textwidth]{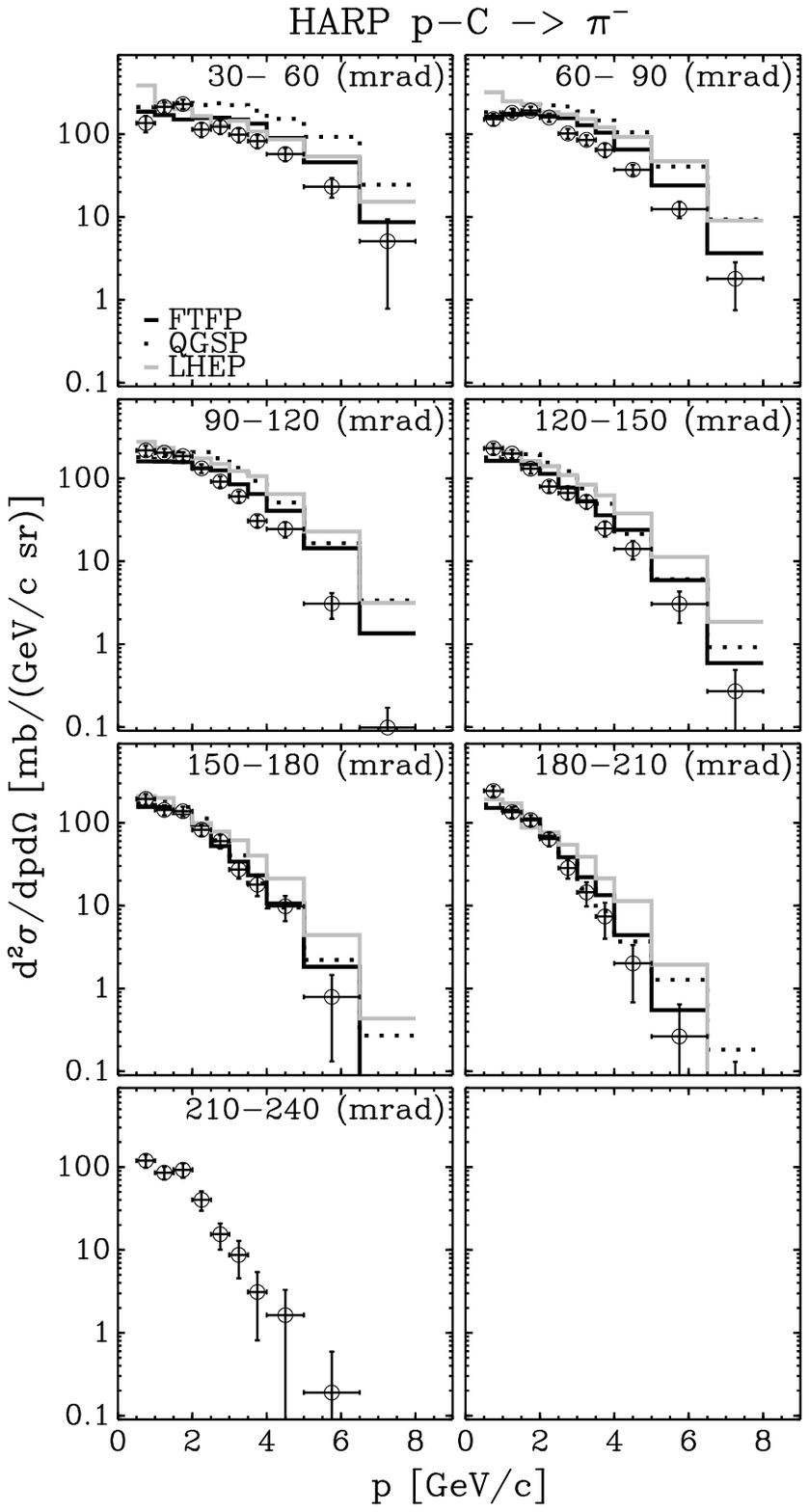}
\caption{
\label{pCpimModelsG4}  
Comparison 
of the measured double-differential production cross-section of $\pi^{-}$
in p--C reactions at 12~{\GeVc} (points with error bars)
with predictions of relevant GEANT4 models.
Seven panels show different angular bins from 30~{mrad}
to 240~{mrad} (the corresponding angular interval is printed on each panel).
}
\end{figure}

\begin{figure}[tbph]
\centering
\includegraphics[width=0.6\textwidth]{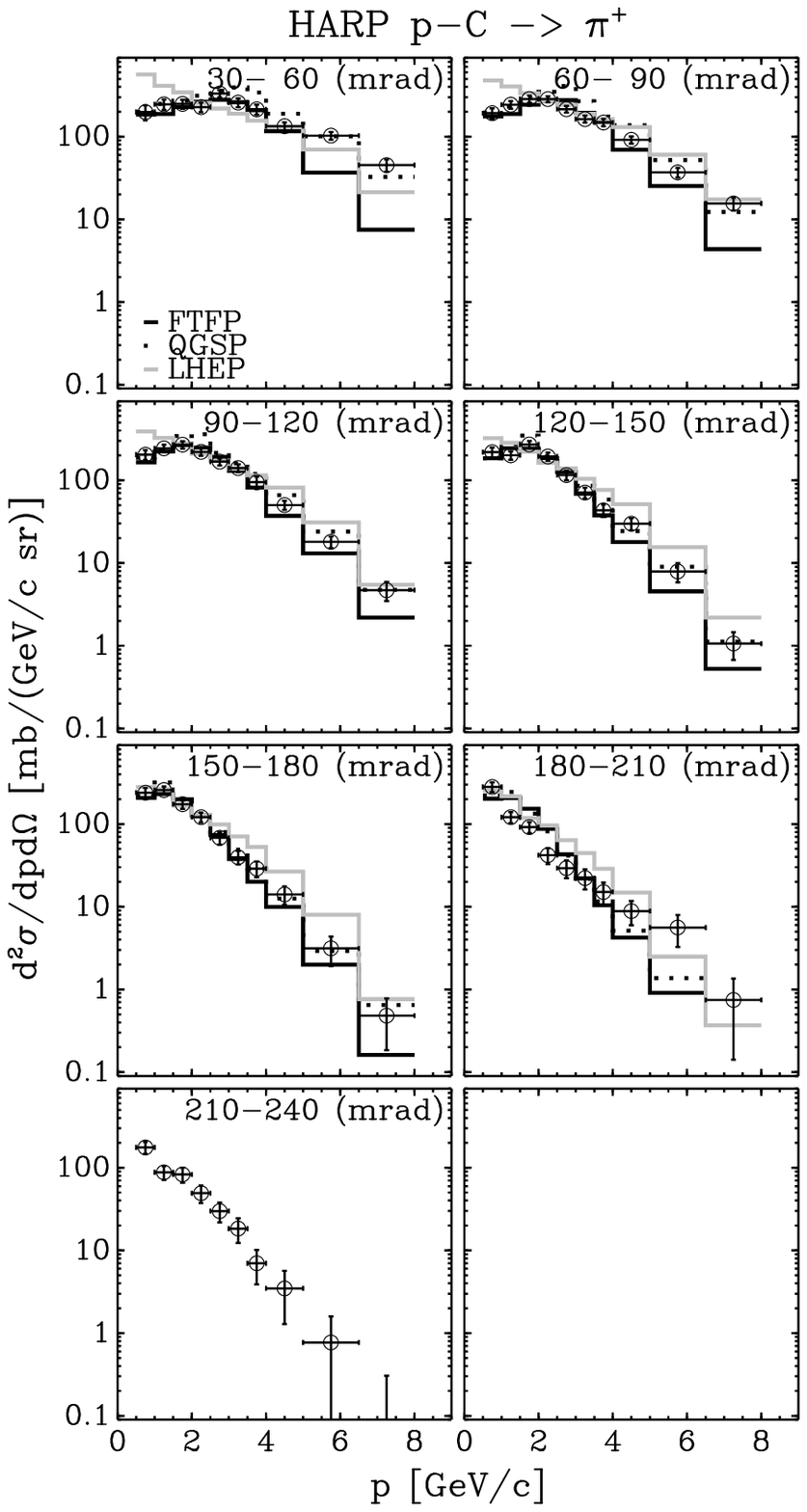}
\caption{
\label{pCpipModelsG4}  
Comparison 
of the measured double-differential production cross-section of $\pi^{+}$
in p--C reactions at 12~{\GeVc} (points with error bars)
with predictions of relevant GEANT4 models.
Seven panels show different angular bins from 30~{mrad}
to 240~{mrad} (the corresponding angular interval is printed on each panel).
}
\end{figure}

\begin{figure}[tbph]
\centering
\includegraphics[width=0.6\textwidth]{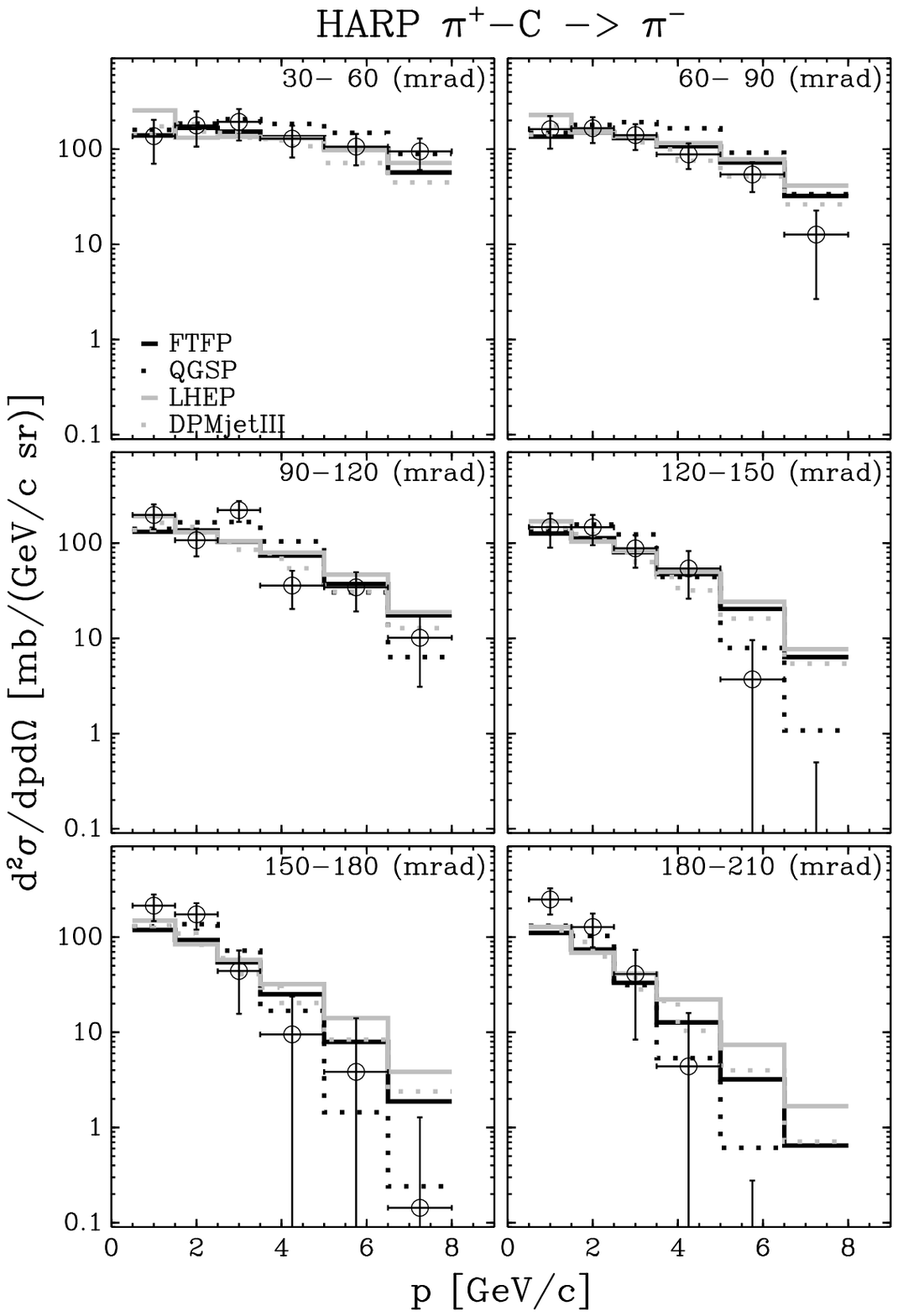}
\caption{
\label{pipCpimModelsG4}  
Comparison 
of the measured double-differential production cross-section of $\pi^{-}$
in $\pip$--C reactions at 12~{\GeVc} (points with error bars)
with predictions of relevant GEANT4 and DPMJET-III models.
Six panels show different angular bins from 30~{mrad}
to 210~{mrad} (the corresponding angular interval is printed on each panel).
}
\end{figure}

\begin{figure}[tbph]
\centering
\includegraphics[width=0.6\textwidth]{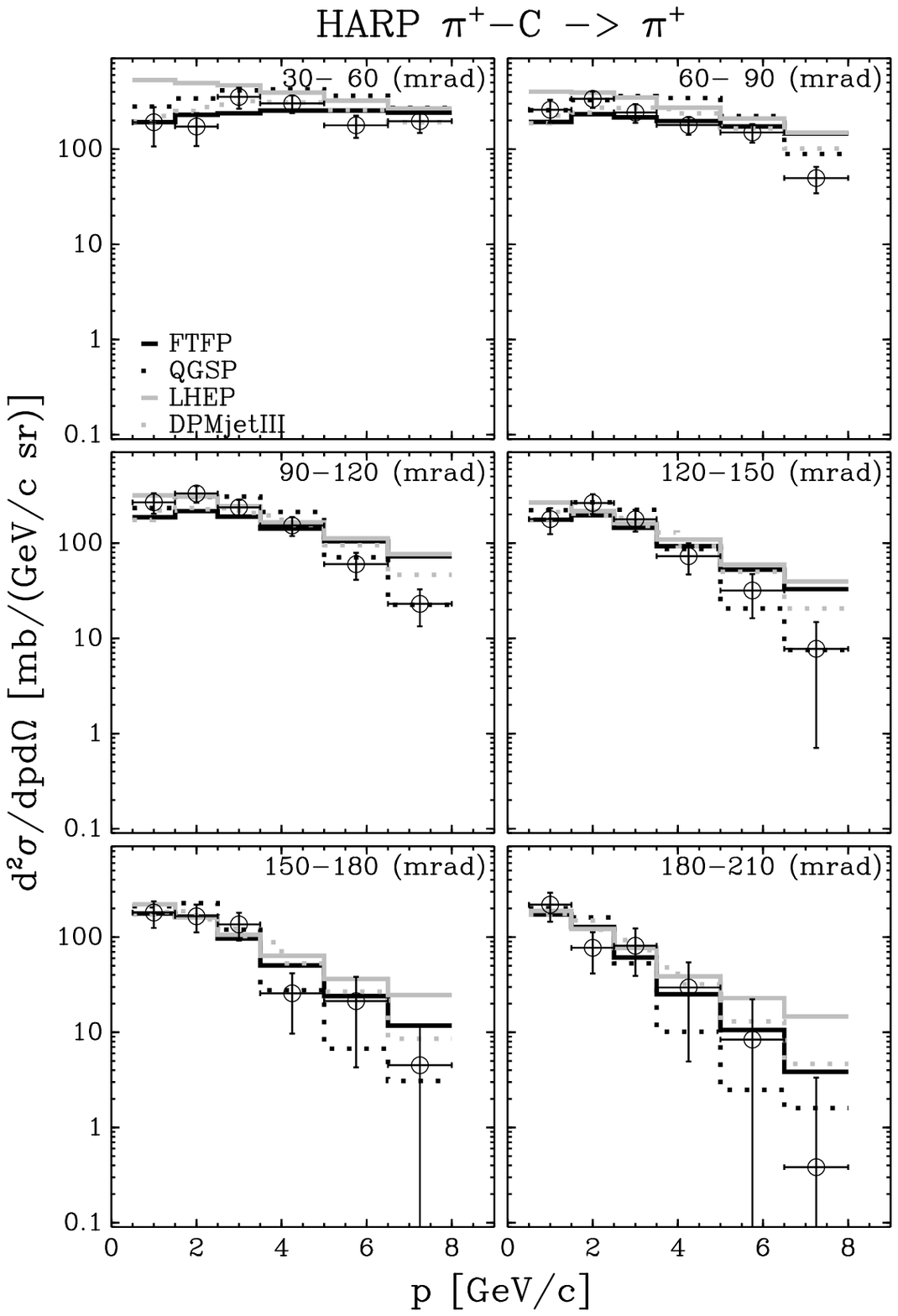}
\caption{
\label{pipCpipModelsG4}  
Comparison 
of the measured double-differential production cross-section of $\pi^{+}$
in $\pip$--C reactions at 12~{\GeVc} (points with error bars)
with predictions of relevant GEANT4 and DPMJET-III models.
Six panels show different angular bins from 30~{mrad}
to 210~{mrad} (the corresponding angular interval is printed on each panel).
}
\end{figure}

\begin{figure}[tbph]
\centering
\includegraphics[width=0.6\textwidth]{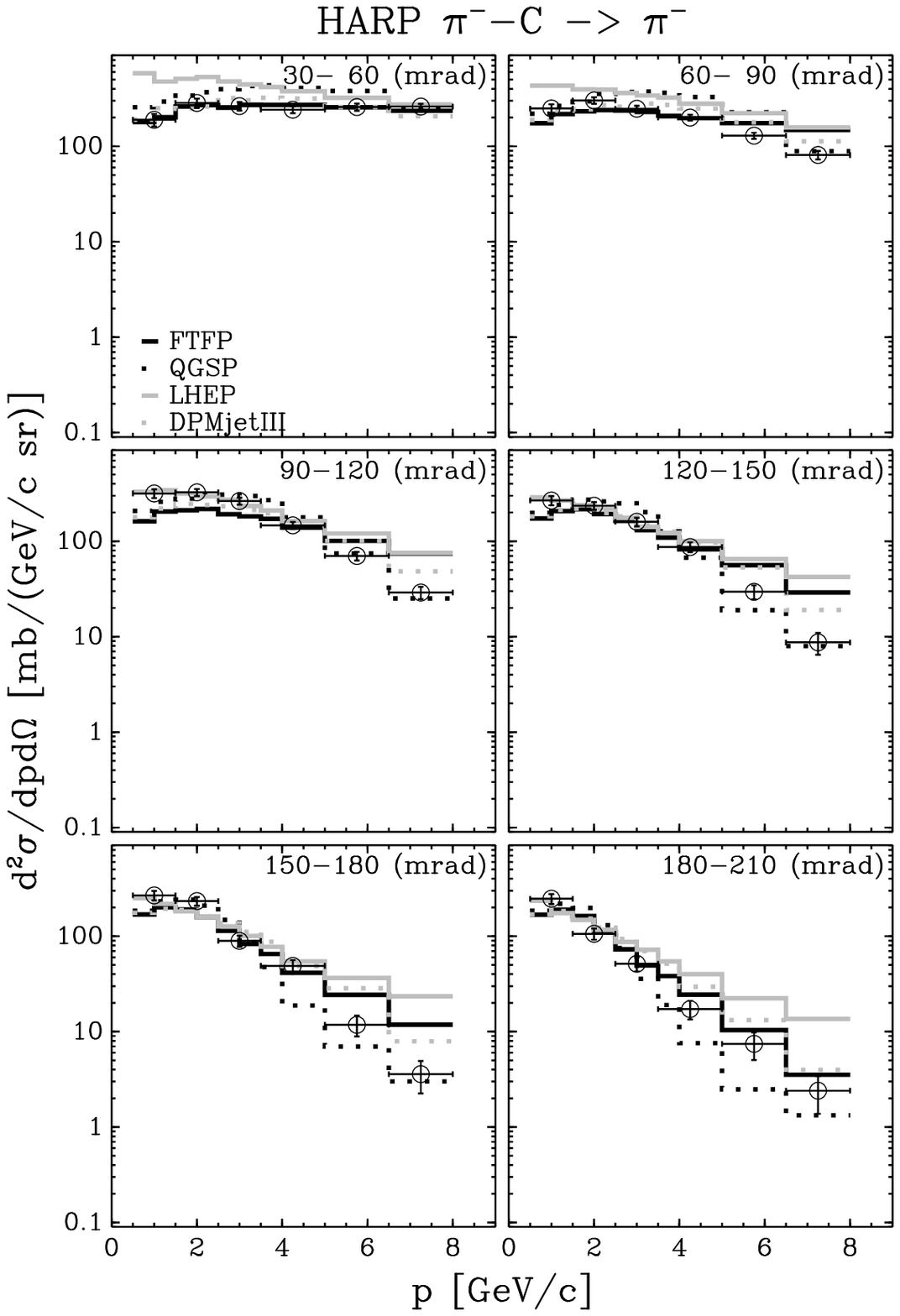}
\caption{
\label{pimCpimModelsG4}  
Comparison 
of the measured double-differential production cross-section of $\pi^{-}$
in $\pim$--C reactions at 12~{\GeVc} (points with error bars)
with predictions of relevant GEANT4 and DPMJET-III models.
Six panels show different angular bins from 30~{mrad}
to 210~{mrad} (the corresponding angular interval is printed on each panel).
}
\end{figure}

\begin{figure}[tbph]
\centering
\includegraphics[width=0.6\textwidth]{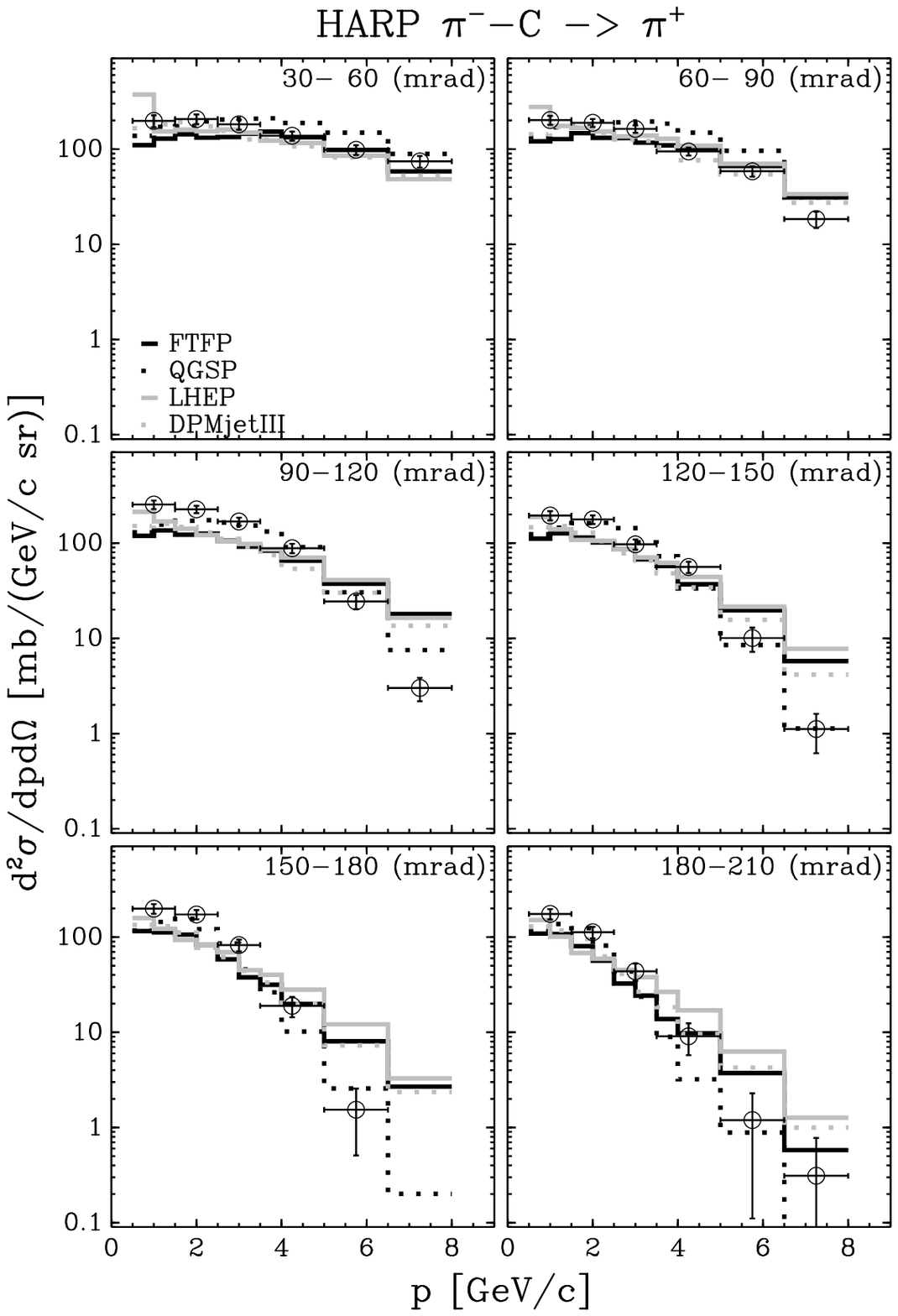}
\caption{
\label{pimCpipModelsG4}  
Comparison 
of the measured double-differential production cross-section of $\pi^{+}$
in $\pim$--C reactions at 12~{\GeVc} (points with error bars)
with predictions of relevant GEANT4 and DPMJET-III models.
Six panels show different angular bins from 30~{mrad}
to 210~{mrad} (the corresponding angular interval is printed on each panel).
}
\end{figure}

We have also compared our measurements with predictions 
of GEANT4~\cite{ref:geant4} models 
relevant in the energy domain studied here
(FTFP~\cite{FolgerWellisch}, QGSP~\cite{FolgerWellisch,WrightEtAl1} 
and LHEP~\cite{ref:geant4,WrightEtAl2}). 
The corresponding plots are presented in Figs.~\ref{pCpimModelsG4} 
and~\ref{pCpipModelsG4} (for incoming protons), in Figs.~\ref{pipCpimModelsG4} 
and~\ref{pipCpipModelsG4} (for incoming \pip) and 
in Figs.~\ref{pimCpimModelsG4} and~\ref{pimCpipModelsG4} (for incoming \pim). 
From these plots one can conclude that the predictions of FTFP and QGSP models 
are closer to the HARP data compared to the LHEP model.
For the \pim and \pip data the DPMJET-III model is shown in the same
figure.
The predictions of the latter model are very close to those of the FTFP model.

We have made a $\chi^2$ comparison between the HARP
data and all the models shown here. 
The full HARP error matrix has been used, and MC statistical errors (small but
non-negligible) have been also taken into account. 
The conclusions of this study are given below. None of the models
describe our data accurately.  However, in general these models tend to
describe the $\pip$ production more correctly than $\pim$ production for all
three incoming particle types. Different models are preferable, depending
on projectile type and on the charge of the pion produced. In particular, 
\begin{itemize}
\item for proton projectiles and $\pip$ production, UrQMD, FTFP and
GHEISHA give the best results; 
\item for proton projectiles and $\pim$ production, FTFP is preferable;
\item for $\pip$ projectiles and $\pip$ production and for $\pim$
projectiles and $\pim$ production, DPMJET-III is best;
\item for $\pip$ projectiles and $\pim$ production and for $\pim$
projectiles and $\pip$ production, QGSP describes the data best.
\end{itemize}

\section{Summary and conclusions}\label{sec:conclusions}

The results reported in this article contribute to the 
precise calculations of atmospheric neutrino fluxes and to the
improvement of our understanding of extended air shower
simulations and hadronic interactions at low
energies. 
A detailed description of uncertainties in atmospheric neutrino flux
calculations due to hadron production can be found in e.g.~\cite{Barr:2006it}.

Simulations show that collisions of protons with a carbon target
are very similar to proton interactions with air. 
That is why these datasets can be used
for tuning models needed in astroparticle physics simulations.

In this paper we presented measurements of the double-differential 
production cross-section of pions in the collisions of 12~\GeVc
protons and charged pions with a carbon 5\% 
 nuclear interaction length
target.  The data were
reported in bins of pion momentum and angle in the kinematic
range from $0.5 \ \GeVc \leq p_{\pi} < 8$~\GeVc and 
$0.030 \ \rad \leq \theta_{\pi} < 0.240 \ \rad$.   
A 
detailed
error analysis has been performed yielding 
integral errors (statistical + systematic) 
of 6.1\% and 7.0\% for \pip and \pim in p-C interactions 
(10.2\% and 8.5\% for \pip and \pim in \pionp-C interactions;
6.5\% and 8.2\% for \pip and \pim in \pionm-C interactions)
and an overall normalization error of 2\% for the proton beam and 3\%
for the pion beams.  

We should stress that the HARP incoming charged pion data are the first
precision measurements 
in this kinematic region.  

To check the reliability of hadronic interaction models which are used
for air shower simulations, 
the HARP measurements have been compared to predictions of
these models. 
Our conclusion is that none of the models is able to describe satisfactorily
and in detail the measured spectra. Discrepancies are found especially
at low and high momenta.

Several models rely on parametrizations of existing accelerator data. 
Therefore a
Sanford-Wang parametrization is given for all measured spectra.
The parametrization is, however, not a good description of the data
in the full phase space region.
From the comparison of the Sanford-Wang fits with model predictions 
we can conclude that such parametrizations have to be used with caution, 
especially if these parametrizations are extrapolated to regions where no
data are available.

\section{Acknowledgments}

We gratefully acknowledge the help and support of the PS beam staff
and of the numerous technical collaborators who contributed to the
detector design, construction, commissioning and operation.  
In particular, we would like to thank
G.~Barichello,
R.~Brocard,
K.~Burin,
V.~Carassiti,
F.~Chignoli,
D.~Conventi,
G.~Decreuse,
M.~Delattre,
C.~Detraz,  
A.~Domeniconi,
M.~Dwuznik,   
F.~Evangelisti,
B.~Friend,
A.~Iaciofano,
I.~Krasin, 
D.~Lacroix,
J.-C.~Legrand,
M.~Lobello, 
M.~Lollo,
J.~Loquet,
F.~Marinilli,
J.~Mulon,
L.~Musa,
R.~Nicholson,
A.~Pepato,
P.~Petev, 
X.~Pons,
I.~Rusinov,
M.~Scandurra,
E.~Usenko,
and
R.~van der Vlugt,
for their support in the construction of the detector.
The collaboration acknowledges the major contributions and advice of
M.~Baldo-Ceolin, 
L.~Linssen, 
M.T.~Muciaccia and A. Pullia
during the construction of the experiment.
The collaboration is indebted to 
V.~Ableev,
F.~Bergsma,
P.~Binko,
E.~Boter,
M.~Calvi, 
C.~Cavion, 
A.~Chukanov,  
M.~Doucet,
D.~D\"{u}llmann,
V.~Ermilova, 
W.~Flegel,
Y.~Hayato,
A.~Ichikawa,
A.~Ivanchenko,
O.~Klimov,
T.~Kobayashi,
D.~Kustov, 
M.~Laveder, 
M.~Mass,
H.~Meinhard,
A.~Menegolli, 
T.~Nakaya,
K.~Nishikawa,
M.~Pasquali,
M.~Placentino,
V.~Serdiouk,
S.~Simone,
S.~Troquereau,
S.~Ueda and A.~Valassi
for their contributions to the experiment.

We acknowledge the contributions of 
V.~Ammosov,
G.~Chelkov,
D.~Dedovich,
F.~Dydak,
M.~Gostkin,
A.~Guskov,
D.~Khartchenko,
V.~Koreshev,
Z.~Kroumchtein,
I.~Nefedov,
A.~Semak,
J.~Wotschack,
V.~Zaets and
A.~Zhemchugov
to the work described in this paper.

 The experiment was made possible by grants from
the Institut Interuniversitaire des Sciences Nucl\'eair\-es and the
Interuniversitair Instituut voor Kernwetenschappen (Belgium), 
Ministerio de Educacion y Ciencia, Grant FPA2003-06921-c02-02 and
Generalitat Valenciana, grant GV00-054-1,
CERN (Geneva, Switzerland), 
the German Bundesministerium f\"ur Bildung und Forschung (Germany), 
the Istituto Na\-zio\-na\-le di Fisica Nucleare (Italy), 
INR RAS (Moscow) and the Particle Physics and Astronomy Research Council (UK).
We gratefully acknowledge their support.

\clearpage

\begin{appendix}

\section{Cross-section data}\label{app:data}

\begin{table}[!h]
  \caption{\label{tab:xsec_results_pC12_pip_pim_2}
    HARP results for the double-differential $\pi^+$ and $\pi^-$ production
    cross-section in the laboratory system,
    $d^2\sigma^{\pi}/(dpd\Omega)$, for p-C interactions at 12~\GeVc. 
    Each row refers to a
    different $(p_{\hbox{\small min}} \le p<p_{\hbox{\small max}},
    \theta_{\hbox{\small min}} \le \theta<\theta_{\hbox{\small max}})$ bin,
    where $p$ and $\theta$ are the pion momentum and polar angle, respectively.
    The central value as well as the square-root of the diagonal elements
    of the covariance matrix are given.}
\begin{center}
 \begin{tabular}{|c|c|c|c|rcr|rcr|} \hline
$\theta_{\hbox{\small min}}$ &
$\theta_{\hbox{\small max}}$ &
$p_{\hbox{\small min}}$ &
$p_{\hbox{\small max}}$ &
\multicolumn{3}{c|}{$d^2\sigma^{\pi^+}/(dpd\Omega)$} &
\multicolumn{3}{c|}{$d^2\sigma^{\pi^-}/(dpd\Omega)$} 
\\
(mrad) & (mrad) & (GeV/c) & (GeV/c) &
\multicolumn{3}{c|}{(mb/(GeV/c sr))} &
\multicolumn{3}{c|}{(mb/(GeV/c sr))}
\\ \hline
 30 &  60 & 0.5 & 1.0 & $198.5$ & $\pm$ & $40.8$ & $135.4$ & $\pm$ & $30.5$ \\
    &     & 1.0 & 1.5 & $245.8$ & $\pm$ & $35.0$ & $212.5$ & $\pm$ & $30.8$ \\
    &     & 1.5 & 2.0 & $248.2$ & $\pm$ & $31.1$ & $230.6$ & $\pm$ & $32.1$ \\
    &     & 2.0 & 2.5 & $227.9$ & $\pm$ & $31.0$ & $113.6$ & $\pm$ & $21.9$ \\
    &     & 2.5 & 3.0 & $331.6$ & $\pm$ & $34.2$ & $122.6$ & $\pm$ & $22.6$ \\
    &     & 3.0 & 3.5 & $258.2$ & $\pm$ & $31.4$ & $98.1$ & $\pm$ & $18.9$ \\
    &     & 3.5 & 4.0 & $214.1$ & $\pm$ & $30.5$ & $82.3$ & $\pm$ & $14.8$ \\
    &     & 4.0 & 5.0 & $133.5$ & $\pm$ & $15.1$ & $57.5$ & $\pm$ & $10.4$ \\
    &     & 5.0 & 6.5 & $102.6$ & $\pm$ & $11.0$ & $23.2$ & $\pm$ & $6.2$ \\
    &     & 6.5 & 8.0 & $45.2$ & $\pm$ & $7.8$ & $5.1$ & $\pm$ & $4.3$ \\ \hline

 60 &  90 & 0.5 & 1.0 & $191.7$ & $\pm$ & $29.1$ & $151.3$ & $\pm$ & $24.9$ \\
    &     & 1.0 & 1.5 & $243.2$ & $\pm$ & $25.4$ & $180.6$ & $\pm$ & $22.0$ \\
    &     & 1.5 & 2.0 & $284.9$ & $\pm$ & $27.9$ & $191.6$ & $\pm$ & $21.3$ \\
    &     & 2.0 & 2.5 & $284.4$ & $\pm$ & $24.3$ & $158.2$ & $\pm$ & $18.0$ \\
    &     & 2.5 & 3.0 & $214.9$ & $\pm$ & $19.8$ & $101.7$ & $\pm$ & $14.0$ \\
    &     & 3.0 & 3.5 & $163.1$ & $\pm$ & $15.8$ & $85.1$ & $\pm$ & $12.0$ \\
    &     & 3.5 & 4.0 & $148.4$ & $\pm$ & $15.2$ & $64.5$ & $\pm$ & $12.2$ \\
    &     & 4.0 & 5.0 & $91.4$ & $\pm$ & $9.3$ & $37.2$ & $\pm$ & $5.5$ \\
    &     & 5.0 & 6.5 & $36.9$ & $\pm$ & $5.0$ & $12.5$ & $\pm$ & $2.8$ \\
    &     & 6.5 & 8.0 & $15.6$ & $\pm$ & $2.7$ & $1.8$ & $\pm$ & $1.0$ \\ \hline

 90 &  120 & 0.5 & 1.0 & $204.0$ & $\pm$ & $27.8$ & $217.4$ & $\pm$ & $31.2$ \\
    &     & 1.0 & 1.5 & $243.7$ & $\pm$ & $26.2$ & $204.7$ & $\pm$ & $23.2$ \\
    &     & 1.5 & 2.0 & $269.4$ & $\pm$ & $27.7$ & $185.1$ & $\pm$ & $21.0$ \\
    &     & 2.0 & 2.5 & $221.3$ & $\pm$ & $23.4$ & $132.1$ & $\pm$ & $16.5$ \\
    &     & 2.5 & 3.0 & $168.0$ & $\pm$ & $17.0$ & $91.8$ & $\pm$ & $13.8$ \\
    &     & 3.0 & 3.5 & $140.5$ & $\pm$ & $15.2$ & $60.5$ & $\pm$ & $9.2$ \\
    &     & 3.5 & 4.0 & $94.8$ & $\pm$ & $15.6$ & $30.7$ & $\pm$ & $5.1$ \\
    &     & 4.0 & 5.0 & $50.2$ & $\pm$ & $6.3$ & $24.4$ & $\pm$ & $5.2$ \\
    &     & 5.0 & 6.5 & $18.0$ & $\pm$ & $2.9$ & $3.1$ & $\pm$ & $1.1$ \\
    &     & 6.5 & 8.0 & $4.7$ & $\pm$ & $1.2$ & $0.1$ & $\pm$ & $0.1$ \\ \hline
\end{tabular}
\end{center}
\end{table}

\begin{table}
\begin{center}
\begin{tabular}{|c|c|c|c|rcr|rcr|} \hline
$\theta_{\hbox{\small min}}$ &
$\theta_{\hbox{\small max}}$ &
$p_{\hbox{\small min}}$ &
$p_{\hbox{\small max}}$ &
\multicolumn{3}{c|}{$d^2\sigma^{\pi^+}/(dpd\Omega)$} &
\multicolumn{3}{c|}{$d^2\sigma^{\pi^-}/(dpd\Omega)$} 
\\
(mrad) & (mrad) & (GeV/c) & (GeV/c) &
\multicolumn{3}{c|}{(mb/(GeV/c sr))} &
\multicolumn{3}{c|}{(mb/(GeV/c sr))}
\\ \hline
 120 &  150 & 0.5 & 1.0 & $218.8$ & $\pm$ & $30.6$ & $230.5$ & $\pm$ & $34.5$ \\
    &     & 1.0 & 1.5 & $200.6$ & $\pm$ & $23.4$ & $198.9$ & $\pm$ & $23.7$ \\
    &     & 1.5 & 2.0 & $271.3$ & $\pm$ & $28.5$ & $130.7$ & $\pm$ & $17.4$ \\
    &     & 2.0 & 2.5 & $194.3$ & $\pm$ & $21.6$ & $79.7$ & $\pm$ & $12.7$ \\
    &     & 2.5 & 3.0 & $115.7$ & $\pm$ & $15.6$ & $66.7$ & $\pm$ & $11.3$ \\
    &     & 3.0 & 3.5 & $71.0$ & $\pm$ & $10.7$ & $52.5$ & $\pm$ & $9.6$ \\
    &     & 3.5 & 4.0 & $43.4$ & $\pm$ & $7.4$ & $24.9$ & $\pm$ & $5.2$ \\
    &     & 4.0 & 5.0 & $29.9$ & $\pm$ & $5.0$ & $14.0$ & $\pm$ & $3.5$ \\
    &     & 5.0 & 6.5 & $7.9$ & $\pm$ & $2.1$ & $3.1$ & $\pm$ & $1.3$ \\
    &     & 6.5 & 8.0 & $1.1$ & $\pm$ & $0.4$ & $0.3$ & $\pm$ & $0.2$ \\ \hline

 150 &  180 & 0.5 & 1.0 & $238.9$ & $\pm$ & $34.1$ & $193.4$ & $\pm$ & $28.9$ \\
    &     & 1.0 & 1.5 & $257.5$ & $\pm$ & $26.9$ & $142.8$ & $\pm$ & $20.0$ \\
    &     & 1.5 & 2.0 & $173.7$ & $\pm$ & $20.8$ & $137.6$ & $\pm$ & $19.3$ \\
    &     & 2.0 & 2.5 & $121.3$ & $\pm$ & $16.7$ & $82.1$ & $\pm$ & $13.1$ \\
    &     & 2.5 & 3.0 & $67.9$ & $\pm$ & $11.8$ & $60.2$ & $\pm$ & $11.2$ \\
    &     & 3.0 & 3.5 & $39.7$ & $\pm$ & $7.4$ & $27.3$ & $\pm$ & $6.2$ \\
    &     & 3.5 & 4.0 & $28.9$ & $\pm$ & $6.3$ & $17.9$ & $\pm$ & $5.0$ \\
    &     & 4.0 & 5.0 & $14.1$ & $\pm$ & $3.5$ & $9.8$ & $\pm$ & $3.3$ \\
    &     & 5.0 & 6.5 & $3.1$ & $\pm$ & $1.2$ & $0.8$ & $\pm$ & $0.7$ \\
    &     & 6.5 & 8.0 & $0.5$ & $\pm$ & $0.3$ & \multicolumn{3}{c|}{---} \\ \hline

 180 &  210 & 0.5 & 1.0 & $280.1$ & $\pm$ & $38.2$ & $242.0$ & $\pm$ & $35.1$ \\
    &     & 1.0 & 1.5 & $121.0$ & $\pm$ & $18.2$ & $134.0$ & $\pm$ & $19.8$ \\
    &     & 1.5 & 2.0 & $91.8$ & $\pm$ & $14.2$ & $107.6$ & $\pm$ & $16.8$ \\
    &     & 2.0 & 2.5 & $42.0$ & $\pm$ & $9.1$ & $63.7$ & $\pm$ & $11.9$ \\
    &     & 2.5 & 3.0 & $29.3$ & $\pm$ & $7.1$ & $28.4$ & $\pm$ & $7.2$ \\
    &     & 3.0 & 3.5 & $22.2$ & $\pm$ & $6.1$ & $14.4$ & $\pm$ & $4.6$ \\
    &     & 3.5 & 4.0 & $15.1$ & $\pm$ & $4.5$ & $7.4$ & $\pm$ & $3.4$ \\
    &     & 4.0 & 5.0 & $8.9$ & $\pm$ & $2.9$ & $2.0$ & $\pm$ & $1.3$ \\
    &     & 5.0 & 6.5 & $5.6$ & $\pm$ & $2.3$ & $0.3$ & $\pm$ & $0.4$ \\
    &     & 6.5 & 8.0 & $0.7$ & $\pm$ & $0.6$ & \multicolumn{3}{c|}{---} \\ \hline

 210 &  240 & 0.5 & 1.0 & $175.8$ & $\pm$ & $29.2$ & $119.4$ & $\pm$ & $21.3$ \\
    &     & 1.0 & 1.5 & $87.9$ & $\pm$ & $16.8$ & $85.4$ & $\pm$ & $14.9$ \\
    &     & 1.5 & 2.0 & $82.8$ & $\pm$ & $17.1$ & $92.6$ & $\pm$ & $18.4$ \\
    &     & 2.0 & 2.5 & $49.1$ & $\pm$ & $11.7$ & $40.3$ & $\pm$ & $10.6$ \\
    &     & 2.5 & 3.0 & $29.9$ & $\pm$ & $8.2$ & $15.5$ & $\pm$ & $5.4$ \\
    &     & 3.0 & 3.5 & $18.3$ & $\pm$ & $6.1$ & $8.7$ & $\pm$ & $4.2$ \\
    &     & 3.5 & 4.0 & $7.0$ & $\pm$ & $3.1$ & $3.1$ & $\pm$ & $2.3$ \\
    &     & 4.0 & 5.0 & $3.5$ & $\pm$ & $2.2$ & $1.6$ & $\pm$ & $1.6$ \\
    &     & 5.0 & 6.5 & $0.8$ & $\pm$ & $0.8$ & $0.2$ & $\pm$ & $0.4$ \\ 
    &     & 6.5 & 8.0 & $0.1$ & $\pm$ & $0.2$ & \multicolumn{3}{c|}{---} \\ \hline

\end{tabular}
\end{center}
\end{table}

\begin{table}
  \caption{\label{tab:xsec_results_pipC12_pip_pim_1}
    HARP results for the double-differential $\pi^+$ and $\pi^-$ production
    cross-section in the laboratory system,
    $d^2\sigma^{\pi}/(dpd\Omega)$, for \pip-C interactions at 12~\GeVc. 
    Each row refers to a
    different $(p_{\hbox{\small min}} \le p<p_{\hbox{\small max}},
    \theta_{\hbox{\small min}} \le \theta<\theta_{\hbox{\small max}})$ bin,
    where $p$ and $\theta$ are the pion momentum and polar angle, respectively.
    The central value as well as the square-root of the diagonal elements
    of the covariance matrix are given.}
\begin{center}
\begin{tabular}{|c|c|c|c|rcr|rcr|} \hline
$\theta_{\hbox{\small min}}$ &
$\theta_{\hbox{\small max}}$ &
$p_{\hbox{\small min}}$ &
$p_{\hbox{\small max}}$ &
\multicolumn{3}{c|}{$d^2\sigma^{\pi^+}/(dpd\Omega)$} &
\multicolumn{3}{c|}{$d^2\sigma^{\pi^-}/(dpd\Omega)$} 
\\
(mrad) & (mrad) & (GeV/c) & (GeV/c) &
\multicolumn{3}{c|}{(mb/(GeV/c sr))} &
\multicolumn{3}{c|}{(mb/(GeV/c sr))}
\\ \hline
30 & 60 & 0.50 & 1.50& 191.5 & $\pm$ &   85.0 & 136.7 & $\pm$ &   66.4\\ 
      &      & 1.50 & 2.50& 173.0 & $\pm$ &   65.1 & 177.6 & $\pm$ &   71.3\\ 
      &      & 2.50 & 3.50& 354.0 & $\pm$ &   88.4 & 193.3 & $\pm$ &   70.2\\ 
      &      & 3.50 & 5.00& 302.1 & $\pm$ &   63.6 & 129.0 & $\pm$ &   47.4\\ 
      &      & 5.00 & 6.50& 177.8 & $\pm$ &   46.9 & 106.1 & $\pm$ &   38.3\\ 
      &      & 6.50 & 8.00& 196.8 & $\pm$ &   48.5 &  94.8 & $\pm$ &   34.4\\ \hline
60 & 90 & 0.50 & 1.50& 259.1 & $\pm$ &   70.7 & 161.9 & $\pm$ &   60.5\\ 
      &      & 1.50 & 2.50& 337.1 & $\pm$ &   65.8 & 166.2 & $\pm$ &   50.2\\ 
      &      & 2.50 & 3.50& 243.0 & $\pm$ &   53.9 & 140.6 & $\pm$ &   42.4\\ 
      &      & 3.50 & 5.00& 179.4 & $\pm$ &   37.4 &  88.2 & $\pm$ &   26.4\\ 
      &      & 5.00 & 6.50& 149.7 & $\pm$ &   32.9 &  54.1 & $\pm$ &   18.8\\ 
      &      & 6.50 & 8.00&  49.7 & $\pm$ &   15.3 &  12.6 & $\pm$ &   10.0\\ \hline
90 & 120 & 0.50 & 1.50& 268.2 & $\pm$ &   64.9 & 197.6 & $\pm$ &   57.5\\ 
      &      & 1.50 & 2.50& 332.0 & $\pm$ &   65.3 & 107.3 & $\pm$ &   34.6\\ 
      &      & 2.50 & 3.50& 237.4 & $\pm$ &   47.4 & 222.3 & $\pm$ &   55.1\\ 
      &      & 3.50 & 5.00& 153.1 & $\pm$ &   34.5 &  35.8 & $\pm$ &   15.5\\ 
      &      & 5.00 & 6.50&  60.2 & $\pm$ &   18.9 &  34.4 & $\pm$ &   15.2\\ 
      &      & 6.50 & 8.00&  23.0 & $\pm$ &    9.7 &  10.1 & $\pm$ &    7.0\\ \hline
120 & 150 & 0.50 & 1.50& 178.9 & $\pm$ &   54.8 & 147.4 & $\pm$ &   57.8\\ 
      &      & 1.50 & 2.50& 264.2 & $\pm$ &   61.8 & 146.7 & $\pm$ &   51.6\\ 
      &      & 2.50 & 3.50& 178.2 & $\pm$ &   46.3 &  88.3 & $\pm$ &   33.0\\ 
      &      & 3.50 & 5.00&  73.0 & $\pm$ &   26.3 &  54.3 & $\pm$ &   28.1\\ 
      &      & 5.00 & 6.50&  31.8 & $\pm$ &   15.5 &   3.7 & $\pm$ &    5.9\\ 
      &      & 6.50 & 8.00&   7.8 & $\pm$ &    7.0 & \multicolumn{3}{c|}{---}\\ \hline
 150 & 180 & 0.50 & 1.50& 181.1 & $\pm$ &   56.3 & 213.8 & $\pm$ &   66.6\\ 
      &      & 1.50 & 2.50& 165.7 & $\pm$ &   53.6 & 173.5 & $\pm$ &   53.7\\ 
      &      & 2.50 & 3.50& 136.2 & $\pm$ &   44.0 &  44.0 & $\pm$ &   28.4\\ 
      &      & 3.50 & 5.00&  25.7 & $\pm$ &   16.0 &   9.5 & $\pm$ &   14.2\\ 
      &      & 5.00 & 6.50&  21.3 & $\pm$ &   17.0 &   3.8 & $\pm$ &   10.2\\ 
      &      & 6.50 & 8.00&   4.5 & $\pm$ &    7.3 & \multicolumn{3}{c|}{---}\\ \hline
 180 & 210 & 0.50 & 1.50& 219.0 & $\pm$ &   73.4 & 248.5 & $\pm$ &   76.1\\ 
      &      & 1.50 & 2.50&  77.1 & $\pm$ &   35.6 & 127.8 & $\pm$ &   49.5\\ 
      &      & 2.50 & 3.50&  81.2 & $\pm$ &   42.1 &  40.9 & $\pm$ &   32.6\\ 
      &      & 3.50 & 5.00&  29.6 & $\pm$ &   24.6 &   4.4 & $\pm$ &   11.6\\ 
      &      & 5.00 & 6.50&   8.4 & $\pm$ &   13.8 &   0.0 & $\pm$ &    0.3\\ 
      &      & 6.50 & 8.00&   0.4 & $\pm$ &    3.0 & \multicolumn{3}{c|}{---}\\ \hline

\end{tabular}
\end{center}
\end{table}

\begin{table}
  \caption{\label{tab:xsec_results_pimC12_pip_pim_1}
    HARP results for the double-differential $\pi^+$ and $\pi^-$ production
    cross-section in the laboratory system,
    $d^2\sigma^{\pi}/(dpd\Omega)$, for \pim-C interactions at 12~\GeVc. 
    Each row refers to a
    different $(p_{\hbox{\small min}} \le p<p_{\hbox{\small max}},
    \theta_{\hbox{\small min}} \le \theta<\theta_{\hbox{\small max}})$ bin,
    where $p$ and $\theta$ are the pion momentum and polar angle, respectively.
    The central value as well as the square-root of the diagonal elements
    of the covariance matrix are given.}
\begin{center}
\begin{tabular}{|c|c|c|c|rcr|rcr|} \hline
$\theta_{\hbox{\small min}}$ &
$\theta_{\hbox{\small max}}$ &
$p_{\hbox{\small min}}$ &
$p_{\hbox{\small max}}$ &
\multicolumn{3}{c|}{$d^2\sigma^{\pi^+}/(dpd\Omega)$} &
\multicolumn{3}{c|}{$d^2\sigma^{\pi^-}/(dpd\Omega)$} 
\\
(mrad) & (mrad) & (GeV/c) & (GeV/c) &
\multicolumn{3}{c|}{(mb/(GeV/c sr))} &
\multicolumn{3}{c|}{(mb/(GeV/c sr))}
\\ \hline

 30 & 60 & 0.50 & 1.50& 198.1 & $\pm$ &   28.7 & 189.6 & $\pm$ &   28.7\\ 
      &      & 1.50 & 2.50& 206.8 & $\pm$ &   24.2 & 284.9 & $\pm$ &   31.4\\ 
      &      & 2.50 & 3.50& 182.0 & $\pm$ &   22.2 & 263.8 & $\pm$ &   27.1\\ 
      &      & 3.50 & 5.00& 138.0 & $\pm$ &   15.3 & 242.0 & $\pm$ &   19.6\\ 
      &      & 5.00 & 6.50&  98.4 & $\pm$ &   11.4 & 257.7 & $\pm$ &   22.1\\ 
      &      & 6.50 & 8.00&  74.4 & $\pm$ &   10.4 & 260.9 & $\pm$ &   17.4\\ \hline
 60 & 90 & 0.50 & 1.50& 201.9 & $\pm$ &   21.7 & 249.0 & $\pm$ &   26.6\\ 
      &      & 1.50 & 2.50& 189.2 & $\pm$ &   18.1 & 302.4 & $\pm$ &   24.5\\ 
      &      & 2.50 & 3.50& 163.1 & $\pm$ &   14.6 & 247.5 & $\pm$ &   18.8\\ 
      &      & 3.50 & 5.00&  94.6 & $\pm$ &    9.1 & 200.3 & $\pm$ &   13.6\\ 
      &      & 5.00 & 6.50&  58.5 & $\pm$ &    7.1 & 129.2 & $\pm$ &    9.4\\ 
      &      & 6.50 & 8.00&  18.4 & $\pm$ &    3.6 &  81.1 & $\pm$ &    8.2\\ \hline
 90 & 120 & 0.50 & 1.50& 254.2 & $\pm$ &   26.1 & 317.1 & $\pm$ &   33.1\\ 
      &      & 1.50 & 2.50& 226.4 & $\pm$ &   20.4 & 325.5 & $\pm$ &   27.5\\ 
      &      & 2.50 & 3.50& 169.0 & $\pm$ &   16.0 & 263.9 & $\pm$ &   22.0\\ 
      &      & 3.50 & 5.00&  88.4 & $\pm$ &   10.0 & 146.9 & $\pm$ &   12.6\\ 
      &      & 5.00 & 6.50&  24.4 & $\pm$ &    4.1 &  70.1 & $\pm$ &    7.3\\ 
      &      & 6.50 & 8.00&   3.0 & $\pm$ &    0.8 &  29.0 & $\pm$ &    4.3\\ \hline
 120 & 150 & 0.50 & 1.50& 195.2 & $\pm$ &   21.6 & 267.8 & $\pm$ &   29.9\\ 
      &      & 1.50 & 2.50& 177.4 & $\pm$ &   19.0 & 235.3 & $\pm$ &   22.1\\ 
      &      & 2.50 & 3.50&  97.1 & $\pm$ &   11.9 & 159.9 & $\pm$ &   16.6\\ 
      &      & 3.50 & 5.00&  56.2 & $\pm$ &    7.7 &  87.1 & $\pm$ &   10.3\\ 
      &      & 5.00 & 6.50&  10.1 & $\pm$ &    2.8 &  29.6 & $\pm$ &    5.0\\ 
      &      & 6.50 & 8.00&   1.1 & $\pm$ &    0.5 &   8.7 & $\pm$ &    2.3\\ \hline
 150 & 180 & 0.50 & 1.50& 198.9 & $\pm$ &   23.2 & 267.8 & $\pm$ &   30.7\\ 
      &      & 1.50 & 2.50& 173.1 & $\pm$ &   19.3 & 233.1 & $\pm$ &   23.7\\ 
      &      & 2.50 & 3.50&  82.6 & $\pm$ &   11.8 &  89.4 & $\pm$ &   12.0\\ 
      &      & 3.50 & 5.00&  19.0 & $\pm$ &    4.6 &  48.9 & $\pm$ &    7.4\\ 
      &      & 5.00 & 6.50&   1.5 & $\pm$ &    1.0 &  11.8 & $\pm$ &    2.9\\ 
      &      & 6.50 & 8.00&  \multicolumn{3}{c|}{---} &   3.6 & $\pm$ &    1.3\\ \hline
 180 & 210 & 0.50 & 1.50& 175.1 & $\pm$ &   22.0 & 246.9 & $\pm$ &   29.6\\ 
      &      & 1.50 & 2.50& 112.6 & $\pm$ &   15.5 & 106.1 & $\pm$ &   14.3\\ 
      &      & 2.50 & 3.50&  43.7 & $\pm$ &    9.0 &  51.4 & $\pm$ &    8.7\\ 
      &      & 3.50 & 5.00&   9.1 & $\pm$ &    3.3 &  17.2 & $\pm$ &    3.8\\ 
      &      & 5.00 & 6.50&   1.2 & $\pm$ &    1.1 &   7.4 & $\pm$ &    2.4\\ 
      &      & 6.50 & 8.00&  \multicolumn{3}{c|}{---} &   2.4 & $\pm$ &    1.0\\ \hline

\end{tabular}
\end{center}
\end{table}

\end{appendix}


\clearpage

\begin{thebibliography}{999}

\bibitem{ref:harp-prop}
  M.~G.~Catanesi {\it et al.}, HARP Collaboration,
    ``Proposal to study hadron production
    for the neutrino factory and for the atmospheric
    neutrino flux'',
    CERN-SPSC/99-35 (1999).

\bibitem{ref:harpTech}
  M.~G.~Catanesi {\it et al.} [HARP Collaboration], 
 ``The HARP Detector at the CERN PS'',
  Nucl.\ Instrum.\ Meth.\ A {\bf 571} (2007) 527.

\bibitem{ref:nufact}
	M. Apollonio {\it et al.}, 
	``Oscillation Physics with a Neutrino Factory'',
	{\rm CERN TH2002-208}, [arXiv:hep-ph/0210192];\\
	A.~Baldini {\it et al.},  BENE Steering Group,
	CERN-2006-005;\\
	A.~Blondel {\it et al.},
	CERN-2004-002, ECFA/04/230.

\bibitem{ref:atm_nu_flux}
  M.~Honda, T.~Kajita, K.~Kasahara and S.~Midorikawa,
  Phys.\ Rev.\ D {\bf 70} (2004) 043008; 
  Phys.\ Rev.\  D {\bf 75} (2007) 043005
  [arXiv:astro-ph/0611201]. \\
  M.~Honda, T.~Kajita, K.~Kasahara, S.~Midorikawa and T.~Sanuki,
  Phys.\ Rev.\  D {\bf 75} (2007) 043006
  [arXiv:astro-ph/0611418]. \\
  G.~D.~Barr, T.~K.~Gaisser, P.~Lipari, S.~Robbins and T.~Stanev,
  Phys.\ Rev.\ D {\bf 70} (2004) 023006. \\
  G.~Battistoni, A.~Ferrari, T.~Montaruli and P.~R.~Sala,
  [arXiv:hep-ph/0305208] \\
  G.~Battistoni, A.~Ferrari, T.~Montaruli and P.~R.~Sala,
  Astropart.\ Phys.\  {\bf 19} (2003) 269
  [Erratum-ibid.\  {\bf 19} (2003) 291] \\
  G.~Battistoni, A.~Ferrari, P.~Lipari, T.~Montaruli, P.~R.~Sala and T.~Rancati,  
  Astropart.\ Phys.\  {\bf 12} (2000) 315

\bibitem{ref:k2k}
 E.~Aliu {\it et al.}  [K2K Collaboration],
 ``Evidence for muon neutrino oscillation in an accelerator-based
 experiment,''
 Phys.\ Rev.\ Lett.\  {\bf 94} (2005) 081802
 [arXiv:hep-ex/0411038].

\bibitem{ref:k2kfinal}
  M.~H.~Ahn {\it et al.}  [K2K Collaboration],
  ``Measurement of neutrino oscillation by the K2K experiment'',
  Phys.\ Rev.\ D {\bf 74} (2006) 072003
  [arXiv:hep-ex/0606032].


\bibitem{ref:miniboone}
  E.~Church {\it et al.}  [BooNe Collaboration],
  ``A proposal for an experiment to measure muon-neutrino $\to$
  electron-neutrino oscillations and muon-neutrino disappearance at the
  Fermilab Booster: BooNE'',  FERMILAB-PROPOSAL-0898. \\
  A.~A.~Aguilar-Arevalo {\it et al.}  [MiniBooNE Collaboration],
  arXiv:0704.1500 [hep-ex].

\bibitem{ref:sciboone}
  A.~A.~Aguilar-Arevalo {\it et al.}  [SciBooNE Collaboration],
  ``Bringing the SciBar detector to the Booster neutrino beam'',
  [arXiv:hep-ex/0601022].

\bibitem{ref:alPaper}
  M.~G.~Catanesi {\it et al.}  [HARP Collaboration],
  ``Measurement of the production cross-section of positive pions in p-Al
  collisions at 12.9~GeV/c'',
  Nucl.\ Phys.\ B {\bf 732} (2006) 1
  [arXiv:hep-ex/0510039].

\bibitem{ref:bePaper}
  M.~G.~Catanesi {\it et al.},  [HARP Collaboration], 
	``Measurement of the production cross-section of positive pions
	in the collision of 8.9 GeV/c protons on beryllium'', 
	Eur. Phys. J. C {\bf 52} (2007) 29
	[arXiv:hep-ex/0702024].

\bibitem{ref:pidPaper}
  M.~G.~Catanesi {\it et al.}  [HARP Collaboration], 
  ``Particle identification algorithms for the HARP forward spectrometer'',
  Nucl.\ Instrum.\ Meth.\ A {\bf 572} (2007) 899.

\bibitem{ref:harp:tantalum}
     M.~G.~Catanesi {\it et al.}, [HARP Collaboration],
  ``Measurement of the Production of charged Pions by Protons on 
    a Tantalum Target'', 
    Eur. Phys. J. C {\bf 51} (2007) 787,
    arXiv:0706.1600 [hep-ex].

\bibitem{ref:harp:carboncoppertin}
     M.~G.~Catanesi {\it et al.}, [HARP Collaboration],
  ``Large-angle production of  charged pions by 3~\GeVc--12~\GeVc protons on  
           carbon, copper and tin  targets'',
  Eur.\ Phys.\ J.\  C {\bf 53} (2008) 177,
  arXiv:0709.3464 [hep-ex].

\bibitem{ref:harp:bealpb}
     M.~G.~Catanesi {\it et al.}, [HARP Collaboration],
  ``Large-angle production of  charged pions by 3~\GeVc--12.9~\GeVc protons on  
           beryllium, aluminium and lead targets'', arXiv:0709.3458 [hep-ex],
to be published in European Physical Journal C.

\bibitem{Baker61}
W.~F.~Baker {\it et~al.},
Phys.\ Rev.\ Lett.\ {\bf 7} (1961) 101.

\bibitem{Dekkers65}
D.~Dekkers {\it et~al.},
Phys.\ Rev.\ {\bf 137} (1965) B962.

\bibitem{Allaby70}
J.~V.~Allaby {\it et~al.},
CERN Yellow Report 70-12,
1970.

\bibitem{Cho71a}
Y.~Cho {\it et~al.},
Phys.\ Rev.\ D {\bf 4} (1971) 1967.

\bibitem{Eichten72}
T.~Eichten {\it et~al.},
Nucl.\ Phys.\ B {\bf 44} (1972) 333.

\bibitem{Antreasyan79}
D.~Antreasyan {\it et~al.},
Phys.\ Rev.\ D {\bf 19} N3 (1979) 764.

\bibitem{E910}
  I.~Chemakin {\it et al.}  [E910 Collaboration], 
  Phys. Rev. C {\bf 77} (2008) 015209, 
  arXiv:0707.2375 [nucl-ex].

\bibitem{Barton83}
D.~S.~Barton {\it et~al.},
Phys.\ Rev.\ D {\bf 27} (1983) 2580.

\bibitem{Alt:2006fr}
C.~Alt {\it et~al.}  [NA49 Collaboration],
Eur. Phys. J. C {\bf 49} (2007) 897 [arXiv:hep-ex/0606028].

\bibitem{MIPP}
  H.~Meyer  [MIPP Collaboration],
  J.\ Phys.\ Conf.\ Ser.\  {\bf 69} (2007) 012025. \\
  R.~Raja,
  Nucl.\ Instrum.\ Meth.\  A {\bf 553} (2005) 225
  [arXiv:hep-ex/0501005].

\bibitem{NA61}
  N.~Antoniou {\it et al.}  [NA61 Collaboration],
  ``Study of hadron production in hadron nucleus and nucleus nucleus
  collisions at the CERN SPS'', CERN-SPSC-P-330, CERN-SPSC-2006-043, 
  CERN-SPSC-2007-004, CERN-SPSC-2007-019.

\bibitem{ref:NOMAD_NIM_DC}
 M.~Anfreville {\it et al.},
 ``The drift chambers of the NOMAD experiment'',
 Nucl.\ Instrum.\ Meth.\ A {\bf 481} (2002) 339
 [arXiv:hep-ex/0104012].

\bibitem{ref:tofPaper}
  M.~Baldo-Ceolin {\it et al.},
  ``The Time-Of-Flight TOFW Detector Of The HARP Experiment: Construction And
  Performance'',
  Nucl.\ Instrum.\ Meth.\ A {\bf 532} (2004) 548.

\bibitem{ref:t9}
	L.~Durieu, A.~Mueller and M.~Martini,
	PAC-2001-TPAH142
	{\it Presented at IEEE Particle Accelerator Conference (PAC2001),
	Chicago, Illinois, 18-22 Jun 2001};\\
	L.~Durieu {\it et al.}, 
	Proceedings of PAC'97, Vancouver, (1997);  \\
	L.~Durieu, O.~Fernando, 
	CERN PS/PA Note 96-38.

\bibitem{ref:NA52} 
	K. Pretzl {\it et al.}, Invited talk at the ``International
	Symposium on Strangeness and Quark Matter'', Crete, (1999) 230.


\bibitem{ref:grossheim}
A.~Grossheim, 
``Particle production yields induced by multi-GeV protons on nuclear
targets'', 
Ph.D. thesis, University of Dortmund, Germany, 2003,
CERN-THESIS-2004-010. 



\bibitem{ref:geant4}
  S.~Agostinelli {\it et al.}  [GEANT4 Collaboration],
  ``GEANT4: A simulation toolkit'',
  Nucl.\ Instrum.\ Meth.\ A {\bf 506} (2003) 250.

\bibitem{ref:DAgostini} 
  G.~D'Agostini, DESY 94-099, ISSN 0418-9833, 1994. \\
  G.~D'Agostini,
  Nucl.\ Instrum.\ Meth.\ A {\bf 362} (1995) 487.



\bibitem{ref:christine_phd}
  Christine Meurer, ``Muon production in extensive air showers and
  fixed target accelerator data'', Ph.D. thesis, Karlsruhe, Germany, 2007,
  CERN-THESIS-2007-078.

\bibitem{ref:Blobel} 
  V.~Blobel and E.~Lohrmann,
  ''Statistische und numerische Methoden der Datenanalyse'', Stuttgart: Teubner,
  1998, ISBN 3-519-03243-0.


\bibitem{kliemant:2004}
M.~Kliemant, B.~Lungwitz, and M.~Gazdzicki,
Phys.\ Rev.\ C {\bf 69} (2004) 044903.


\bibitem{ref:CORSIKA}
  D.~Heck {\it et al.}, Report FZKA 6019 (1998)


\bibitem{SanfordWang1967}
  J.~R. Sanford and C.~L. Wang,
  ''Empirical formulas for particle production in p-Be collisions 
  between 10 and 35 BeV/c'',
  Brookhaven National Loboratory, AGS internal report (1967).


\bibitem{Fesefeldt85a}
H.~Fesefeldt,
report PITHA-85/02, RWTH Aachen,
1985.

\bibitem{Bleicher99a}
M.~Bleicher {\it et~al.},
J. Phys. G: Nucl. Part. Phys. 25 (1999) 1859.

\bibitem{Roesler00a}
S.~Roesler, R.~Engel, and J.~Ranft,
in Proc. of Int. Conf. on Advanced Monte Carlo for Radiation Physics, Particle
  Transport Simulation and Applications (MC 2000), Lisbon, Portugal, 23-26 Oct
  2000, A. Kling, F. Barao, M. Nakagawa, L. Tavora, P. Vaz eds.,
  Springer-Verlag Berlin, p. 1033-1038 (2001),
2000.

\bibitem{FolgerWellisch}
  G.~Folger and H.~P.~Wellisch, String parton models in Geant4,
CHEP'03 (La Jolla, California, USA, 24-28 March 2003); Preprints
CHEP-2003-MOMT007, e-Print physics/0306007

\bibitem{WrightEtAl1}
D.~H.~Wright, T.~Koi, G.~Folger, V.~Ivanchenko, M.~Kossov, N.~Starkov,
A.~Heikkinen and H.~P.~Wellisch, 2007 AIP Conf. Proc. 896 11

\bibitem{WrightEtAl2}
D.~H.~Wright, T.~Koi, G.~Folger, V.~Ivanchenko, M.~Kossov, N.~Starkov,
A.~Heikkinen and H.~P.~Wellisch, 2006 AIP Conf. Proc. 867 479

\bibitem{Barr:2006it}
  G.~D.~Barr, T.~K.~Gaisser, S.~Robbins and T.~Stanev,
  Phys.\ Rev.\  D {\bf 74} (2006) 094009
  [arXiv:astro-ph/0611266].

  
\end{thebibliography}
\end{document}